\documentclass[pre,aps ,superscriptaddress,showpacs]{revtex4-1}
\usepackage{natbib}
\usepackage{amsmath}
\usepackage{epsfig}
 \usepackage{color}
\begin{document}

\title{Heat and work in Markovian quantum master equations: concepts, fluctuation theorems, and computations  }
\author{Fei Liu}
\email[Email address: ]{feiliu@buaa.edu.cn}
\affiliation{School of Physics and Nuclear Energy Engineering, Beihang University, Beijing 100191, China}
\date{\today}

\begin{abstract}
{Markovian quantum master equations (MQMEs) were established nearly half a century ago. They have often been used in the study of irreversible thermodynamics. However, the previous results were mainly concerned about ensemble averages; the stochastic thermodynamics of these systems went unnoticed for a very long time. This situation remained unchanged until a variety of fluctuation theorems in classical and quantum regimes were found in the past two decades. In this paper, we systematically summarize the current understanding on the stochastic heat and work in MQMEs using two distinct strategies. One strategy is to treat the system and its surrounding heat bath as a closed quantum system, to suppose that the evolution of the composite system is unitary under a time-dependent total Hamiltonian and to define the heat and work as the changes in energy by applying two energy measurements scheme to the composite system. The other strategy is to unravel these MQMEs into random quantum jump trajectories (QJTs) and to define the stochastic heat and work along the individual trajectories. 
Many important physical concepts, mathematical techniques, and fluctuation theorems at different descriptive levels are given in as  detailed a manner as possible. We also use concrete models to illustrate these results. }

\end{abstract}
\pacs{05.70.Ln, 05.30.-d}
\maketitle

\section{Introduction}
\label{section1}
In the past two decades, there has been growing interest~\cite{Mukamel2003,DeRoeck2004,DeRoeck2007,Esposito2009,Horowitz2013,
Esposito2006,Crooks2008,Talkner2009,Subasi2012,Horowitz2012,Liu2012,Chetrite2012,Jaksic2013,Hekking2013,Horowitz2013,Leggio2013,
Liu2014,Liu2014a,Silaev2014,Suomela2014,Suomela2015,Suomela2016,Suomela2016a,Manzano2015,Gasparinetti2014,Cuetara2015,Alonso2016,
Gong2016,Strasberg2017,Pigeon2015,Pigeon2016,Garrahan2010,Hickey2012,nidari2014,Carrega2015,Carrega2016,Manzano2015,Manzano2016,
Schmidt2015,Chiara2015,Wang2017,Blattmann2017,Elouard2017,Perarnau-Llobet2017,Zhu2016,Wang2017,Smith2017} in the stochastic heat and work of nonequilibrium quantum processes. These studies include investigations of the definitions of these fundamental thermodynamic quantities, their statistical characteristics in general or specific physical models, computational methods, and experimental measurements and realizations. This research boom is driven mainly by two causes. One cause is the practical possibility of manipulating and controlling small quantum systems~\cite{Gemmer2005,Boixo2014,Kosloff2013,ShuomingAn2015,Rossnagel2016}. Energy, one of the most important physical quantities, and its transfer are always of great concern, while randomness is an intrinsic characteristic of quantum systems. The other motivation is from theoretical interests. Over almost the same period, a breakthrough in non-equilibrium physics was the discovery of a variety of fluctuation theorems (FTs)~\cite{Bochkov1977,Evans1993,Gallavotti1995,Jarzynski1997,Kurchan1998,
Lebowitz1999,Maes1999,Crooks2000,Hatano2001,Seifert2005,Kawai2007,Sagawa2010}. These formulas concerning heat, work, and total entropy production were proved to be exact even in far-from-equilibrium regions, whereas in near-equilibrium regions, they are usually reduced to the famous fluctuation-dissipation theorems~\cite{deGroot1964,Zubarev1995,KuboStatistical}. Because these remarkable results were usually obtained in classical physical systems, there has been  intensive theoretical interest to extend them into the quantum regime~\cite{Bochkov1977,Kurchan2000,Tasaki2000,
Yukawa2000,Allahverdyan2005,Talkner2007,Andrieux2008,Campisi2011,Batalhao2014,ShuomingAn2015,Jarzynski2015}.

Among these theoretical efforts, quantum systems that can be described by Markovian quantum master equations (MQMEs) have attracted considerable attention~\cite{Mukamel2003,DeRoeck2004,DeRoeck2007,Esposito2009,Horowitz2013,Zinidarifmmodeheckclseci2014,
Esposito2006,Crooks2008,Talkner2009,Horowitz2012,Liu2012,Chetrite2012,Jaksic2013,Hekking2013,Horowitz2013,Leggio2013,
Liu2014,Liu2014a,Silaev2014,Suomela2014,Suomela2015,Suomela2016,Suomela2016a,Manzano2015,Gasparinetti2014,Cuetara2015,Alonso2016,Gong2016,
Pigeon2015,Pigeon2016,Manzano2016,Wang2017,Elouard2017}. We believe that this is not coincidental. On the one hand, as the closest extension of Hamiltonian systems, MQMEs~\cite{Davies1974,Lindblad1976,Gorini1976} have solid mathematical and physical foundations~\cite{Breuer2002,Alicki2010,Rivas2012}. On the other hand, in the statistical physics community, there has long been a tradition of studying irreversible thermodynamics using the  MQMEs~\cite{Spohn1978,Spohn1978a,Davies1978,Alicki1979,Kohn2001,Boukobza2006,Kosloff2013,Alicki2017}. Mukamel~\cite{Mukamel2003} proved a quantum Jarzynski equality (JE) for specific MQMEs that can be transformed to the classic MEs~\cite{Jarzynski1997a,Crooks2000,Seifert2011} in the adiabatic basis. He also presented a formal analogy between the JE and the dephasing of quantum coherences in spectral line shapes. Using an analogous idea, for general Lindblad-type QMEs, Esposito and Mukamel~\cite{Esposito2006} obtained several FTs by recasting these equations into the classical birth-death MEs in a time-dependent basis that diagonalizes the reduced density matrix of the system. These authors then unraveled these MEs by classical stochastic jump processes and called them quantum trajectories. Although Esposito and Mukamel defined heat and work at these quantum trajectories as done in the classical cases~\cite{Seifert2005,Seifert2011}, the relations of these trajectories to measurement scheme remain unclear~\cite{Teich1992,Wiseman1993}. Additionally, to establish a quantum JE, De Roeck and Maes defined the quantum history of a subsystem (they called the subsystem and heat bath the total system)~\cite{DeRoeck2004}. A realization in their paper is a repeated discrete projective measurement on an open quantum system; between two successive measurements, the subsystem is weakly coupled to an infinite free reservoir and is subjected to an external time-dependent driving potential varying on the time scale of dissipation. After that, De Roeck and Maes~\cite{DeRoeck2006}, De Roeck~\cite{DeRoeck2007}, and Derezinski et al~\cite{Derezinski2008} systematically investigated the quantum extension of the Callavotti-Cohen FT for entropy production~\cite{Gallavotti1995}. Their results showed that this task could be realized either in microscopic Hamiltonian systems or in time-homogenous Lindblad QME with a stationary state. Importantly, the former was rigorously proved to converge to the latter in the weak-coupling limit. The important ingredients in defining the entropy production of the specific MQMEs are to unravel them into quantum jump trajectories (QJTs)~\cite{Srinivas1981,Plenio1998,Carmichael1993,Breuer2002,Gardiner2004,Wiseman2010,Ueda1990}, to give probability measures on these trajectories and to interpret the fluctuating heat current as energy quanta that are transferred between the subsystem and reservoirs. Note that the QJTs here are not the same as those unravelled from the classical MEs as Espositor and Mukamel used~\cite{Esposito2006}. Interestingly, Breuer~\cite{Breuer2003} independently defined the same heat current when he investigated the entropy production rate of the stochastic state vector dynamics. For Hamiltonian systems, heat was microscopically defined as energy eigenvalue differences between the reservoirs at the end and beginning of non-equilibrium processes. This is the famous two-energy-measurement (TEM) scheme defining stochastic thermodynamic quantities~\cite{Kurchan2000}. Talkner et al.~\cite{Talkner2009} proved that JE is always true for general open quantum systems if the system weakly interacts with its environment. In their paper, work was microscopically defined as energy eigenvalue differences between the composite system and its environment at the end and beginning of non-equilibrium processes. Their conclusion does not depend on whether the dynamics of the system is Markovian. Although MQMEs are typical open systems and fulfill the weakly coupling condition, since additional approximations are involved in these equations, whether the strong statement given by Talkner et al. remains true in these equations is not very clear. Crooks~\cite{Crooks2008a} discussed how to express the JE for open quantum systems whose dynamics can be described by general quantum dynamical semigroups~\cite{Breuer2002}. To concisely represent the measurements of the heat flow from environment to the systems, a Hermitian map superoperator was used. In his another paper, Crooks~\cite{Crooks2008} defined the time reversal of a quantum dynamical semigroup having a positive-definite invariant state. At arbitrary time the dynamical semigroup is a completely positive, trace-preserving (CPTP) quantum map, and the map admits a operator-sum representation in terms of a collection of Kraus operators. Because each Kraus operator represents an interaction with the environment that may be measured and recorded by an external observer, Crooks defined a quantum history by noting the observed sequence of Kraus operators. Under these two definitions, Crooks found that the probability of a quantum history and that of the time-reversed history are related by the heat exchanged with the environment. This very insightful paper was further generalized by Manzano et al.~\cite{Manzano2015} recently, and a very general FT was obtained.

The reader might note that the above studies were mainly concerned with the validity of various quantum FTs, and their considerations were usually abstract. This situation remained unchanged until Esposito et al.~\cite{Esposito2009} derived a generalized quantum master equation (GQME) that can concretely calculate the characteristic function (CF, or the generating function in their paper) of a heat or matter current in time-homogenous MQMEs. The heat or matter current therein was also defined by the TEM scheme on the reservoirs. Importantly, the statistics of these stochastic quantities were implied in the GQME and its solution. This key equation quickly became popular in the literature. For instance, Silaev et al.~\cite{Silaev2014} used the GQME to investigate the work and heat statistics in weakly driven open quantum systems~\cite{Rivas2012}. Gasparinetti et al~\cite{Gasparinetti2014} and Cuetara et al.~\cite{Cuetara2015} further extended the idea of Esposito et al. to  periodically driven MQMEs~\cite{Bluemel1991,Kohler1997,Breuer1997,Szczygielski2013}. A very analogous GQME was derived as well. Garrahan and his coauthors~\cite{Garrahan2010,Garrahan2011,Hickey2012,Hickey2013,Lesanovsky2013} applied the large-deviation method~\cite{Touchette2008} to study the ``thermodynamics" of the QJTs of Lindblad QMEs, e.g., the dynamical crossover and dynamical phase transitions. Their discussions were also based on the GQME. In one of their papers, Garrahan and Lesanovsky~\cite{Garrahan2010} noted that the GQME has been in the literature of quantum optics for a long time~\cite{Gardiner2004}. However, in contrast to what Esposito et. al~\cite{Esposito2009} did, the earlier version was based on the concept of QJTs and was obtained by finding a set of evolution equations for the reduced density matrix with given jump events~\cite{Mollow1975,Zoller1987}. Hence, their observation could be thought of as a concrete demonstration of the statement given by Derezinski et al.~\cite{Derezinski2008}. Horowitz~\cite{Horowitz2012} formulated heat and work for the QJTs of a continuously monitored forced harmonic oscillator coupled to a thermal reservoir. A detailed FT for QJTs was derived. In contrast to previous results using QJTs~\cite{DeRoeck2006,Garrahan2010}, in his model, the Hamiltonian is time dependent. In the same spirit, Hekking and Pelako~\cite{Hekking2013} investigated the statistics of work for the QJTs in a weakly driven two-level system coupled to a heat reservoir. They calculated the work distributions of the simple model and proved a quantum JE. Horowitz and Parrondom~\cite{Horowitz2013} and Horowitz and Sagawa~\cite{Horowitz2014} studied the quantum adiabatic and non-adiabatic entropy productions for the general Lindblad master equations. A quantum dual process analogous to the classical dual Markov process~\cite{Esposito2010} was defined by these authors. They found that these two entropy productions could be defined as ratios of the probabilities of QJTs in the original and dual processes. Chetrite and Mallick~\cite{Chetrite2012} obtained a quantum Jarzynski-Hatano-Sasa equality for a general Lindblad equation having an instantaneous stationary state solution. To prove this result, they introduced a modified Lindblad equation and utilized a quantum Feynman-Kac (FK) formula~\cite{Liu2012}. Their results were abstract and lacked probability interpretations. We~\cite{Liu2014,Liu2014a} noted that, in weakly driven and adiabatically driven MQMEs, the quantum FK formula was a book-keeping expression of the moment-generating function for work; it has a probability explanation if applying the measure of the QJTs. Additionally, in these two papers, in addition to proving a Bochkov-Kuzovlev equality (BKE) and JE, we presented two CF methods for calculating the exclusive and inclusive work distributions~\cite{Jarzynski2007,Campisi2011a}. These equations are very analogous to those utilized in classical stochastic systems~\cite{Imparato2005,Imparato2007}. We~\cite{Liu2016} studied whether the CF of the exclusive work is equivalent to the CF based on the TEM work defined on the composite system and heat reservoir~\cite{Talkner2007,Campisi2011a}. This affirmative result further stimulated us~\cite{Liu2016a} to develop the CF methods of heat and work based on the QJT concept for several time-dependent MQMEs. Their connections with the CFs of heat and work based on the TEM scheme were established as well. Very recently, Suomela et al.~\cite{Suomela2016,Suomela2016a} developed a modified QJT model for systems interacting with a finite-size heat bath. This work was mainly motivated by a calorimetric measurement of work in driven open quantum system~\cite{Pekola2013}. Elouard et al.~\cite{Elouard2017} considered the possibility of defining quantum heat that is purely measurement induced. In addition to QJTs, their formalism also includes the quantum sate diffusion (QSD) unravelling of MQMEs. Finally, Gong et al.~\cite{Gong2016} considered these feedback effects on quantum fluctuation effects using the QJT concept.

After the above long sketch of the literature, we clearly see that two main strategies are emerging in the studies of the quantum stochastic thermodynamics (QST) of MQMEs. One strategy is based on the TEM scheme defining thermodynamic quantities on the composite system and environments~\cite{Kurchan2000,DeRoeck2006,Esposito2009,Campisi2011}; the basic tool of computations and statistical analysis is the GQME~\cite{Esposito2009,Silaev2014,Gasparinetti2014,Cuetara2015}. The other strategy is to utilize the fact that the MQMEs can be unraveled into QJTs~\cite{Srinivas1981,Ueda1990,Carmichael1993,Plenio1998,Gardiner2004,Breuer2002,Wiseman2010}; along each QJT, heat, work, and entropy production can be well defined~\cite{Breuer2003,DeRoeck2006,Horowitz2012,Hekking2013,Liu2014,Liu2014a}; the computational and statistical analysis method could be straightforward Monte-Carlo simulation~\cite{Horowitz2012,Hekking2013} or the CFs based on the QJT concept~\cite{Liu2014,Liu2014a,Liu2016a}. In this paper, we attempt to give a self-contained review of the QST using these two strategies. The concrete contents are as follows. (1) We systematically survey a variety of MQMEs that were often used in the study of quantum thermodynamics. Although the MQMEs have been criticized because many conditions or approximations are involved in their derivations, these equations indeed have elegant mathematical structures and are also easily accessible. (2) We provide a careful overview of the CFs for heat defined by the TEM scheme in these equations, and we check whether there is a unified GQME. In particular, we apply the same effort to the case of work. To our knowledge, few studies have examined this issue in a systematic manner. (3) We interpret QJTs in terms of the TEM scheme on the heat bath atoms. Many papers have applied this energy interpretation~\cite{DeRoeck2006,Horowitz2013,Hekking2013,Liu2014}, but few of them clearly explained their reasoning. Additionally, the concept of QJTs remains unfamiliar to many researchers who are concerned about quantum FTs, although there have been such concepts since as early as the 70s of the last century~\cite{Davies1969}; now, they are broadly applied in quantum optics~\cite{Gardiner2004,Wiseman2010}. To attract additional attention, a detailed explanation seems to be essential. (4) We systematically establish the CFs of heat and work defined at QJTs and utilize the result to prove several finite-time FTs. In contrast to previous work, where almost all formulas were restricted to the MQMEs whose QJT has a wave-vector description~\cite{DeRoeck2006,Horowitz2013,Hekking2013,Liu2014}, our current results are sufficiently general to account for the cases in which the concept of QJTs is only available in the density matrix spaces of the quantum systems.

Let us present the scope of this paper's discussion. First, we only  consider one thermal heat bath interacting with a quantum open system. The particles of the system are not allowed to be exchanged with the environment. Hence, we do not discuss the issues about quantum heat engines,  quantum refrigerators, or quantum transport here~\cite{Kosloff2013,Schaller2014,Vinjanampathy2016}. Second, we are only concerned with the finite-time statistics for heat and work rather than their long-time behaviors. For the latter, the interested reader is referred to the excellent paper by Espositor et al~\cite{Esposito2009} or the other relevant references~\cite{Zinidarifmmodeheckclseci2014,Gasparinetti2014,Cuetara2015,Pigeon2015,Pigeon2015a}. Finally, we focus on the concrete MQMEs rather than provide abstract discussions using the general CPTP quantum maps~\cite{Crooks2008,Manzano2015}.

The paper is organized as follows. In Sec.~\ref{section2}, we sketch the general mathematical structure of MQMEs. Then, four types of equations that have often been used in the literature are derived. We show that all types can be unified into a general formula. The reader who knows well about the MQMEs may skip this section and start with the next one. In Sec.~\ref{section3}, we introduce the definition of work based on the TEM scheme in closed quantum systems. The time evolution equation of the work characteristic operator (WCO) is established. In Sec.~\ref{section4}, we investigate the derivations of the time evolution equations of the heat characteristic operator (HCO) and WCO for individual MQMEs. We show that they have unified forms as well. In addition, we compare the means of the stochastic heat and work to the ensemble heat and work that are typically proposed and used in ensemble quantum thermodynamics (EQT). In Sec.~\ref{section5}, we utilize the evolution equations of the HCO and WCO to prove several integral and detailed FTs. In Sec.~\ref{section6}, we introduce the concept of QJTs using a repeated interaction model in detail. We explicitly apply the TEM scheme to heat bath atoms. Based on this concept, we define the stochastic heat and work at the QJTs in Sec.~\ref{section7}, and their corresponding COs and CFs are investigated in Sec.~\ref{section8}. In Sec.~\ref{section9}, the FTs at the QJT level are proved. Section~\ref{section10} concerns the total entropy production of open quantum systems. In Sec.~\ref{section11}, we use a periodically driven two-level system to numerically illustrate several results obtained in this paper. Section~\ref{section12} is the conclusion of this paper.

\section{Markovian quantum master equations}
\label{section2}
\subsection{General mathematical structure}
\label{section2A}
An open quantum system is a quantum system that interacts with another quantum system. The latter is usually called the environment. We denote the system and the environment by capital letters A and B, respectively. Because the environment B in which we are interested has an infinite number of degrees of freedom and is assumed to be in a thermal equilibrium state initially, we also call it the heat bath. The system A evolves under its own dynamics, the interaction with the heat bath B, and some external classical forces or fields; therefore, its dynamics is usually very complex. A conventional method of treating such a complicated situation is to regard the open quantum system A and heat bath B as a global system. Because the composite system is closed, we can describe its dynamics by a unitary evolution with a total Hamiltonian
\begin{eqnarray}
\label{compositeHamiltonian}
H(t)=H_A(t)+H_B+V,
\end{eqnarray}
where $H_A(t)$ and $H_B$ are the Hamiltonians of system A and heat bath B, respectively. The interaction Hamiltonian $V$ between these two subsystems is
\begin{eqnarray}
\label{interactionHamiltonianVdetailedexpression}
V=\sum_a A_a\otimes B_a,
\end{eqnarray}
where $A_a$ and $B_a$ are the Hermitian operators of these two respective subsystems. This interaction is supposed to be very weak to ensure the validity of MQMEs. In addition, throughout this paper we also suppose that the composite system has an uncorrelated initial density matrix,
\begin{eqnarray}
\label{initialdensitymatrix}
\rho(0)=\rho_A(0)\otimes\rho_B=\rho_A(0)\otimes \frac{e^{-\beta H_B}}{Z_B}.
\end{eqnarray}
Here, $\rho_A(0)$ and $\rho_B$ are the initial density matrix of system A and heat bath B, respectively. Because  heat bath B is in a thermal equilibrium state at time 0, $\rho_B$ has a canonical form: $\beta$ is the inverse temperature, $1/k_BT$, and $Z_B={\rm Tr}_B[e^{-\beta H_B}]$ is the partition function. Note that ${\rm Tr}_B$ is the trace taken over the heat bath. In the remainder of this paper, we always use the subscript to denote the degree of freedoms over which the trace is being taken. At a later time $t$, the density matrix of the composite system, $\rho(t)$, satisfies the Liouville-von Neumann equation
\begin{eqnarray}
\label{timeevolutionequationdensitymatrix}
\partial_t \rho(t)=-i[H(t),\rho(t)].
\end{eqnarray}
Planck's constant $\hbar$ has been set equal to 1 here. If one could solve the equation, the dynamics of the open quantum system A is then obtained by taking the trace over the degree of freedom of the heat bath, i.e.,
\begin{eqnarray}
\label{reduceddensitymatrix}
\rho_A(t)={\rm Tr}_B[\rho(t)].
\end{eqnarray}
Because this is a reduction procedure of the global density matrix, $\rho_A(t)$ is also called the reduced density matrix of the reduced system A.

The solution of Eq.~(\ref{timeevolutionequationdensitymatrix}) can always be written formally as
\begin{eqnarray}
\label{formalsolutionofdensitymatrix}
\rho(t) = U(t)\rho(0) U(t)^\dag,
\end{eqnarray}
by introducing the time evolution operator of the composite system,
\begin{eqnarray}
\label{timeevolutionoperator}
U(t)=T_\leftarrow e^{-i\int_0^t dsH(s)},
\end{eqnarray}
where the symbol $T_\leftarrow$ denotes the chronological time-ordering operator. Considering that the open system A and heat bath B are initially uncorrelated, as in Eq.~(\ref{initialdensitymatrix}), the reduced density matrix of  system A at time $t$ can also be expressed using a dynamical map ${\cal M}(t)$:
\begin{eqnarray}
\label{Krausrepresentation}
\rho_A(t)=\sum_{lk} K_{lk}(t,0) \rho_A(0) K_{lk}^\dag(t,0)\equiv{\cal M}(t,0)[\rho_A(0)],
\end{eqnarray}
where
\begin{eqnarray}
\label{Krausoperator}
K_{lk}(t,0)=\sqrt{p_k}\langle \chi_l|U(t)|\chi_k\rangle,
\end{eqnarray}
is called the Kraus operator. To obtain Eq.~(\ref{Krausrepresentation}), we have used a spectral decomposition of the density matrix of the heat bath:
\begin{eqnarray}
\label{densitymatrixheatbath}
\rho_B=\sum_l p_l|\chi_l\rangle\langle \chi_l|,
\end{eqnarray}
where $\chi_l$ and $|\chi_l\rangle$ are the eigenvalue and eigenvector of the Hamiltonian $H_B$, respectively, and
\begin{eqnarray}
\label{canonicaldistributionheatbath}
p_l=\frac{e^{-\beta \chi_l}}{Z_B}.
\end{eqnarray}
The dynamical map ${\cal M}(t)$ has three important properties: convex linearity, trace preservation, and complete positivity~\cite{Breuer2002,Alicki2006,Rivas2012}. The dynamic map at arbitrary time $t\ge 0$ forms a one-parameter family $\{{\cal M}(t)\}$. Although in principle this family describes the entire evolution of the open quantum system A, due to the complexity of the composite unitary evolution $U(t,0)$ for a general $H(t)$, it could be very involved. The simplest one-parameter family should be a Markovian family, namely,
\begin{eqnarray}
\label{compositionlaw}
{\cal M}(t_2,0)={\cal M}(t_2,t_1){\cal M}(t_1,0), \hspace{0.5cm} 0\le t_1\le t_2.
\end{eqnarray}
Such a family is also called quantum dynamical semigroups~\cite{Breuer2002,Alicki2006,Rivas2012}. Importantly, one can alternatively express the dynamical semigroup using a linear differential equation or a MQME:
\begin{eqnarray}
\label{formalQME}
\partial_t \rho_A(t)=\lim_{\epsilon\rightarrow 0} \frac{[{\cal M}(t+\epsilon,t)-1]}{\epsilon}\rho_A(t)={\cal L}_t \rho_A(t),
\end{eqnarray}
where the superoperator ${\cal L}_t$ is called the semigroup generator. Obviously, Eq.~(\ref{compositionlaw}) is the quantum analogue to the Chapman-Kolmogorov equation in classical Markovian processes~\cite{Gardiner1983}. In particular, for a finite $D$-dimensional semigroup, the most general form of the generator was shown to be~\cite{Rivas2012}
\begin{eqnarray}
\label{formalmasterequation1}
\partial_t\rho_A(t)&=&-i[H_A'(t),\rho_A(t)]+\sum_{a,b=1}^M \gamma_{ab}(t)\left[F_b\rho_A(t) F_a^\dag -\frac{1}{2}\left\{ F_a^\dag F_b,\rho_A(t)\right \}\right],
\end{eqnarray}
where $H_A'(t)$ is the effective Hamiltonian, which is not necessarily identical to the system's free Hamiltonian, $H_A(t)$;  $\{F_a,a=1,\cdots,M=D^2-1\}$ is a complete orthonormal basis with respect to the Hilbert-Schmidt inner product; $\langle F_a,F_b\rangle={\rm Tr}[F_a^\dag F_b]=\delta_{ab}$ in the space of operators for the open quantum system A; and the $M\times M$ matrix ${\cal A}(t)=\{ \gamma_{ab}(t) \}$ is Hermitian and positive semidefinite. We call the commutator in the above equation the coherent dynamics and call the summation term the dissipator. Because ${\cal A}(t)$ is positive, one may always introduce a new set of operators $\{E_a(t)\}$ to replace the set of operators $\{F_a\}$ and rewrite Eq.~(\ref{formalmasterequation1}) in the ``diagonal" form:
\begin{eqnarray}
\label{formalmasterequation2}
\partial_t\rho_A(t)&=&-i[H_A'(t),\rho_A(t)]+\sum_{a=1}^M r_{a}(t)\left[E_a(t)\rho_A(t) E_a^\dag(t) -\frac{1}{2}\left\{ E_a^\dag(t) E_a(t),\rho_A(t)\right \}\right],
\end{eqnarray}
where the non-negative coefficients $r_a(t)$ are the eigenvalues of ${\cal A}(t)$. Historically, Gorini, Kossakowski, and Sudarshan~\cite{Gorini1976} were credited with establishing Eq.~(\ref{formalmasterequation1}) for a time-homogenous quantum dynamic map, that is, ${\cal M}(t_1,t_2)$ depends only on the difference $t_2-t_1$. In this situation, the matrix ${\cal A}$ is time independent. Independently, also for the homogenous case, Lindblad~\cite{Lindblad1976} obtained Eq.~(\ref{formalmasterequation2}) for the infinite-dimensional semigroups with bounded generators~\cite{Alicki2006}. In the literature, Eqs.~(\ref{formalmasterequation1}) and~(\ref{formalmasterequation2}) are called  the Gorini-Kossakowski-Sudarshan-Lindblad (GKSL) equation.

\subsection{Microscopic models}
\label{section2B}
Eq.~(\ref{formalmasterequation1}), and its equivalent Eq.~(\ref{formalmasterequation2}), is only formally exact; its derivation does not depend on the concrete physical models. From the viewpoint of physics, it shall be more desirable if we could derive it from first principles, e.g., the microscopic expressions for $F_k$ and $\gamma_{ab}(t)$. Indeed, under the weak-coupling assumption, that is, the interaction term $V$ in Eq.~(\ref{compositeHamiltonian}) is so weak that the influence of the open quantum system A on the heat bath B is small, and under several other key approximations, including the Born-Makovian approximation and the rotating wave approximation, one can indeed obtain a variety of MQMEs having the form of Eq.~(\ref{formalmasterequation1}). In this paper, we will discuss four types of equations. The first type is the time-homegeneous MQMEs~\cite{Davies1974,Lindblad1975,Gorini1976,Rivas2012}, which we occasionally call the standard master equations  in this paper. These equations have mainly been applied to issues related to how a system relaxes into a thermal equilibrium state~\cite{Breuer2002,Alicki2010} or arrives at non-equilibrium steady state~\cite{DeRoeck2004,Derezinski2008,Esposito2009,Garrahan2010,Zinidarifmmodeheckclseci2014,Schaller2014}.
The second type has time-dependent coherent dynamics, but the dissipator is static~\cite{Geva1994,Hekking2013,Liu2014,Silaev2014}. These equations have often been used in quantum optics~\cite{Breuer2002,Carmichael1993,Wiseman2010}, e.g., a two-level atom interacting with a radiation field while being driven by a classical time-varying electric field~\cite{Mollow1975}. Their physical validity is ensured if the external driving field is so weak that its effect on the environment is negligible. The equations of the last two types fully depend on time, including the adiabatically (slowly) driven~\cite{Davies1978,Alicki1979,Albash2012,Horowitz2012,Liu2014a,Suomela2015} and periodically driven MQME~\cite{Bluemel1991,Kohler1997,Breuer1997,Szczygielski2013,Gasparinetti2014,Cuetara2015}. Although the applicable regions of these four types of master equations are very distinct~\cite{Alicki2006}, all of them are very analogue to Eq.~(\ref{formalmasterequation1}). In the remainder of this section, we will survey their derivations.

We first write Eq.~(\ref{timeevolutionequationdensitymatrix}) in the interaction picture:
\begin{eqnarray}
\label{timeevolutionequationopensystemorginteractionpicture}
\partial_t {\widetilde {\rho}}(t)&=&-i[\widetilde{V}(t), {\widetilde {\rho}(t)}]\equiv\widetilde{\cal V}(t) {\widetilde {\rho}(t)}.
\end{eqnarray}
These notations with tildes indicate that they are interaction picture operators with respect to the free Hamiltonians, namely, an arbitrary operator $O$,
\begin{eqnarray}
\label{interactionpicture}
\widetilde{O}(t)=U_0^\dag(t) O U_0(t),
\end{eqnarray}
where the time evolution operator for the free Hamiltonians is
\begin{eqnarray}
\label{freetimeevolutionoperator}
U_0(t)=U_A(t)\otimes U_B(t)={T}_\leftarrow e^{-i\int_0^t ds H_A(s)}\otimes e^{-it H_B}.
\end{eqnarray}
The explicit expression of the interaction picture operator for ${\widetilde V}(t)$ is
\begin{eqnarray}
\label{interactionHamiltonianinteractionpicture}
\sum_a {\widetilde A}_a(t) \otimes {\widetilde B}_a(t)=\sum_a [U_A^\dag (t){A}_a U_A(t)]
 \otimes [U_B^\dag (t){B}_a U_B(t)].
\end{eqnarray}
According to the Nakajima-Zwanzig method~\cite{Nakajima1958On,Zwanzig1960}, we introduce the projection operators
\begin{eqnarray}
\label{projectionoperatordefinitions}
&&{\cal P}O={\rm Tr}_B[O ]\otimes\rho_B,\\
&&{\cal Q}O=(1-{\cal P})O.
\end{eqnarray}
It is not difficult to prove that
\begin{eqnarray}
\label{projectionoperatorproperty1}
{\cal P}\widetilde {\cal V}{\cal P}=0.
\end{eqnarray}
Given the assumption ${\rm Tr}_B[B_a\rho_B]=0$ and using the property~(\ref{projectionoperatorproperty1}), we  apply $\cal P$ and $\cal Q$ to Eq.~(\ref{timeevolutionequationopensystemorginteractionpicture}) to obtain the following two equations:
\begin{eqnarray}
\label{projectedeq1densitymatrixcase}
\partial_t {\cal P}{\widetilde {\rho}(t )}&=&{\cal P}\widetilde{\cal V}(t){\cal Q}{\widetilde {\rho}(t)},\\
\label{projectedeq2densitymatrixcase}
\partial_t {\cal Q}{\widetilde {\rho}(t)}&=&{\cal Q} \widetilde{\cal V}(t){\cal Q}{\widetilde { \rho}(t )}+{\cal Q}\widetilde{\cal V}(t){\cal P}{\widetilde {\rho}(t)}.
\end{eqnarray}
The second equation has a formal solution:
\begin{eqnarray}
\label{solutionprojectedeq2densitymatrixcase}
{\cal Q}{\widetilde { \rho }(t)}= \int_0^t ds {\cal G}(t,s){\cal Q}\widetilde{\cal V} (s){\cal P}{\widetilde { \rho}(s)}.
\end{eqnarray}
Here, we have considered the vanishing initial condition, ${\cal Q}{\widetilde {\rho}}(0)=0$, and have defined the propagator
\begin{eqnarray}
\label{propagatorsolutionprojectedeq2densitymatrixcase}
{\cal G}(t,s)\equiv T_\leftarrow e^{\int_s^t du {\cal Q}\widetilde{\cal V}(u) }.
\end{eqnarray}
Considering that the Liouville-von Neumann equation~(\ref{timeevolutionequationdensitymatrix}) is unitary, we may connect ${\widetilde {\rho}}$ at two times $s$ and $t$ ($s\le t$) by
\begin{eqnarray}
\label{densitymatrixattwotimes}
{\widetilde {\rho}(s)}&=& U_V(s) U_V(t)^\dag {\widetilde {\rho}(t)}  U_V(t) U_V(s)^\dag \equiv\widetilde{ {\cal U}}(s,t){\widetilde {\rho}(t)},
\end{eqnarray}
where $U_V(t)=U_0(t)^\dag U(t)$ is the time evolution operator of the interaction picture operator ${\widetilde V}(t)$,
\begin{eqnarray}
\label{timeevolutionoperatorinteractionpictureoperator}
\widetilde{ {\cal U}}(s,t)=T_\rightarrow e^{-\int_s^t du  \widetilde{\cal V}(u)},
\end{eqnarray}
and the symbol $T_\rightarrow$ denotes the antichronological time-ordering operator. Substituting Eqs.~(\ref{solutionprojectedeq2densitymatrixcase}) and~(\ref{densitymatrixattwotimes}) into Eq.~(\ref{projectedeq1densitymatrixcase}), we obtain
\begin{eqnarray}
\label{timeevolutionequationopensystemorginteractionpictureprojected}
\partial_t{\cal P}{\widetilde { \rho}(t)} = \int_0^t ds   {\cal P} \widetilde{\cal V}  (t){\cal G}(t,s){\cal Q} \widetilde{\cal V} (s){\cal P}\widetilde{ {\cal U}}(s,t){\widetilde {\rho}(t)},
\end{eqnarray}
The reader is reminded that this result is a time-convolutionless form with respect to ${\widetilde {\rho}}$: the density matrix on the right-hand side (RHS) depends on the time $t$ instead of the earlier time $s$~\cite{Breuer2002}.

To simplify the complex integral in Eq.~(\ref{timeevolutionequationopensystemorginteractionpictureprojected}), we must resort to several approximations. Let us redefine $V$ as $\alpha V$, where $\alpha$ is a small perturbation parameter. We expand the above equation up to the second order of $\alpha$ and then arrive at
\begin{eqnarray}
\label{timeevolutionequationopen5}
\partial_t{\cal P}{\widetilde { \rho}(t)}= \alpha^2\int_0^t ds   {\cal P}\widetilde{\cal V}(t) \widetilde{\cal V}(t-s){\cal P} {\widetilde {\rho}(t)}.
\end{eqnarray}
Here, we have performed a change of variables, $s\rightarrow t-s$, and used an approximation
\begin{eqnarray}
\widetilde{ {\cal U}}(s,t){\widetilde {\rho }(t)}={\widetilde {\rho}(t)}+{\cal O}(\alpha).
\end{eqnarray}
Inserting the concrete formulas of the projector operator $\cal P$ and $\widetilde {\cal V}$ into Eq.~(\ref{timeevolutionequationopen5}), we obtain
\begin{eqnarray}
\label{timeevolutionequationopensystemorginteractionpictureprojectedexplicitdensitymatrixcase}
\partial_t {\widetilde {\rho}_A(t)}=&&-\alpha^2\sum_{a,b}\int_0^t ds {\widetilde A}_a^\dag(t) {\widetilde A}_b(t-s){\widetilde {\rho}_A(t)} {\rm Tr}_B[ \widetilde B_a(t) \widetilde B_b(t-s)\rho_B]\nonumber\\
&&+\alpha^2\sum_{a,b}\int_0^t ds {\widetilde A}_b(t-s){\widetilde { \rho}_A(t)} {\widetilde A}_a^\dag (t)  {\rm Tr}_B[ \widetilde B_a(t) \widetilde B_b(t-s )\rho_B] \nonumber\\
&&-\alpha^2\sum_{a,b}\int_0^t ds {\widetilde {\rho}_A(t)} {\widetilde A}_b(t-s){\widetilde A}^\dag_a(t)  {\rm Tr}_B[ \widetilde B_b(t-s) \widetilde B_a(t)\rho_B] \nonumber\\
&&+\alpha^2\sum_{a,b}\int_0^t ds {\widetilde A}^\dag _a(t){\widetilde {\rho}_A(t)} {\widetilde A}_b(t-s) {\rm  Tr}_B[ \widetilde B_b(t-s) \widetilde B_a(t)\rho_B],
\end{eqnarray}
where ${\widetilde {\rho}_A(t)}={\rm Tr}_B[{\widetilde {\rho}(t)}]$ and its initial condition is $\rho_A(0)$. To proceed further, we need to perform a spectrum decomposition on the interaction picture operators ${\widetilde A}_a(t)$ or ${\widetilde A}_a^\dag(t)$. Unfortunately, this is not always well-defined for general $H_A(t)$. We have to restrict our discussion to the four types of MQMEs. After individually investigating them, we will see that these equations have a unified formula.

\subsubsection{Static Hamiltonian}
\label{section2B1}
This is the standard case and is also the simplest case in which the Hamiltonian of the quantum system is time independent~\cite{Lindblad1975,Gorini1976}. The time evolution operator of system A is
\begin{eqnarray}
\label{timeevolutionoperatorstaticHamiltonian}
U_A(t)= e^{-itH_A}=\sum_n |\varepsilon_n\rangle \langle  \varepsilon_n| e^{-it\varepsilon_n },
\end{eqnarray}
where $|\varepsilon_n\rangle$ and $\varepsilon_n$ are the discrete energy eigenvector and eigenvalue of $H_A$, respectively. Then, the interaction picture operator ${\widetilde A}_a(t)$ has the following spectrum decomposition~\cite{Breuer2002}:
\begin{eqnarray}
\label{decompositionofAoperatorstaticHamiltonian}
{\widetilde A}_a(t)=\sum_{\omega} A_a^\dag (\omega)e^{i\omega t}=\sum_\omega A_a(\omega )e^{-i\omega t},
\end{eqnarray}
where
\begin{eqnarray}
\label{LoperatorsstaticHamiltoniancase}
A_a^\dag (\omega)&=&\sum_{n,m } \delta_{\omega,\varepsilon_n- \varepsilon_m} \langle \varepsilon_n|A_a|\varepsilon_m\rangle   |\varepsilon_n\rangle \langle  \varepsilon_m|,\\
A_a(\omega)&=&A_a^\dag(-\omega),
\end{eqnarray}
The energy differences $\omega$ are called Bohr frequencies, which may be positive or negative but always appear in pairs, and $\delta$ is the Kronecker delta. These operators satisfy simple properties:
\begin{eqnarray}
\label{eigenoperatorstaticHamiltonian}
[H_A, A_a^\dag(\omega)]=\omega A_a^\dag(\omega),\hspace{1cm}[H_A, A_a(\omega)]=-\omega A_a(\omega).
\end{eqnarray}
Therefore, $A_a(\omega)$ and $A_a^\dag(\omega)$ are also called the eigenoperators of $H_A$. The reader is particularly reminded that different operators $A_a(\omega)$ denoted by different indices $a$ may have the same frequency $\omega$, as with the  operators $A_a^\dag(\omega)$. This point is very relevant to the descriptions of QJT using either wave vectors or density matrices. Substituting the decomposition~(\ref{decompositionofAoperatorstaticHamiltonian}) into Eq.~(\ref{timeevolutionequationopensystemorginteractionpictureprojectedexplicitdensitymatrixcase}), we obtain
\begin{eqnarray}
\label{timeevolutionequationopensystemorginteractionpictureprojectedexplicitstaticHamiltoniandensitymatrixcase}
\partial_t {\widetilde{\rho}}_A(t)=&&- \alpha^2 \sum_{a,b,\omega,\omega'} e^{i(\omega'-\omega)t }A_a^\dag(\omega') { A}_b(\omega){\widetilde {\rho}_A(t)} \int_0^t ds e^{i\omega s}{\rm Tr}_B[\widetilde B_a(s)B_b\rho_B] \nonumber\\
&&+\alpha^2\sum_{a,b,\omega,\omega'} e^{i(\omega'-\omega)t }{A}_b(\omega){\widetilde {\rho}_A(t)} A_a^\dag(\omega')\int_0^t ds e^{i\omega s}{\rm Tr}_B[\widetilde B_a(s)B_b\rho_B] \nonumber\\
&&-\alpha^2\sum_{a,b,\omega,\omega'} e^{i(\omega'-\omega)t }{\widetilde {\rho}_A(t)} { A}_b(\omega)A_a^\dag(\omega') \int_0^t ds e^{i\omega s}{\rm Tr}_B[\widetilde B_b(-s)B_a\rho_B]\nonumber\\
&&+\alpha^2 \sum_{a,b,\omega,\omega'} e^{i(\omega'-\omega)t }A_a^\dag(\omega'){\widetilde {\rho}_A(t)} { A}_b(\omega)  \int_0^t ds e^{i\omega s}{\rm Tr}_B[\widetilde B_b(-s)B_a\rho_B].
\end{eqnarray}
Here, we have used the time-homogeneous property of the two-time correlation function,
\begin{eqnarray}
\label{correlationfunction}
{\rm Tr}_B[\widetilde B_a(t)\widetilde B_b(t-s)\rho_B]={\rm Tr}_B[\widetilde B_a(s)B_b\rho_B],
\end{eqnarray}
which is a result of the stationary heat bath $\rho_B$.

There are two further approximations involved in the following derivations~\cite{Breuer2002}. The first approximation is that the upper integral limit $t$ is replaced by  $+\infty$. This approximation is justified if the evolution time scale $t$ of  system A is
far larger than the decay time of the heat bath. Specifically, significant non-zero contributions of these correlation functions in the  integrals are around $s=0$. Under this approximation, it shall be convenient to write these one-side Fourier integrals by the double-side Fourier integrals, e.g.,
\begin{eqnarray}
\label{onesideFouriertransformcorrelationfunc}
\int_0^\infty ds e^{i\omega s}{\rm Tr}_B[{\widetilde B}_a(s)B_b\rho_B]=
\frac{1}{2}r_{ab}(\omega) +iS_{ab}(\omega),
\end{eqnarray}
where
\begin{eqnarray}
\label{FouriertransformCorrlationfunctions}
r_{ab}(\omega)&=&\int_{-\infty}^\infty ds e^{i\omega s} {\rm Tr}_B[{\widetilde B}_a(u)B_b\rho_B], \\
S_{ab}(\omega)&=&\frac{1}{2\pi i}P.V. \int_{-\infty}^\infty d\omega' \frac{r_{ab}(\omega')}{\omega-\omega'},
\end{eqnarray}
and $P.V.$ denotes the Cauchy principal value of the integral. The other integrals can be rewritten analogously. Because the heat bath is at thermal equilibrium, these correlation functions satisfy the so-called Kubo-Martin-Schwinger (KMS) condition~\cite{Kubo1957,Martin1959}. In particular, this condition on the Fourier transform is
\begin{eqnarray}
\label{DetailedBalanceCondition}
r_{ab}(\omega) =r_{ba}(-\omega)e^{\beta\omega}.
\end{eqnarray}
We will show later that this relation plays a key role in proving various FTs. The second approximation is the rotation wave approximation, or the secular approximation. Specifically, the terms with different $\omega$ and $\omega'$ on the RHS of Eq.~(\ref{timeevolutionequationopensystemorginteractionpictureprojectedexplicitstaticHamiltoniandensitymatrixcase}) are neglected. This approximation would be valid if a typical value of $\omega'-\omega$ is far larger than the reciprocal of time, $t^{-1}$. Given these two key approximations, doing a simple algebraic manipulation, we finally obtain
\begin{eqnarray}
\label{timeevolutionequationopensystemorginteractionpictureprojectedexplicitstaticHamiltonianfinaldensitymatrixcase}
\partial_t {\widetilde {\rho}_A(t)} =&-&i[{H}_{LS},{\widetilde {\rho}_A(t)} ]\nonumber \\
&+& \sum_{\omega,a,b}r_{ab}(\omega)\left[ {A}_b(\omega){\widetilde {\rho}_A(t)} A_a^\dag(\omega)-\frac{1}{2}\left\{A_a^\dag(\omega) {A}_b(\omega),{\widetilde {\rho}_A(t)}\right \} \right],
\end{eqnarray}
where the Lamb-shift term is
\begin{eqnarray}
\label{LambshiftH}
{H}_{LS}= \sum_{a,b,\omega}S_{ab}(\omega)A^\dag_a(\omega)A_b(\omega).
\end{eqnarray}
Here, we have reabsorbed the small parameter $\alpha$ into the interaction Hamiltonian $V$. Note that these two approximations can be rigorously justified by taking the weak coupling limit, $\alpha\rightarrow 0$, $t\rightarrow\infty$, and keeping $\alpha^2 t$ constant~\cite{Davies1974,Gorini1976,Alicki2006,Rivas2012}.

Eq.~(\ref{timeevolutionequationopensystemorginteractionpictureprojectedexplicitstaticHamiltonianfinaldensitymatrixcase}) can be transformed back into the Schr$\ddot{o}$dinger picture. First, we know that
\begin{eqnarray}
\label{masterequationstaticHamiltoniandensitymatrixcase}
\partial_t {\rho}_A(t)&=&-i[H_A, {\rho}_A(t)]+U_A(t)\partial_t {\widetilde {\rho}_A(t)} U_A(t)^\dag.
\end{eqnarray}
Because of
\begin{eqnarray}
\label{interaction2schrodingerstaticHamiltonian}
U_A(t)A_a^\dag(\omega)U_A(t)^\dag=A_a^\dag(\omega)e^{-i\omega t},
\end{eqnarray}
we immediately find that the second term on the RHS of Eq.~(\ref{masterequationstaticHamiltoniandensitymatrixcase}) is  exactly the same as the RHS of Eq.~(\ref{timeevolutionequationopensystemorginteractionpictureprojectedexplicitstaticHamiltonianfinaldensitymatrixcase}) except that all ${\widetilde {\rho}}_A(t)$ therein are replaced by ${\rho}_A(t)$. After the above discussion, we have an conclusion that, if the decomposition~(\ref{decompositionofAoperatorstaticHamiltonian}) is true, the following equations,~(\ref{timeevolutionequationopensystemorginteractionpictureprojectedexplicitstaticHamiltoniandensitymatrixcase})-(\ref{LambshiftH}), would be available. This observation will guide our discussion about the time-dependent MQMEs.


\subsubsection{Weakly driven Hamiltonian}
\label{weaklydrivenHamiltonian}
The first time-dependent case is the weakly driven Hamiltonian,
\begin{eqnarray}
\label{weakdrivenHamiltonian}
H_A(t)=H_0 + \gamma H_1(t),
\end{eqnarray}
where $H_0$ is the constant Hamiltonian of the bare system, $H_1(t)$ represents the coupling term of the bare system and some time-varying external force or fields, and the parameter $\gamma$ accounts for the coupling strength. We assume that $\gamma$ is of the same order of magnitude as  $\alpha$ in the interaction Hamiltonian $V$. This is the precise meaning of being weakly driven. Under this assumption, the time evolution operator of  system A is approximated to be
\begin{eqnarray}
\label{approximatedtimeevolutionoperatorweaklydriving}
U_A(t)=  U_A^0(t) + {\cal O}(\gamma) = e^{-it H_0} + {\cal O}(\gamma).
\end{eqnarray}
When we substitute Eq.~(\ref{approximatedtimeevolutionoperatorweaklydriving}) into Eq.~(\ref{timeevolutionequationopensystemorginteractionpictureprojectedexplicitdensitymatrixcase}), the higher orders should be ${\cal O}(\alpha^2\gamma)$. Because we only retain the terms up to second order and neglect the other higher order terms, we re-obtain Eq.~(\ref{timeevolutionequationopensystemorginteractionpictureprojectedexplicitstaticHamiltonianfinaldensitymatrixcase}). However, we shall bear in mind that the decomposition, Eq.~(\ref{decompositionofAoperatorstaticHamiltonian}), is with respect to the eigenvectors and eigenvalues of the bare Hamiltonian, $H_0$. Additionally, when we transform the time evolution equation into the Schr$\ddot{o}$dinger picture, $\gamma H_1(t)$ will be recovered because this Hamiltonian has a contribution to first order in $\gamma$.

\subsubsection{Periodically driven Hamiltonian}
It will be more desirable if we have a decomposition of Eq.~(\ref{decompositionofAoperatorstaticHamiltonian}) even if the Hamiltonian of the quantum system A is strongly driven. This is indeed possible if $H_A(t)$ is periodic with a frequency $\Omega$~\cite{Bluemel1991,Kohler1997,Breuer1997,GRIFONI1998,  Breuer2000,Kohn2001,Szczygielski2013,Langemeyer2014,Gasparinetti2014,Cuetara2015}, namely,
\begin{eqnarray}
\label{perodicallydrivenHamiltonian}
H_A(t)=H_A(t+2\pi/\Omega).
\end{eqnarray}
In this case, the famous Floquet theorem can be applied~~\cite{Shirley1965,Zeldovich1967,Sambe1973}:
\begin{eqnarray}
\label{timeevolutionoperatorperiodiccase}
U_A(t)=\sum_n |\epsilon_n(t)\rangle \langle  \epsilon_n(0)| e^{-it\epsilon_n},
\end{eqnarray}
where these time-independent $\epsilon_n$ are called quasi-energies, and the Floquet basis satisfies
\begin{eqnarray}
\label{Floquetbasis}
[H_A(t)-i\partial_t ]|\epsilon_n(t)\rangle=\epsilon_n |\epsilon_n(t)\rangle.
\end{eqnarray}
The Floquet basis is periodic, that is,
\begin{eqnarray}
\label{periodicFloquetbasis}
|\epsilon_n(t+2\pi/\Omega)\rangle=|\epsilon_n(t)\rangle.
\end{eqnarray}
It is not difficult to find that the solution
\begin{eqnarray}
\label{shiftedFloquetbasis}
|\epsilon_{n,q}(t)\rangle=e^{iq\Omega t}|\epsilon_n(t)\rangle
\end{eqnarray}
is still in the Floquet basis but with a shifted quasi-energy
\begin{eqnarray}
\label{shiftedFloquetquasienergy}
\epsilon_{n,q}=\epsilon_n + q\Omega,
\end{eqnarray}
where $q$ is an arbitrary integer. Therefore, for the periodic Hamiltonian, the Floquet bases connected by Eq.~(\ref{shiftedFloquetbasis}) are equivalent in physics. We may project them into one Brillouin zone of width $\Omega$. Using Eq.~(\ref{timeevolutionoperatorperiodiccase}), we may see that ${\widetilde A}_a(t)$ still has a decomposition similar to Eq.~(\ref{decompositionofAoperatorstaticHamiltonian})~\cite{Breuer2004}:
\begin{eqnarray}
\label{decompositionofAoperatorperiodicallydrivenHamiltonian}
{\widetilde A}_a(t)=\sum_{\omega} A_a^\dag (\omega,0)e^{i\omega t}=\sum_\omega A_a(\omega,0)e^{-i\omega t},
\end{eqnarray}
where
\begin{eqnarray}
\label{Loperatorspreodicallydrivencase}
A_a^\dag (\omega,0)&=&\sum_{n,m,q } \delta_{\omega,\epsilon_n- \epsilon_m+q\Omega} \langle\langle \epsilon_{n,q}|A_a|\epsilon_m\rangle \rangle  |\epsilon_{n,q}(0)\rangle \langle  \epsilon_m(0)|, \nonumber\\
A_a(\omega,0)&=&A_a^\dag(-\omega,0),
\end{eqnarray}
and
\begin{eqnarray}
\langle \langle\epsilon_{n,q}|A_a|\epsilon_m\rangle\rangle =\frac{\Omega}{2\pi}\int_0^{2\pi/\Omega} ds e^{-iq\Omega s} \langle\epsilon_n(s)|A_a|\epsilon_m(s)\rangle
\end{eqnarray}
or the Fourier coefficient with frequency $q\Omega $.  Notice that in contrast to the static Hamiltonian case, the Bohr frequency $\omega$ here depends on the external periodicity $\Omega$.  Substituting the new decomposition into Eq.~(\ref{timeevolutionequationopensystemorginteractionpictureprojectedexplicitdensitymatrixcase}), applying the same approximations as those in the static Hamiltonian case, and going back to the Schr$\ddot{o}$dinger picture, we finally obtain
\begin{eqnarray}
\label{masterequationperiodicallydrivenHamiltoniandensitymatrixcase}
\partial_t {{\rho}}_A(t)&=&-i[H_A(t)+H_{LS}(t), {{\rho}}_A(t)]\nonumber\\
&&+\sum_{a,b,\omega}r_{ab}(\omega)\left[A_b(\omega,t) {\rho}_A(t) A_a^\dag(\omega,t)-\frac{1}{2}\left\{ A_a^\dag(\omega,t)A_b(\omega,t), {{\rho}}_A(t )\right \}\right],
\end{eqnarray}
where the Lamb shift is now
\begin{eqnarray}
H_{LS}(t)= \sum_{a,b,\omega}S_{kl}(\omega)A^\dag_a(\omega,t)A_b(\omega,t).
\end{eqnarray}
In this derivation, we have used the formula~\cite{Breuer2004}
\begin{eqnarray}
\label{AtransformationbetweenSandHperiodicallydrivenHamiltonian}
U_A(t)A_a^\dag (\omega,0)U_A(t)^\dag=A_a^\dag (\omega,t)e^{-i\omega t}.
\end{eqnarray}
Compared with the evolution equation for the static Hamiltonian case, Eq.~(\ref{masterequationstaticHamiltoniandensitymatrixcase}), the time dependence in Eq.~(\ref{masterequationperiodicallydrivenHamiltoniandensitymatrixcase}) is through the operators $A_a(\omega,t)$ and $A_a^\dag(\omega,t)$ in addition to the time-dependent coherent dynamics. Nevertheless, the coefficients $r_{ab}(\omega)$ remain static.

\subsubsection{Adiabatically driven Hamiltonian}
\label{section2B4}
If the Hamiltonian varies in time very slowly, the adiabatical approximation can be applied to the operators ${\widetilde A}_k(t)$~\cite{Davies1978,Alicki1979,Amin2008,Albash2012}. In this case, the time evolution operator of system A is
\begin{eqnarray}
\label{timeevolutionoperatoradiabaticalcase}
U_A(t)\simeq \sum |\epsilon_n(t)\rangle \langle  \epsilon_n(0)| e^{-i\mu_n(t)},
\end{eqnarray}
where $|\varepsilon_n(t)\rangle$ and $\varepsilon_n(t)$ are the instantaneous energy eigenvector and eigenvalues of $H_A(t)$, respectively, and the phase in the exponential function is
\begin{eqnarray}
&&\mu_n(t)=\int_0^t ds [\varepsilon_n(s)-i\langle \varepsilon_n(s)|\partial_s|\varepsilon_n(s)\rangle ].
\end{eqnarray}
With the above approximation, we have the following decomposition:
\begin{eqnarray}
\label{decompositionofAoperatoradiabaticallydrivenHamiltonian1}
{\widetilde A}_a(t)=\sum_{nm} A_{a,nm}^\dag(t,0) e^{i\mu_{nm}(t)}=\sum_{nm} A_{a,mn}(t,0) e^{-i\mu_{mn}(t)},
\end{eqnarray}
where
\begin{eqnarray}
\label{LoperatorsAdiabaticallydrivencase}
&&A_{a,nm}^\dag(t,0)=\langle\varepsilon_n(t)|A_a|\varepsilon_m(t)   \rangle |\varepsilon_n(0)    \langle \varepsilon_m(0) |,\\
&&A_{a,mn}(t,0)=A_{a,nm}^\dag(t,0),\\
&&\mu_{nm}(t)=\mu_n(t)-\mu_m(t).
\end{eqnarray}
Note that there still exist two identities analogous to Eq.~(\ref{eigenoperatorstaticHamiltonian}):
\begin{eqnarray}
\label{eigenoperatorAdiabaticallydrivenHamiltonian}
[H_A(t),A_{a,mn}(t)]=-\omega_{mn}(t) A_{a,mn}(t),\hspace{1cm}
[H_A(t),A^\dag_{a,mn}(t)]=\omega_{mn}(t) A^\dag_{a,mn}(t),
\end{eqnarray}
where $A_{a,mn}(t)$ is brief notation of $A_{a,mn}(t,t)$ and the Bohr frequency is
\begin{eqnarray}
\label{Bohrfrequencyadiabaticallydrivencase}
\omega_{nm}(t)=\varepsilon_n(t)-\varepsilon_m(t).
\end{eqnarray}
Different from previous cases, these frequencies are generally time dependent. For notational simplicity, we suppose that both $\mu_{mn}(t)$ and $\omega_{mn}(t)$ are uniquely determined by the quantum numbers $(m,n)$. The decomposition~(\ref{decompositionofAoperatoradiabaticallydrivenHamiltonian1}) remains inadequate; we still need another key approximation:
\begin{eqnarray}
\label{approximatetimeevolutionoperatorinadiabaticallydrivenHamiltonian}
U_A(t-s)\simeq e^{isH(t)}U_A(t).
\end{eqnarray}
Under this approximation, we have the following decomposition:
\begin{eqnarray}
\label{decompositionofAoperatoradiabaticallydrivenHamiltonian2}
{\widetilde A}_a(t-s)&=&\sum_{n,m} A_{a,nm}^\dag(t,0) e^{i\mu_{nm}(t)-is\omega_{nm}(t)}\\
&=&\sum_{n,m} A_{a,mn}(t,0) e^{-i\mu_{mn}(t)+is\omega_{mn}(t)}.
\end{eqnarray}
The approximation~(\ref{approximatetimeevolutionoperatorinadiabaticallydrivenHamiltonian}) shall be justified if $H_A(t)$ is almost unchanged in the time interval $(t-s,t)$. Obviously, $s\sim 0$ is preferred. Importantly, because of the very short decay time of these two-time correlation functions in Eq.~(\ref{timeevolutionequationopensystemorginteractionpictureprojectedexplicitdensitymatrixcase}), this condition is always fulfilled.
We may relax the seemly strict condition of $s\sim 0$. To ensure the validity of the adiabatical approximation, Eq.~(\ref{timeevolutionoperatoradiabaticalcase}), the duration $t_f$ of the non-equilibrium process shall be sufficiently large~\cite{Messiah1962}.
In this situation, the change in the Hamiltonian $H_A(t)$ would be very slow. Hence, in the interval $(t-s,t)$ with nonzero $s$, the Hamiltonian could still be reasonably thought to be a constant operator. In particular, we expect that in the quantum adiabatic limit, or $t_f\rightarrow \infty$, $s$ may be any finite value.

Inserting Eqs.~(\ref{decompositionofAoperatoradiabaticallydrivenHamiltonian1}) and~(\ref{decompositionofAoperatoradiabaticallydrivenHamiltonian2}) into Eq.~(\ref{timeevolutionequationopensystemorginteractionpictureprojectedexplicitdensitymatrixcase}) and repeating the previous derivations, we obtain
\begin{eqnarray}
\label{masterequationadiabaticallydrivenHamiltonianinteractionpicturedensitymatrixcase}
\partial_t {\widetilde {{\rho}}_A(t)} =&-&i[H_{LS}(t,0),{\widetilde {{\rho}}_A(t)} ]+\sum_{a,b,m,n}r_{ab}(\omega_{mn}(t))\left[{A}_{b,mn}(t,0){\widetilde {{\rho}}_A(t)} A_{a,mn}^\dag(t,0)\right. \nonumber \\
&&\left. -\frac{1}{2}\left\{A_{a,mn}^\dag(t,0) {A}_{b,mn}(t,0),{\widetilde {{\rho}}_A(t)}\right \} \right],
\end{eqnarray}
where the Lamb-shift term is
\begin{eqnarray}
{\widetilde H}_{LS}(t,0)= \sum_{a,b,m,n}S_{ab}(\omega_{mn}(t))A^\dag_{a,mn}(t,0)A_{b,mn}(t,0).
\end{eqnarray}
Considering that
\begin{eqnarray}
U_A(t)A_{a,mn}^\dag(t,0) U_A(t)^\dag=A_{a,mn}^\dag (t)e^{-i\mu_{mn}(t)},
\end{eqnarray}
where the operator $A_{a,mn}^\dag(t)$ is in fact the $A_{a,mn}^\dag(t,t)$ defined in Eq.~(\ref{LoperatorsAdiabaticallydrivencase}), when we return to the Schr$\ddot{o}$dinger picture, the time evolution equation for the reduced density matrix of the quantum open system A is
\begin{eqnarray}
\label{masterequationadiabaticallydrivenHamiltonian}
\partial_t {{\rho}}_A(t)&=&-i[H_A(t)+H_{LS}(t), {{\rho}}_A(t)] \nonumber \\
&+& \sum_{a,b,m,n}r_{ab}(\omega_{mn}(t))\left[{A}_{b,mn}(t){ {\rho}_A(t)} A_{a,mn}^\dag(t) -\frac{1}{2}\left\{A_{a,mn}^\dag(t) {A}_{b,mn}(t),{ {{\rho}}_A(t)} \right\} \right].
\end{eqnarray}
Distinct from the three previous cases, the coefficients $r_{ab}$ in the current case generally vary with time through the time-dependent Bohr frequencies, $\omega_{mn}(t)$. Finally, we want to emphasize that Eq.~(\ref{masterequationadiabaticallydrivenHamiltonian}) can be proved to be rigorous if one considers the weak coupling limit and quantum adiabatic limit simultaneously~\cite{Davies1978,Thunstroem2005,Oreshkov2010}.\\

After a survey of the above derivations of these four types of MQMEs, we find that all of them can be unified into a general formula:
\begin{eqnarray}
\label{quantummasterequationgeneralformdensitymatrixcase}
\partial_t\rho_A(t)&=&-i[H_A(t)+H_{LS}(t),\rho_A(t)]\nonumber\\
&&+\sum_{\omega_t,a,b}r_{ab}(\omega_t)\left[A_b(\omega_t,t)\rho_A(t) A_a^\dag(\omega_t,t)-\frac{1}{2}\left\{A_a^\dag(\omega_t,t)A_b(\omega_t,t),\rho_A(t)\right\}\right]\nonumber\\
&\equiv&{\cal L}(t)\rho_A(t),
\end{eqnarray}
where
\begin{eqnarray}
\label{AoperatorpropertygeneralMQME}
A_a(\omega_t,t)&=&A_a^\dag(-\omega_t,t),
\end{eqnarray}
and the positive and negative Bohr frequencies appear in pairs. Obviously, this formula follows the structure of Eq.~(\ref{formalmasterequation1}). We can write its solution formally as
\begin{eqnarray}
\label{propagatorsolutionofheatequation}
\rho_A(t)= T_{\leftarrow} e^{\int_0^t ds{{\cal L}}(s) }\rho_A(0)\equiv G(t,0)[\rho_A(0)].
\end{eqnarray}
The whole exponential term, or $G(t,0)$, is called the superpropagator of Eq.~(\ref{quantummasterequationgeneralformdensitymatrixcase}), or the dynamical map; see Eq.~(\ref{Krausrepresentation}).

\section{Work in closed quantum systems}
\label{section3}
Let us start with the stochastic work of closed quantum systems. This concept is relatively simple and is also the foundation of studying the stochastic heat and work in open quantum systems. For convenience, throughout this paper, we suppose that the time dependence of the Hamiltonian of the quantum system A is always realized by a protocol, $\lambda(t)$, namely,
\begin{eqnarray}
\label{closedsystemHamiltonian}
H(t)=H[\lambda(t)],
\end{eqnarray}
as is the case for the time dependence of the instantaneous energy eigenvectors and eigenvalues of the Hamiltonian,
\begin{eqnarray}
\label{eigenvectorandvaluesofclosedHamiltonian}
H(t)|\varepsilon_n(t)\rangle=\varepsilon_n(t) |\varepsilon_n(t)\rangle.
\end{eqnarray}
Because there is only one quantum system, we do not explicitly write the subscript A here. Consider that the non-equilibrium process of a closed quantum system is performed within the time interval $(0,t_f)$. According to the TEM scheme~\cite{Kurchan2000,Tasaki2000}, the stochastic work done by some external agents on the quantum system is defined as the difference in the instantaneous energy eigenvalues between the end and beginning of the process:
\begin{eqnarray}
\label{inclusiveworkdefinition}
W_{nm}=\varepsilon_n(t_f)-\varepsilon_m(0).
\end{eqnarray}
Following the terms named by Jarzynski~\cite{Jarzynski2007}, we call Eq.~(\ref{inclusiveworkdefinition}) the inclusive work. The reason for this will be given shortly. By repeating this process many times, we can construct the probability distribution of the work as
\begin{eqnarray}
\label{distributioninclusiveworkclosedsystem}
P(W)=\sum_{n,m} \delta( W-W_{nm}) \left |\langle \varepsilon_n(t_f)|U(t_f)|\varepsilon_m(0)\rangle \right |^2 P_m(0),
\end{eqnarray}
where $U(t)$ is the time evolution operator of the system and $P_m(0)$ is the probability of finding the system at the eigenvector $|\varepsilon_m(0)\rangle$ at time $0$. Because of the involvement of the Dirac function, Eq.~(\ref{eigenvectorandvaluesofclosedHamiltonian}) is not the most convenient form if we want to analyze the statistical properties of the work distribution. An alternative method is to resort to its Fourier transform or CF~\cite{Gardiner1983,Risken1984,Kurchan1998,Lebowitz1999,Imparato2007,Talkner2007,Esposito2009,Campisi2011}:
\begin{eqnarray}
\label{CFinclusivework}
\Phi_w(\eta)&=&\int_{-\infty}^{+\infty}  e^{i\eta W}P(W)dW \nonumber \\
&=&\sum_{n,m} e^{i\eta[\varepsilon_n(t_f)-\varepsilon_m(0)]} \left |\langle \varepsilon_n(t_f)|U(t_f)|\varepsilon_m(0)\rangle \right |^2 P_m(0).
\end{eqnarray}
At first glance, both Eqs.~(\ref{distributioninclusiveworkclosedsystem}) and (\ref{CFinclusivework}) depend on the time evolution operator of the quantum system, $U(t)$; no apparent advantages are shown in the latter expression. However, the CF can be rewritten as taking the trace over
some operators, that is,
\begin{eqnarray}
\label{CFworkclosedsystem}
\Phi_w(\eta)&=&{\rm Tr}\left[ e^{i\eta H(t_f)} U(t_f)e^{-i\eta H(0)}\rho_0U^\dag(t_f)\right]\nonumber\\
&\equiv&{\rm Tr}[e^{i\eta H(t_f)}\hat {\rho}(t_f,\eta)]\equiv{\rm Tr}[K(t_f,\eta)],
\end{eqnarray}
where
\begin{eqnarray}
\label{initialdensitymatrixmeasuredinclusivework}
\rho_0=\sum_m P_m(0)|\varepsilon_m(0)\rangle \langle \varepsilon_m(0)|.
\end{eqnarray}
We call $\hat{\rho}(t,\eta)$ and $K(t,\eta)$ the heat and work characteristic operators (HCO and WCO), respectively. The reader is reminded that  naming $\hat{\rho}(t,\eta)$ as HCO is purely formal; see Eq.~(\ref{CFinclusiveworkopensystemgeneralbasedheat}). After all, there is no heat production in closed quantum cases. It is easy to prove that $\hat{\rho}$ satisfies the Liouville-von Neumann equation~(\ref{timeevolutionequationdensitymatrix}) with a new initial condition $e^{-i\eta H(0)}\rho_0$, whereas the evolution equation for the newly introduced operator $K$ is
\begin{eqnarray}
\label{timeevolutionequationclosedsysteminclusivework}
\partial_t K(t,\eta) =&&-i[H(t),K(t,\eta)]+\partial_t \left[e^{i\eta H(t)}\right ] e^{-i\eta H(t)} K(t,\eta) \nonumber \\
\equiv&&-i[H(t),K(t,\eta)]+ {\cal W}_{\eta}(t)K(t,\eta).
\end{eqnarray}
Note that the initial condition of $K(t,\eta)$ is $\rho_0$, and the symbol ${\cal W}_\eta(t)$ is a superoperator; its action on an operator is a simple multiplication from the left-hand side (LHS) of the operator. In addition, we must emphasize that due to the first energy measurement, the density matrix, $\rho_0$, may  not necessarily be identical to the original initial density matrix of the system, $\rho(0)$. This is an intrinsic difficulty of the work definition based on the TEM scheme~\cite{Allahverdyan2005,Perarnau-Llobet2017}. To avoid this additional complexity, throughout this paper, we assume that the initial density matrix of the system, $\rho(0)$, is commutative with $H(0)$. Then, $\rho_0=\rho(0)$. A typical example is the canonical density matrix, $\rho(0)=e^{-\beta H(0)}/Z(0)$, where $Z(0)$ is the partition function of the system at time $0$ and is equal to ${\rm Tr}[e^{-\beta H(0)}]$.

The work definition is not unique, for instance, if the Hamiltonian of the system is composed of two parts:
\begin{eqnarray}
\label{freeandinteractionHamiltonian}
H(t)=H_0+ H_1(t).
\end{eqnarray}
Following the terms in Eq.~(\ref{weakdrivenHamiltonian}), we call $H_0$ the bare system. Because the coupling term, $H_1(t)$, is not required to be weak, the parameter $\gamma$ is absent here. A common physical model described by this Hamiltonian in textbooks is an electric dipole driven by a time-varying electric field~\cite{Ballentine2014}. Assuming that $H_1(t)$ vanishes at time $0$, we may alternatively define the work done on the bare system using the TEM scheme as
\begin{eqnarray}
\label{exclusiveworkdefinition}
W_{nm}^0=\varepsilon_n-\varepsilon_m,
\end{eqnarray}
and its distribution is obtained by
\begin{eqnarray}
\label{distributionexclusivework}
P_0(W)=\sum \delta(W-W_{nm}^0) \left |\langle \varepsilon_n|U(t)|\varepsilon_m\rangle \right |^2 P_m^0,
\end{eqnarray}
where
\begin{eqnarray}
\label{eigenequationsfreeHamiltonian}
H_0|\varepsilon_n\rangle=\varepsilon_n |\varepsilon_n\rangle,
\end{eqnarray}
and $P_m^0$ is the probability of finding the bare system at the eigenvector $|\varepsilon_m\rangle$ at time $0$. Because this work definition excludes the coupling term, $H_1(t)$, it was called the exclusive work by Jarzynski~\cite{Jarzynski2007} to distinguish it from the inclusive work, Eq.~(\ref{inclusiveworkdefinition}). Unless $H_1$ exactly vanishes at the end time $t_f$, these two work distributions are generally different.
In analogy to Eqs.~(\ref{CFinclusivework}) and~(\ref{CFworkclosedsystem}), we define the CF of the exclusive work and introduce an alternative WCO:
\begin{eqnarray}
\label{CFexclusivework}
\Phi_{w_0}(\eta)&=&\int_{-\infty}^{+\infty} dW e^{i\eta W}P_0(W)\nonumber\\
&=&\sum e^{i\eta[\varepsilon_n -\varepsilon_m ]} \left |\langle \varepsilon_n|U(t_f)|\epsilon_m\rangle \right |^2 P_m^0\nonumber\\
&=&{\rm Tr}\left[ e^{i\eta H_0} U(t_f)e^{-i\eta H_0}\rho_0 U(t_f)^\dag\right]\nonumber\\
&\equiv &{\rm Tr}[e^{i\eta H_0} \hat\rho_0(t_f,\eta)]\equiv{\rm Tr}[K_0(t_f,\eta)].
\end{eqnarray}
We can easily prove that the operator ${\hat \rho}_0(t_f,\eta)$ still satisfies the Liouville-von Neumann equation~(\ref{timeevolutionequationdensitymatrix}) with an initial condition $e^{-i\eta H_0}\rho_0$, while $K_0(t,\eta)$ satisfies a time-evolution equation given by
\begin{eqnarray}
\label{timeevolutionequationclosedsystemexclusivework}
\partial_t K_0(t,\eta) =&&-i[H(t),K_0(t,\eta)]-i[e^{i\eta H_0},H_1(t)] e^{-i\eta H_0}K_0(t,\eta) \nonumber \\
=&&-i[H(t),K_0(t,\eta)]+ {\cal W}^0_{\eta}(t)K_0(t,\eta).
\end{eqnarray}
Its initial condition remains $\rho_0$, and ${\cal W}^0_\eta(t)$ is a superoperator acting on operators on its LHS.

Let us give several comments on Eq.~(\ref{timeevolutionequationclosedsysteminclusivework}). First, if the parameter $\eta$ is set to be zero, Eq.~(\ref{timeevolutionequationclosedsysteminclusivework}) is reduced to the Liouville-von Neumann equation~(\ref{timeevolutionequationdensitymatrix}), and $K(t,\eta)$ is simply the density matrix of the system, $\rho(t)$. Second, if $\rho_0$ is a thermal equilibrium state with inverse temperature $\beta$, Eq.~(\ref{timeevolutionequationclosedsysteminclusivework}) with $\eta=i\beta$ has a simple solution:
\begin{eqnarray}
K(t,i\beta)=\frac{e^{-\beta H(t)}}{Z(0)}.
\end{eqnarray}
According to the probability interpretation of the CF, Eq.~(\ref{CFinclusivework}), we proved the quantum JE in the closed quantum system. On the other hand, this thermal equilibrium condition also indicates that $\hat\rho(0,i\beta)$ is proportional to an identity operator. Hence, the Liouville-von Neumann equation of $\hat\rho(t,i\beta)$ has a trivial solution, $\hat\rho(t,i\beta)=1/Z(0)$. Using the second equation in Eq.~(\ref{CFworkclosedsystem}), we are led to the JE again. Of course, the most straightforward proof of this equality is to apply  the first equation in Eq.~(\ref{CFworkclosedsystem}) and set $\eta=i\beta$ therein. Very analogous results can be obtained in Eq.~(\ref{timeevolutionequationclosedsystemexclusivework}),
\begin{eqnarray}
K_0(t,i\beta)=\frac{e^{-\beta H_0}}{Z(0)},
\end{eqnarray}
and the quantum Bochkov-Kuzovlev equality (BKE)~\cite{Bochkov1977,Bochkov1981,Campisi2011a} is proved in this case. These formulas were previously found by us, but the backward-time versions of Eqs.~(\ref{timeevolutionequationclosedsysteminclusivework}) and~(\ref{timeevolutionequationclosedsystemexclusivework}) were utilized therein~\cite{Liu2012,Liu2014}. Third, Eq.~(\ref{timeevolutionequationclosedsysteminclusivework}) can be regarded as the quantum version of the time evolution equation about the CF of the classical inclusive work~\cite{Hummer2001,Imparato2005,Imparato2007,Jarzynski1997a}. According to the quantum-classical correspondence principle~\cite{Ballentine2014}, as Planck's constant $\hbar$ tends to zero, the commutator is reduced to the Poisson bracket, and operators are replaced by ordinary functions~\cite{Polkovnikov2010}. Hence, the classical correspondence of Eq.~(\ref{timeevolutionequationclosedsysteminclusivework}) is
\begin{eqnarray}
\label{timeevolutionequationclosedclassicalsystem}
\partial_t K(z,t,\eta)=\{H(z,t),K(z,t,\eta)\}_{PB} +i\eta \partial_t H(z,t) K(z,t,\eta), 
\end{eqnarray}
where $z$ represents the phase coordinates of the closed classical system and $\{$,   $\}_{PB}$ denotes the Poisson bracket. It is not difficult to prove that the integral of the function $K(z,t,\eta)$ is simply the CF of the classical work~\cite{Jarzynski2007}. On the other hand, based on the celebrated Feynman-Kac (FK) formula~\cite{Kac1949}, Eq.~(\ref{timeevolutionequationclosedclassicalsystem}) has a path integral representation. Because of this, we called Eq.~(\ref{timeevolutionequationclosedsysteminclusivework}) the quantum FK formula~\cite{Liu2012,Liu2014}. Analogous results are also true for the exclusive work case. A further explanation is presented in Appendix A. Finally, in addition to the straightforward definition of $K(t,\eta)$, Eq.~(\ref{CFworkclosedsystem}), the time evolution equation~(\ref{timeevolutionequationclosedsysteminclusivework}) also reminds us that it has an alternative expression using the Dyson series~\cite{Sakurai1994}:
\begin{eqnarray}
\label{Kattwotimests}
K(t,\eta)=U(t)\left [ {T}_{\leftarrow}e^{\int_s^t d\tau U(\tau)^\dag{\cal W}_{\eta}(\tau)U(\tau)}\right ]U(s)^\dag K(s,\eta)  U(s) U(t)^\dag.
\end{eqnarray}
This formula connects two WCOs at two time points, $s$ and $t$, where $s\le t$. We may re-express the above equation by exchanging the positions of  $K(s,\eta)$ and $K(t,\eta)$ and obtain
\begin{eqnarray}
\label{Kattwotimesst}
K(s,\eta)=U(s) \left[{T}_{\rightarrow}e^{-\int_s^t d\tau U(\tau)^\dag{\cal W}_{\eta}(\tau)U(\tau)}\right] U(t)^\dag K(t,\eta)  U(t) U(s)^\dag.
\end{eqnarray}
This mimics Eq.~(\ref{densitymatrixattwotimes}). The primary advantage of using this seemly complex expression for $K(t,\eta)$ is that we may approximate it by expanding the time-ordered exponential term in terms of $\cal W_\eta$. We will show its applications in the next section.

\begin{figure}
\includegraphics[width=1.\columnwidth]{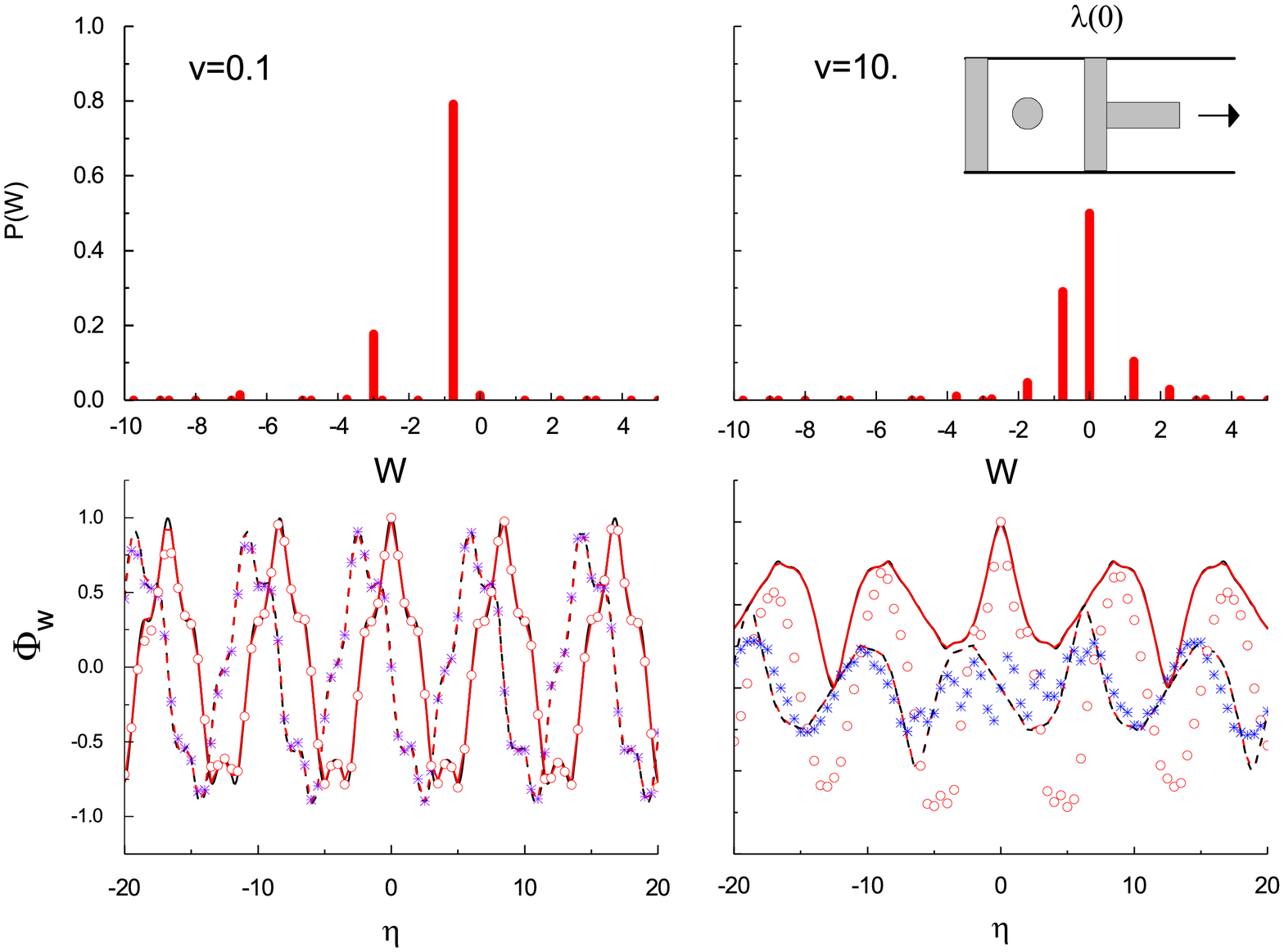}
\caption{Inset: the quantum piston model. At time 0, the position of the moving wall is $\lambda(0)$. The system is in thermal equilibrium with a heat bath whose inverse temperature is $\beta=1.0\varepsilon_1^{-1}$, where $\varepsilon_1$ is the first energy level of the infinite square-well potential, which equals $\hbar^2\pi^2/2m\lambda(0)^2$. The system is then disengaged from the heat bath, and the wall moves at a uniform speed up to a new position, $\lambda(t_f)=2\lambda(0)$. The left and right columns are the probability distributions of the inclusive work (in units of $\epsilon_1$) and their corresponding CFs at two moving speeds: the speed of the left column is $v=0.1$ $\varepsilon_1\lambda(0)/\hbar$, while that of the right column is $v=10.0$ $\varepsilon_1\lambda(0)/\hbar$. The black solid and dashed lines in the lower panels are the real and imaginary parts of these CFs, respectively. They are calculated by directly applying Fourier transforms to the work distributions in the upper panels. The circle and star symbols are the real and imaginary parts of the CFs obtained by solving the time evolution equation~(\ref{timeevolutionequationclosedsysteminclusivework}). The red solid and dashed lines are the real and imaginary parts of the CFs obtained by solving~(\ref{timeevolutionequationdensitymatrix}) with an initial condition $e^{-i\eta H(0)}\rho_0$. Because these red lines are exactly overlapped with the black lines, we obviously cannot read them. For
convenience, we have let $\hbar=1$, $\varepsilon_1=1$, and $\lambda(0)=1$ in the computations.    }\label{figure0}
\end{figure}

To illustrate the application of Eq.~(\ref{timeevolutionequationclosedsysteminclusivework}), we concretely calculate the CF of a quantum piston with a moving wall~\cite{Quan2012}; see the inset of Fig.~(\ref{figure0}). Suppose that the uniform speed of the wall is $v$. We are interested in this model because the piston Hamiltonian appears to depend only on the kinetic energy, $p^2/2m$. Because the superoperator $W_\eta$ is always zero, Eq.~(\ref{timeevolutionequationclosedsysteminclusivework}) seems problematic in calculating the CF of the inclusive work. However, this is not the whole story: the Hamiltonian acts on the Hilbert space of the system of the quantum piston, and the space is changing in time due to the moving wall; therefore, $H(t)$ indeed depends on time. This point would become clear if we explicitly wrote and solved this equation in the instantaneous energy representation. Figure~(\ref{figure0}) shows two probability distributions of the inclusive work at two moving speeds and their corresponding CFs. These distributions are directly calculated using the definition, Eq.~(\ref{distributioninclusiveworkclosedsystem}). The CFs are obtained in two ways: performing Fourier transforms of these work distributions(see the black solid and dashed lines) or solving Eq.~(\ref{timeevolutionequationclosedsysteminclusivework}) by a simple mathematical program and then utilizing Eq.~(\ref{CFworkclosedsystem}) (see the symbols in the same figure). We have assumed that at time $0$, the piston is in a thermal equilibrium state. We see that at a low pulling speed, the results obtained under these two different methods are in good agreement, but an apparent deviation has occurred at the higher speed. We think that this discrepancy is induced by the numerical errors in solving the differential equation~(\ref{timeevolutionequationdensitymatrix}); $W_\eta(t)$ is a rapidly oscillating term with respect to time in the energy representation. To avoid additional numerical efforts, we find that solving the Liouville-von Neumann equation~(\ref{timeevolutionequationdensitymatrix}) of $\hat\rho(t,\eta)$ with an initial condition $e^{-i\eta H_0}\rho_0$ is a simple and efficient method. The calculated results are shown in the same panels using the red solid and dashed lines. We find that they are highly consistent with those obtained by directly performing Fourier transforms of the work distributions.

\section{Heat and work in open quantum systems}
\label{section4}
\subsection{TEM scheme applied to the composite system}
\label{subsection4A}
\begin{figure}
\label{figure1}
\includegraphics[width=1.\columnwidth]{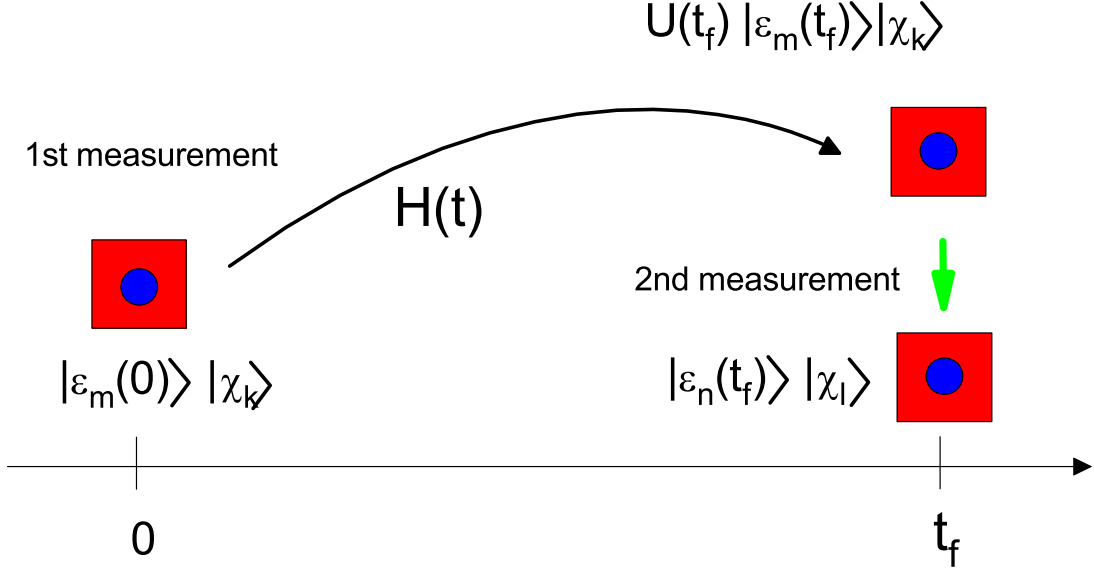}
\caption{A schematic diagram describing the work definition~(\ref{inclusiveworkdefinitionopensystem}) based on the TEM scheme applied to the combined quantum system (the blue circles) and heat bath (red squares). The evolution of the global system is unitary under the Hamiltonian in Eq.~(\ref{compositeHamiltonian}). The green arrow on the right-hand side denotes the projected energy measurement on the system and bath at the end time $t_f$.}
\end{figure}
As we have mentioned in Sec.~(\ref{section2}), an open quantum system A and its heat bath B can be combined as a closed composite system. Hence, the TEM scheme can be naturally extended to define stochastic thermodynamic quantities in the open situation. The global Hamiltonian is given by  Eq.~(\ref{compositeHamiltonian}). Note that we have assumed that its time dependence is through a protocol $\lambda(t)$ acting only on the quantum system A.

We are concerned about the changes in energy of subsystem A and heat bath B. Following Kurchan~\cite{Kurchan2000}, Talkner et al.~\cite{Talkner2009} and Esposito et al.~\cite{Esposito2009}, we define the difference of the energy eigenvalues of the heat bath Hamiltonian, $H_B$, between the end and beginning of the non-equilibrium process as the heat {\it released} by quantum system A. Formally, the stochastic heat is
\begin{eqnarray}
\label{heatdefintionopensystem}
Q_{lk}&=&\chi_l-\chi_k.
\end{eqnarray}
Hence, its probability distribution is
\begin{eqnarray}
\label{distributioninheat}
P(Q)&=&\sum_{l,k} \delta(Q-Q_{lk}){\rm Tr}\left [  |\chi_l\rangle \langle \chi_l| U(t_f)\rho_A(0)\otimes  |\chi_k \rangle \langle \chi_k| U(t_f)^\dag \right ]  p_{k},
\end{eqnarray}
where $p_{k}$ is the probability of finding the bath at the eigenvector $|\chi_k\rangle$; see Eqs.~(\ref{densitymatrixheatbath}) and~(\ref{canonicaldistributionheatbath}). Correspondingly, the Fourier transform of the distribution or the CF of the heat is
\begin{eqnarray}
\label{CFheatopensystemorg}
\Phi_h(\eta)
&=&{\rm Tr}\left[e^{i\eta H_B}U(t_f) e^{-i\eta H_B}\rho_A(0)\otimes\rho_B U(t_f)^\dag\right].
\end{eqnarray}
The reader is reminded that we did not impose any restriction on the initial density matrix of the open system A.

Considering that the global quantum system~(\ref{compositeHamiltonian}) is closed, we may define the inclusive or exclusive work as we did in the preceding section. In particular, considering that the interaction between system A and heat bath B is very weak, we can approximate the inclusive work as~\cite{Talkner2009,Crooks2008a}
\begin{eqnarray}
\label{inclusiveworkdefinitionopensystem}
W_{lnkm}=[\chi_l+\varepsilon_n(t_f)]-[\chi_k+\varepsilon_m(0)]
\end{eqnarray}
by disregarding the effect of the interaction term. Here, $\varepsilon_m(t)$ is the instantaneous energy eigenvalue of $H_A(t)$. To ensure the validity of the work definition, we have to impose the condition
\begin{eqnarray}
[\rho_A(0),H_A(0)]=0.
\end{eqnarray}
Therefore, in terms of the condition of the initial density matrix, the case of work is distinct from the case of heat. Given these notations, we may easily obtain the distribution of the inclusive work of the open quantum system:
\begin{eqnarray}
\label{distributioninclusiveworkopensystem}
P(W)=\sum_{k,n,m,l} \delta(W-W_{lnkm}) {\rm Tr}\left [ |\varepsilon_n(t_f)\rangle |\chi_l\rangle  \langle \chi_l |\langle \varepsilon_n(t_f) | U(t_f)  |\varepsilon_m(0)\rangle \langle \varepsilon_m(0)|  \otimes  |\chi_k \rangle \langle \chi_k| U(t_f)^\dag \right ]P_{km}(0),
\end{eqnarray}
where $P_{km}(0)$ is the joint probability of finding the composite system at a state $|\varepsilon_m(0)\rangle\otimes |\chi_l\rangle $,
\begin{eqnarray}
P_{km}(0)={\rm Tr}\left[|\varepsilon_m(0)\rangle |\chi_k\rangle \langle \chi_k |\langle \varepsilon_m(0)| \rho(0)\right].
\end{eqnarray}
Note that $\rho(0)$ is assumed to be uncorrelated; see Eq.~(\ref{initialdensitymatrix}).
Correspondingly, we may write the CF of the inclusive work of the open quantum system as
\begin{eqnarray}
\label{CFinclusiveworkopensystemorg}
\Phi_w(\eta)={\rm Tr}[e^{i\eta [H_A(t_f)+H_B] }U(t_f) e^{-i\eta [H_A(0)+H_B]}\rho_A(0)\otimes\rho_B U(t_f)^\dag]
\end{eqnarray}

According to the work terms proposed by Jarzynski~\cite{Jarzynski2007}, if the Hamiltonian of quantum system A is composed of two parts, as in Eq.~(\ref{freeandinteractionHamiltonian}), i.e.,
\begin{eqnarray}
\label{freeopenquantumsystemHamiltonian}
H_A(t)=H_0+H_1(t).
\end{eqnarray}
we can define the exclusive work and its CF as Eqs.~(\ref{inclusiveworkdefinitionopensystem})-(\ref{CFinclusiveworkopensystemorg}) with slight modifications. For instance, if we remove the time $t_f$ and the circle brackets on both sides on the RHS of  Eq.~(\ref{inclusiveworkdefinitionopensystem}), this equation is simply the formal definition of the exclusive work,
\begin{eqnarray}
\label{exclusiveworkdefinitionopensystem}
W^0_{lnkm}=(\chi_l+\varepsilon_n)-(\chi_k+\varepsilon_m).
\end{eqnarray}
We may apply similar treatments to the other equations, and thus, we do not write them out explicitly. An important observation is that although the heat and work are different in energy transfer forms, both of their CFs have mathematical formulas analogous to Eq.~(\ref{CFworkclosedsystem}). Hence, we are able to introduce HCO and WCO and derive their time evolution equations. Finally, we want to mention that we may also define the stochastic inner energy of  system A and its CF. The interested reader may realize this straightforwardly.

\subsection{Heat characteristic operator}
\label{section4B}
According to the structure of Eq.~(\ref{CFheatopensystemorg}), we call the whole term in its trace the HCO of the quantum open system A, i.e.,
\begin{eqnarray}
\label{heatcharacteristicoperatoropensystemorg}
\hat{\rho}(t,\eta)=e^{i\eta H_B}U(t) e^{-i\eta H_B}\rho_A(0)\otimes\rho_B U(t)^\dag.
\end{eqnarray}
This newly defined operator satisfies a simple time evolution equation,
\begin{eqnarray}
\label{timeevolutionequationheatopensystemorg}
\partial_t{\hat{\rho}(t,\eta) } =-i[H_A(t)+H_B+V,\hat{\rho}(t,\eta)] -i[e^{i\eta H_B},V] e^{-i\eta H_B} \hat{\rho}(t,\eta).
\end{eqnarray}
and its initial condition is $\rho_A(0)\otimes\rho_B$. We see that there are similarities between Eqs.~(\ref{timeevolutionequationheatopensystemorg}) and~(\ref{timeevolutionequationclosedsystemexclusivework}). Although the above equation is exact, it is inconvenient because of the involvement of the complicated global evolution. The idea of QMEs may be used to address this issue~\cite{Breuer2002}. To our knowledge, Esposito et al.~\cite{Esposito2009,Esposito2014} were the first to carry out such an effort. Because they used a modified Hamiltonian, they derived an evolution equation that is distinct from our equation.

Following the work in Sec.~(\ref{section2B}), we first write Eq.~(\ref{timeevolutionequationheatopensystemorg}) in the interaction picture with respect to the free Hamiltonian, $H_A(t)+H_B$:
\begin{eqnarray}
\label{timeevolutionequationheatopensystemorginteractionpicture}
\partial_t {\widetilde {\hat{\rho}}}(t,\eta)&=&-i[\widetilde{V}(t), {\widetilde {\hat{\rho}}(t,\eta)}] -i[e^{i\eta H_B}, \widetilde {V}(t)] e^{-i\eta H_B}{\widetilde {\hat{\rho}}(t,\eta)} \nonumber \\
&\equiv&\widetilde{\cal V}(t) {\widetilde {\hat{\rho}}(t,\eta)} + \widetilde{\cal L}_h(t) {\widetilde {\hat{\rho}}(t,\eta)},
\end{eqnarray}
Based on the projection operators defined in Eq.~(\ref{projectionoperatordefinitions}), it is not difficult to prove that
\begin{eqnarray}
\label{projectionoperatorproperty11}
{\cal P}\widetilde {\cal L}_h{\cal P}=0.
\end{eqnarray}
Applying this property and  applying  $\cal P$ and $\cal Q$ to Eq.~(\ref{timeevolutionequationheatopensystemorginteractionpicture}), we obtain
\begin{eqnarray}
\label{projectedheateq1}
\partial_t {\cal P}{\widetilde {\hat{\rho}}(t,\eta)}&=&{\cal P}\left(\widetilde{\cal V}+\widetilde{\cal L}_h\right)(t){\cal Q}{\widetilde {\hat{\rho}}(t,\eta)},\\
\label{projectedheateq2}
\partial_t {\cal Q}{\widetilde {\hat{\rho}}(t,\eta)}&=&{\cal Q}\left(\widetilde{\cal V}+\widetilde{\cal L}_h\right)(t){\cal Q}{\widetilde {\hat{\rho}}(t,\eta)}+{\cal Q}\left(\widetilde{\cal V}+\widetilde{\cal L}_h\right)(t){\cal P}{\widetilde {\hat{\rho}}(t,\eta)}.
\end{eqnarray}
The second equation has the formal solution
\begin{eqnarray}
\label{solutionprojectedheateq2}
{\cal Q}{\widetilde {\hat{\rho}}(t,\eta)}= \int_0^t ds {\cal G}_h(t,s){\cal Q}\left(\widetilde{\cal V}+\widetilde{\cal L}_h\right)(s){\cal P}{\widetilde {\hat{\rho}}(s,\eta)},
\end{eqnarray}
where we have considered the condition ${\cal Q}{\widetilde {\hat{\rho}}}(0,\eta)=0$ and defined the propagator
\begin{eqnarray}
{\cal G}_h(t,s)=T_\leftarrow e^{\int_s^t du {\cal Q}\left(\widetilde{\cal V}+\widetilde{\cal L}_h\right)(u) }.
\end{eqnarray}
On the other hand, the similarity between Eqs.~(\ref{timeevolutionequationheatopensystemorginteractionpicture}) and~(\ref{timeevolutionequationclosedsysteminclusivework}) reminds us that ${\widetilde {\hat{\rho}}}$ at two time points $s$ and $t$ ($s\le t$) are connected by
\begin{eqnarray}
\label{heatKattwotimesst}
{\widetilde {\hat{\rho}}(s,\eta)}&=& U_V(s)\left[T_{\rightarrow}e^{-\int_s^t du U_V(u)^\dag \widetilde{\cal L}_h(u)U_V(u)}\right]U_V(t)^\dag {\widetilde {\hat{\rho}}(t,\eta)}  U_V(t) U_V(s)^\dag \equiv\widetilde{ {\cal U}}_h(s,t){\widetilde {\hat{\rho}}(t,\eta)},
\end{eqnarray}
where $U_V(t)=U_0(t)^\dag U(t)$ is the time evolution operator of the interaction picture operator ${\widetilde V}(t)$; see Eq.~(\ref{timeevolutionoperatorinteractionpictureoperator}). Substituting Eqs.~(\ref{solutionprojectedheateq2}) and~(\ref{heatKattwotimesst}) into Eq.~(\ref{projectedheateq1}), we obtain
\begin{eqnarray}
\label{timeevolutionequationheatopensystemorginteractionpictureprojected}
\partial_t{\cal P}{\widetilde {\hat{\rho}}(t,\eta)} = \int_0^t ds   {\cal P}\left(\widetilde{\cal V}+\widetilde{\cal L}_h\right)(t){\cal G}(t,s){\cal Q}\left(\widetilde{\cal V}+\widetilde{\cal L}_h\right)(s){\cal P}\widetilde{ {\cal U}}_h(s,t){\widetilde {\hat{\rho}}(t,\eta)},
\end{eqnarray}
The reader is reminded that this equation also has a time-convolutionless form that is analogous to Eq.~(\ref{timeevolutionequationopensystemorginteractionpictureprojected}).

To eliminate the complex integral in Eq.~(\ref{timeevolutionequationheatopensystemorginteractionpictureprojected}), we must resort to some approximations, as we did in deriving MQMEs previously. First, redefining $V$ as $\alpha V$, we expand the above equation up to second order in $\alpha$ and then obtain
\begin{eqnarray}
\label{timeevolutionequationKHeatOpen5}
\partial_t{\cal P}{\widetilde {\hat{\rho}}(t,\eta)}= \alpha^2\int_0^t ds   {\cal P}\left(\widetilde{\cal V}+\widetilde{\cal L}_h\right)(t)\left(\widetilde{\cal V}+\widetilde{\cal L}_h\right)(t-s){\cal P} {\widetilde {\hat{\rho}}(t,\eta)} .
\end{eqnarray}
This approximation is based on
\begin{eqnarray}
\widetilde{ {\cal U}}_h(s,t){\widetilde {\hat{\rho}}(t,\eta)}={\widetilde {\hat{\rho}}(t,\eta)}+{\cal O}(\alpha),
\end{eqnarray}
where we made the change of variables $s\rightarrow t-s$. To explicitly write the projector operator $\cal P$, the superoperators $\widetilde {\cal L}_h$ and $\widetilde {\cal V}$, we obtain
\begin{eqnarray}
\label{timeevolutionequationheatopensystemorginteractionpictureprojectedexplicit}
\partial_t {\widetilde {\hat{\rho}}_A(t,\eta)}=&&-\alpha^2\sum_{a,b}\int_0^t ds {\widetilde A}_a^\dag(t) {\widetilde A}_b(t-s){\widetilde {\hat{\rho}}_A(t,\eta)} {\rm Tr}_B[ \widetilde B_a(s) \widetilde B_b(0)\rho_B]\nonumber\\
&&+\alpha^2\sum_{a,b}\int_0^t ds {\widetilde A}_b(t-s){\widetilde {\hat{\rho}}_A(t,\eta)} {\widetilde A}_a^\dag (t)  {\rm Tr}_B[ \widetilde B_a(s-\eta) \widetilde B_b(0)\rho_B] \nonumber\\
&&-\alpha^2\sum_{a,b}\int_0^t ds {\widetilde {\hat{\rho}}_A(t,\eta)} {\widetilde A}_b(t-s){\widetilde A}_a^\dag(t) {\rm Tr}_B[ \widetilde B_b(-s) \widetilde B_a(0)\rho_B] \nonumber\\
&&+\alpha^2\sum_{a,b}\int_0^t ds {\widetilde A}_a^\dag(t) {\widetilde {\hat{\rho}}_A(t,\eta)} {\widetilde A}_b(t-s) {\rm  Tr}_B[ \widetilde B_b(-s-\eta) \widetilde B_a(0)\rho_B],
\end{eqnarray}
where
\begin{eqnarray}
 {\widetilde {\hat{\rho}}_A(t,\eta)}={\rm Tr}_B[{\widetilde {\hat{\rho}}(t,\eta)}],
\end{eqnarray}
and its initial condition is $\rho_A(0)$. Because taking the trace of the operator $ {\widetilde {\hat{\rho}}_A(t,\eta)}$ (the interaction picture operator) over the degree of freedom of  system A is just the CF~(\ref{CFheatopensystemorg}), we also call it the HCO of the open quantum system A. The next step is to check whether Eq.~(\ref{timeevolutionequationheatopensystemorginteractionpictureprojectedexplicit}) has a Markovian form with respect to concrete open quantum systems. We have to do this on a case-by-case basis.

\subsubsection{Static Hamiltonian}
Inserting the decomposition~(\ref{decompositionofAoperatorstaticHamiltonian}) into Eq.~(\ref{timeevolutionequationheatopensystemorginteractionpictureprojectedexplicit}), we obtain
\begin{eqnarray}
\label{timeevolutionequationheatopensystemorginteractionpictureprojectedexplicitstaticHamiltonian}
\partial_t {\widetilde{ \hat{\rho}}}_A(t,\eta)=-&& \alpha^2 \sum_{a,b,\omega,\omega'} e^{i(\omega'-\omega)t }A_a^\dag(\omega') { A}_b(\omega){\widetilde {\hat{\rho}}_A(t,\eta)} \int_0^t ds e^{i\omega s}{\rm Tr}_B[\widetilde B_a(s)B_b\rho_B] \nonumber\\
&&+\alpha^2\sum_{a,b,\omega,\omega'} e^{i(\omega'-\omega)t }{A}_b(\omega){\widetilde {\hat{\rho}}_A(t,\eta)} A_a^\dag(\omega')\int_0^t ds e^{i\omega s}{\rm Tr}_B[\widetilde B_a(s-\eta)B_b\rho_B] \nonumber\\
&&-\alpha^2\sum_{a,b,\omega,\omega'} e^{i(\omega'-\omega)t }{\widetilde {\hat{\rho}}_A(t,\eta)} { A}_b(\omega)A_a^\dag(\omega') \int_0^t ds e^{i\omega s}{\rm Tr}_B[\widetilde B_b(-s)B_a\rho_B]\nonumber\\
&&+\alpha^2 \sum_{a,b,\omega,\omega'} e^{i(\omega'-\omega)t }A_a^\dag(\omega'){\widetilde {\hat{\rho}}_A(t,\eta)} { A}_b(\omega)  \int_0^t ds e^{i\omega s}{\rm Tr}_B[\widetilde B_b(-s-\eta)B_a\rho_B].
\end{eqnarray}
According to the procedure for deriving the MQEMs, what follows is to apply the two approximations. The rotation wave approximation seems to still be reasonable. However, the replacement of the upper integral limit $t$ by an infinity or the Markovian approximation given by Esposito et al.~\cite{Esposito2009} seems to be problematic here. As we have mentioned before, the validity of the replacement is based on the fact that the significant non-zero contribution of the two-time-correlation functions, the trace terms ${\rm Tr}_B$ in Eq.~(\ref{timeevolutionequationheatopensystemorginteractionpictureprojectedexplicitstaticHamiltonian}), shall be around $s=0$~\cite{Breuer2002,Rivas2012}. Obviously, the presence of $\eta$ here changes the situation. For instance, if we choose $\eta$ to be far larger than the time $t$, the second integral in the equation is negligible. In contrast, the same integral but with an upper limit of infinity  is not at all zero. However, we must emphasize that this approximation will become exact in the weak-coupling limit, namely, $\alpha\rightarrow 0$, $t\rightarrow\infty$ and keeping $\alpha^2 t$ constant~\cite{Davies1974,Gorini1976,Alicki2006,Rivas2012}. For a physically relevant finite $\alpha$, the Markovian approximation should work well for the cases in which $\eta$ is approximately 0 or the time $t$ is very large; for other cases, it is at most considered as an ansatz to obtain a simple Markovian evolution equation. Note that the validity of the MQMEs shall be understood in the same manner~\cite{Rivas2012}. Therefore, if we admit these two approximations, introduce the double-sided Fourier integrals (some details are presented in Appendix B), and perform a simple calculation, we obtain the following result:
\begin{eqnarray}
\label{timeevolutionequationheatopensystemorginteractionpictureprojectedexplicitstaticHamiltonianfinal}
\partial_t {\widetilde {\hat{\rho}}_A(t,\eta)} =&-&i[{H}_{LS},{\widetilde {\hat{\rho}}_A(t,\eta)} ]\nonumber \\
&+& \sum_{\omega,a,b}r_{ab}(\omega)\left[ e^{i\eta\omega}{ A}_b(\omega){\widetilde {\hat{\rho}}_A(t,\eta)} A_a^\dag(\omega)-\frac{1}{2}\left\{A_a^\dag(\omega) {A}_b(\omega),{\widetilde {\hat{\rho}}_A(t,\eta)}\right \} \right],
\end{eqnarray}
Eq.~(\ref{timeevolutionequationheatopensystemorginteractionpictureprojectedexplicitstaticHamiltonianfinal}) can be transformed back into the Schr$\ddot{o}$dinger picture by applying a procedure analogous to Eq.~(\ref{masterequationstaticHamiltoniandensitymatrixcase}).

\subsubsection{Weakly driven Hamiltonian}
\label{weaklydrivenHamiltonian}
This case is almost the same as the static Hamiltonian case. We may obtain Eqs.~(\ref{timeevolutionequationheatopensystemorginteractionpictureprojectedexplicitstaticHamiltonian}) and~(\ref{timeevolutionequationheatopensystemorginteractionpictureprojectedexplicitstaticHamiltonianfinal}) again except that the Bohr frequency, $\omega$,  and the spectral decomposition, $A_b(\omega)$ and $A_a^\dag(\omega)$, are all with respect to the bare Hamiltonian $H_0$; see Eq.~(\ref{weakdrivenHamiltonian}).

\subsubsection{Periodically driven Hamiltonian}
Because the spectral decomposition, Eq.~(\ref{decompositionofAoperatorperiodicallydrivenHamiltonian}), is exactly the same as that of the case of the static Hamiltonian, we obtain a time evolution equation analogous to Eq.~(\ref{masterequationperiodicallydrivenHamiltoniandensitymatrixcase}) under the two approximations in the Schr$\ddot{o}$dinger picture; the only  difference is that an additional exponential phase factor, $e^{i\eta\omega}$, is present in front of the first operator $A_b(\omega,t)$ here. To our knowledge, Gasparinetti et al.~\cite{Gasparinetti2014} was the first to obtain such a time evolution equation. Rather than the very general form that we obtained here, they were concerned with a specific two-level system and wrote it in the Floquet basis representation.

\subsubsection{Adiabatically driven Hamiltonian}
Compared to the previous three cases, the adiabatically driven situation is slightly complicated because the presence of $\eta$ in Eq.~(\ref{timeevolutionequationheatopensystemorginteractionpictureprojectedexplicit}) seems to render the approximation in Eq.~(\ref{approximatetimeevolutionoperatorinadiabaticallydrivenHamiltonian}) invalid in general: the correlation function is significantly non-zero around $s=\eta$, whereas $\eta$ is arbitrary, and $s$ does not need to be a small number. However, this difficulty will disappear if we take the weak coupling limit and the quantum adiabatical limit simultaneously~\cite{Davies1978,Thunstroem2005,Oreshkov2010}. We have given an explanation in Sec.~(\ref{section2B4}). Hence, we can obtain a time evolution equation analogous to Eq.~(\ref{masterequationadiabaticallydrivenHamiltonian}) except for the presence of the factor $e^{i\eta\omega_{mn}(t)}$ in front of the first ${A}_{b,mn}(t)$. \\

Based on the above discussions, we find that for the four types of QMEs that we have studied so far, their evolution equations of the HCO can be unified as specific cases of the generic case:
\begin{eqnarray}
\label{Heatmasterequationgeneralform}
\partial_t  \hat\rho_A(t,\eta)&=&-i[H_A(t)+H_{LS}(t), \hat\rho_A(t,\eta)]+\nonumber\\
&&\sum_{\omega_t,a,b}r_{ab}(\omega_t)\left[ e^{i\eta\omega_t}A_b(\omega_t,t)\hat\rho_A(t,\eta) A_a^\dag(\omega_t,t)-\frac{1}{2}\left\{A_a^\dag(\omega_t,t)A_b(\omega_t,t),\hat\rho_A(t,\eta)\right\}\right]\nonumber\\
&\equiv&\check{{\cal L}}(t,\eta) \hat\rho_A(t,\eta).
\end{eqnarray}
We may write the solution of Eq.~(\ref{Heatmasterequationgeneralform}) formally as
\begin{eqnarray}
\label{propagatorsolutionofheatequation1}
\hat\rho_A(t,\eta)= T_{\leftarrow} e^{\int_0^t ds{\check{\cal L}}(s,\eta) }[\rho_A(0)],
\end{eqnarray}
where the whole time-ordered exponential term is a superpropagator. The CF of the heat is then
\begin{eqnarray}
\label{CFheatopensystemgeneral}
\Phi_h(\eta)={\rm Tr}_A[\hat\rho_A(t_f,\eta)].
\end{eqnarray}
We must bear in mind that Eq.~$(\ref{Heatmasterequationgeneralform})$ has a rigorous meaning only in the weak coupling limit and quantum adiabatical limit (only for the adiabatically driven case). Otherwise, it shall be regarded as an approximation in solving the HCO of the open quantum system. We expect that this equation shall be sound under the conditions of smaller $\eta$ and larger $t$.

\subsection{Work characteristic operators}
\label{subsection4C}
There are two equivalent methods for obtaining the CF~(\ref{CFinclusiveworkopensystemorg}) of the inclusive work. We first call the whole term in its trace the WCO $K(t,\eta)$ of the open quantum system, namely,
\begin{eqnarray}
\label{inclusiveworkcharacteristicoperatoropensystemorg}
K(t,\eta)&=& e^{i\eta H_A(t) } \left[ e^{i\eta H_B}U(t) e^{-i\eta H_B}\left( e^{-i\eta H_A(0)} \rho_A(0)\right)\otimes\rho_B U(t)^\dag\right].
\end{eqnarray}
Here, we added two square brackets. We immediately find that the term in the brackets is almost the same as the HCO, Eq.(\ref{heatcharacteristicoperatoropensystemorg}), except that the initial ``density matrix" of the system A has become $ e^{-i\eta H_A(0)} \rho_A(0)$. We observed the same structure when we investigated the WCO in the closed quantum system; see the second equation in Eq.~(\ref{CFworkclosedsystem}). This also explains why we called it the HCO in closed quantum systems even though there is no heat concept therein. Therefore, the CF of the inclusive work can be evaluated by first solving the time evolution Eq.~(\ref{Heatmasterequationgeneralform}) with the new initial condition and then taking the trace. This procedure is shown in the following equation:
\begin{eqnarray}
\label{CFinclusiveworkopensystemgeneralbasedheat}
\Phi_w(\eta)={\rm Tr}_A\left[e^{i\eta H_A(t)}T_{\leftarrow} e^{\int_0^t ds{\check{\cal L}}(s,\eta)}\left(e^{-i\eta H_A(0)}\rho_A(0)\right)\right].
\end{eqnarray}
This method has been used by Silaev et al.~\cite{Silaev2014} and Cuetara et al.~\cite{Cuetara2015} in several special MQMEs. On the other hand, we are curious as to whether there exists a time evolution equation of the WCO that is defined only on the degree of freedom of system A similar to Eq.~(\ref{Heatmasterequationgeneralform}): when we solve it, we can obtain the CF of the inclusive work by simply taking the trace over the open quantum system A. This consideration is very natural because the roles of the work and heat are equal. This is not difficult to realize if we regard the whole term in the trace of Eq.~(\ref{CFinclusiveworkopensystemgeneralbasedheat}) as the WCO, $K_A(t,\eta)$, of the open quantum system A. Taking the time derivative and using the known Eq.~(\ref{Heatmasterequationgeneralform}), we obtain
\begin{eqnarray}
\label{inclusiveworkmasterequationgeneralform}
\partial_t K_A(t,\eta)&=&e^{i\eta H_A(t)}{\check{\cal L}}(\eta,t)\left [e^{-i\eta H_A(t)}K_A(t,\eta)\right] +
\partial_t \left[ e^{i\eta H_A(t)}\right ]e^{-i\eta H_A(t)}K_{A}(t,\eta).
\end{eqnarray}
We must emphasize that the initial condition of $K_A(t,\eta)$ is the genuine density matrix, $\rho_A(0)$. When we solve the equation, the CF of the inclusive work is then
\begin{eqnarray}
\Phi_w(\eta)={\rm Tr}_A[K_A(t_f,\eta)].
\end{eqnarray}
From the perspective of numerical calculations, the former method is more attractive than the latter. We see that these two methods are parallel to those that we used for the inclusive work in the closed quantum systems; see Eqs.~(\ref{CFworkclosedsystem}) and~(\ref{timeevolutionequationclosedsysteminclusivework}).

In addition to the inclusive work, for the given Hamiltonian~(\ref{freeopenquantumsystemHamiltonian}), we have defined the stochastic exclusive work, Eq.~(\ref{exclusiveworkdefinitionopensystem}), of the open quantum system. Mimicking Eq.~(\ref{inclusiveworkcharacteristicoperatoropensystemorg}), we also define the WCO
\begin{eqnarray}
\label{exclusiveworkcharacteristicoperatoropensystemorg}
K_0(t,\eta)&=& e^{i\eta H_0} \left[ e^{i\eta H_B}U(t) e^{-i\eta H_B}\left( e^{-i\eta H_0} \rho_A(0)\right)\otimes\rho_B U(t)^\dag\right].
\end{eqnarray}
Using very similar arguments, we can calculate the CF of the exclusive work by solving Eq.~(\ref{Heatmasterequationgeneralform}) with the initial condition $e^{-i\eta H_0} \rho_A(0)$ and then taking the trace over the degree of freedom of system A, namely,
\begin{eqnarray}
\label{CFexclusiveworkopensystemgeneralbasedheat}
\Phi_{w_0}(\eta)={\rm Tr}_A\left[e^{i\eta H_0}T_{\leftarrow} e^{\int_0^t ds{\check{\cal L}}(s,\eta)}\left(e^{-i\eta H_0}\rho_A(0)\right)\right].
\end{eqnarray}
In addition, we can also define the whole term in the trace of the above equation as the WCO, $K_{A}^0(t,\eta)$, on the open quantum system A. Then, we have an alternative method to obtain the CF by directly solving the following equation:
\begin{eqnarray}
\label{exclusiveworkmasterequationgeneralform}
\partial_t K_{A}^0(t,\eta)&=&e^{i\eta H_0}{\check{\cal L}}(\eta,t)\left [e^{-i\eta H_0}K_{0_A}(t,\eta)\right],
\end{eqnarray}
where
\begin{eqnarray}
\Phi_{w_0}(\eta)={\rm Tr}_A[K_A^0(t_f,\eta)].
\end{eqnarray}
Note that its initial condition is $\rho_A(0)$.

These pretty results raise an interesting question: can we directly obtain Eqs.~(\ref{inclusiveworkmasterequationgeneralform}) or~(\ref{exclusiveworkmasterequationgeneralform}) using the idea of QMEs as we have done in the case of heat? We have mentioned that the roles of work and heat are equal. Hence, one should be able to obtain the time equations concerning the work first and then derive the equations for the heat later. To answer this question, we write the time evolution of the operator defined in Eq.~(\ref{inclusiveworkcharacteristicoperatoropensystemorg}):
\begin{eqnarray}
\label{timeevolutionequationworkopensystemorg}
\partial_t K(t,\eta) =&&-i[H_A(t)+H_B+V,K(t,\eta)]\nonumber\\
&&-i\left[e^{i\eta[H_A(t)+H_B]},V\right]e^{-i\eta[H_A(t)+H_B]} K(t,\eta) +\partial_t \left [e^{i\eta H_A(t)}\right] e^{-i\eta H_A(t)} K(t,\eta).
\end{eqnarray}
We transform it into the interaction picture with respect to the free Hamiltonian $H_A(t)+H_B$,
\begin{eqnarray}
\label{timeevolutionequationworkopensystemorginteractionpicture}
\partial_t {\widetilde K}(t,\eta)&=&\widetilde{\cal V}(t) {\widetilde K}(t,\eta) + \widetilde{\cal L}_w(t) {\widetilde K}(t,\eta)+{\widetilde{\cal W}}_A(t) {\widetilde K}(t,\eta),
\end{eqnarray}
where
\begin{eqnarray}
&&\widetilde{\cal L}_w(t) \equiv -i\left[e^{i\eta[\widetilde{H}_A(t)+H_B]},\widetilde{V}(t)\right]e^{-i\eta[\widetilde{H}_A(t)+H_B]},\\
&&{\widetilde{\cal W}}_A(t)\equiv\partial_t \left [e^{i\eta \widetilde{H}_A(t)}\right] e^{-i\eta \widetilde{H}_A(t)}.
\end{eqnarray}
In analogy to Eq.~(\ref{projectionoperatorproperty1}), we prove
\begin{eqnarray}
\label{projectionoperatorproperty2}
{\cal P}{\widetilde {\cal L}}_w(t){\cal P}=0\hspace{0.5cm}{\rm and}\hspace{0.5cm}
{\cal P}{\widetilde {\cal W}}_A(t)={\widetilde {\cal W}}_A(t){\cal P}.
\end{eqnarray}
Hence,  applying the projection operators $\cal P$ and $\cal Q$ to Eq.~(\ref{timeevolutionequationworkopensystemorginteractionpicture}) once more, we obtain
\begin{eqnarray}
\label{projectedworkeq1}
\partial_t {\cal P}{\widetilde K}(t,\eta)&=&{\cal P}\left(\widetilde{\cal V}+\widetilde{\cal L}_w\right)(t){\cal Q}{\widetilde K}(t,\eta)+{\widetilde {\cal W}}_A(t){\cal P}{\widetilde K}(t,\eta),\\
\label{projectedworkeq2}
\partial_t {\cal Q}{\widetilde K }(t,\eta)&=&\left[{\cal Q}\left(\widetilde{\cal V}+\widetilde{\cal L}_w\right)(t)+{\widetilde {\cal W}}_A(t)\right] {\cal Q}{\widetilde K}(t,\eta)+{\cal Q}\left(\widetilde{\cal V}+\widetilde{\cal L}_w\right)(t){\cal P}{\widetilde K}(t,\eta).
\end{eqnarray}
Considering the vanishing initial condition, ${\cal Q}{\widetilde K}(0,\eta)=0$, we write the formal solution of the second equation as
\begin{eqnarray}
\label{solutionprojectedworkeq2}
{\cal Q}{\widetilde {K}(t,\eta)}= \int_0^t ds {\cal G}_w(t,s){\cal Q}\left(\widetilde{\cal V}+\widetilde{\cal L}_w\right)(s){\cal P}{\widetilde K(s,\eta)},
\end{eqnarray}
where the superpropagator is
\begin{eqnarray}\label{G1superoperator}
{\cal G}_w(t,s)=T_\leftarrow e^{\int_s^t du [{\cal Q}\left(\widetilde{\cal V}+\widetilde{\cal L}_w\right)(u)+{\widetilde {\cal W}}_A(u) ]}.
\end{eqnarray}
Substituting the solution into Eq.~(\ref{projectedworkeq1}), we have
\begin{eqnarray}
\label{timeevolutionequationworkopensystemorginteractionpictureprojected}
\partial_t {\cal P}{\widetilde K}(t,\eta)-{\widetilde {\cal W}}_A(t){\cal P}{\widetilde K}(t,\eta)&=& \int_0^t ds   {\cal P}\left(\widetilde{\cal V}+\widetilde{\cal L}_w\right)(t){\cal G}_w(t,s){\cal Q}\left(\widetilde{\cal V}+\widetilde{\cal L}_w\right)(s){\cal P}\widetilde{ {\cal U}}_w(s,t){\widetilde K}(t,\eta)
\end{eqnarray}
where the superoperator $\widetilde{ {\cal U}}_w(s,t)$ is the same as in Eq.~(\ref{heatKattwotimesst}) except that $\widetilde{\cal L}_h$ therein is now replaced with $\widetilde{\cal L}_w+{\widetilde{\cal W}}_A$. This is expected if one notices the similarity between Eqs.~(\ref{timeevolutionequationworkopensystemorginteractionpicture}) and~(\ref{timeevolutionequationheatopensystemorginteractionpicture}).
Redefining $V\rightarrow \alpha V$, we expand the RHS of the above equation up to second order in $\alpha$ and arrive at an equation given by
\begin{eqnarray}
\label{timeevolutionequationworkopensystemorginteractionpictureprojectedleftside}
\alpha^2\int_0^t ds   {\cal P}\left(\widetilde{\cal V}+\widetilde{\cal L}_w\right)(t) T_\leftarrow e^{\int_{t-s}^t du \widetilde{W}_A(u)} \left(\widetilde{\cal V}+\widetilde{\cal L}_w\right)(t-s) T_\rightarrow e^{-\int_{t-s}^t du\widetilde{W}_A(u)} {\cal P} {\widetilde K}(t,\eta).
\end{eqnarray}
To obtain this result, we have used the following two approximations:
\begin{eqnarray}
\label{zeroorderG1}
{\cal G}_w(t,s)=T_\leftarrow e^{\int_s^t du {\widetilde {\cal W}}_A(u) }+{\cal O}(\alpha),
\end{eqnarray}
and
\begin{eqnarray}
\label{zeroorderK}
\widetilde{ {\cal U}}_w(s,t){\widetilde K}(t,\eta)&=& T_{\rightarrow}e^{-\int_s^t du \widetilde W_A(u) } {\widetilde K}(t,\eta)+{\cal O}(\alpha).
\end{eqnarray}
Combining Eqs.~(\ref{timeevolutionequationworkopensystemorginteractionpictureprojected}) and~(\ref{timeevolutionequationworkopensystemorginteractionpictureprojectedleftside}), explicitly writing the projection operator $\cal P$, the superoperators $\widetilde {\cal L}_w$ and $\widetilde {\cal V}$, and performing a simple algebraic manipulation, we have
\begin{eqnarray}
\label{timeevolutionequationworkopensystemorginteractionpictureprojectedexplicitmiddle}
&&\partial_t {\widetilde K}_A(t,\eta)-{\widetilde {\cal W}}_A(t){\widetilde K}_A(t,\eta)\nonumber\\
=&&-\alpha^2\sum_{a,b}\int_0^t ds e^{i\eta{\widetilde H}_A(t)}{\widetilde A}_a^\dag(t) e^{-i\eta{\widetilde H}_A(t)}
T_\leftarrow e^{\int_{t-s}^t du {\widetilde {\cal W}}_A(u) }
e^{i\eta{\widetilde H}_A(t-s)} {\widetilde A}_b(t-s)e^{-i\eta{\widetilde H}_A(t-s)}\nonumber\\&& \hspace{0cm}  T_{\rightarrow}e^{-\int_{t-s}^t du \widetilde W_A(u) } {\widetilde K}_A(t,\eta) {\rm  Tr}_B[ \widetilde B_a(s) \widetilde B_b(0)\rho_B]+\nonumber\\
&&\alpha^2\sum_{a,b}\int_0^t ds T_\leftarrow e^{\int_{t-s}^t du {\widetilde {\cal W}}_A(u) } e^{i\eta{\widetilde H}_A(t-s)}{\widetilde A}_b(t-s)e^{-i\eta{\widetilde H}_A(t-s)}T_{\rightarrow}e^{-\int_{t-s}^t du \widetilde W_A(u) }{\widetilde K}_A(t,\eta) {\widetilde A}_a ^\dag(t)\nonumber \\&&\hspace{0cm} {\rm Tr}_B[ \widetilde B_a(s-\eta) \widetilde B_b(0)\rho_B]- \alpha^2\sum_{a,b}\int_0^t ds {\widetilde K}_A(t,\eta) {\widetilde A}_b(t-s){\widetilde A}_a^\dag(t) {\rm Tr}_B[ \widetilde B_b(-s) \widetilde B_a(0)\rho_B]+ \nonumber\\
&&\alpha^2\sum_{a,b}\int_0^t ds e^{i\eta{\widetilde H}_A(t)}{\widetilde A}_a^\dag(t) e^{-i\eta{\widetilde H}_A(t)} {\widetilde K}_A(t,\eta) {\widetilde A}_b(t-s)  {\rm Tr}_B[ \widetilde B_b(-s-\eta) \widetilde B_a(0)\rho_B],
\end{eqnarray}
where ${\widetilde K}_A(t,\eta)$ is the interaction picture operator of $K_A(t,\eta)$. This lengthy equation looks so complicated that it is difficult to see the equivalence between Eqs.~(\ref{timeevolutionequationworkopensystemorginteractionpictureprojectedexplicitmiddle}) and~(\ref{inclusiveworkmasterequationgeneralform}). However, if we notice the following two identities~\cite{Liu2012,Liu2014b},
\begin{eqnarray}
&&T_{\leftarrow} e^{\int_{t'}^t  du \widetilde W_A(u) }= e^{i\eta \widetilde H_A(t)}e^{-i\eta \widetilde H_A(t')}\\
&&T_{\rightarrow} e^{-\int_{t'}^t  du \widetilde W_A(u) }= e^{i\eta \widetilde H_A(t')}e^{-i\eta \widetilde H_A(t)},
\end{eqnarray}
which can be easily seen from the two equivalent expressions of $K(t,\eta)$ in Eqs.~(\ref{CFworkclosedsystem}) and~(\ref{Kattwotimests}), the RHS of Eq.~(\ref{timeevolutionequationworkopensystemorginteractionpictureprojectedexplicitmiddle}) can be dramatically simplified into
\begin{eqnarray}
\label{timeevolutionequationworkopensystemorginteractionpictureprojectedexplicit}
e^{i\eta{\widetilde H}_A(t)}&& \left\{-\alpha^2\sum_{a,b}\int_0^t ds {\widetilde A}_a^\dag(t) {\widetilde A}_b(t-s)e^{-i\eta{\widetilde H}_A(t)}{\widetilde K}_A(t,\eta){\rm  Tr}_B[ \widetilde B_a(s) \widetilde B_b(0)\rho_B]\right.\nonumber\\
&&+\alpha^2\sum_{a,b}\int_0^t ds {\widetilde A}_b(t-s)e^{-i\eta{\widetilde H}_A(t)}{\widetilde K}_A(t,\eta) {\widetilde A}_a^\dag(t)  {\rm Tr}_B[ \widetilde B_a(s-\eta) \widetilde B_b(0)\rho_B] \nonumber\\
&&-\alpha^2\sum_{a,b}\int_0^t ds e^{-i\eta{\widetilde H}_A(t)}{\widetilde K}_A(t,\eta) {\widetilde A}_b(t-s){\widetilde A}_a^\dag(t) {\rm Tr}_B[ \widetilde B_b(-s) \widetilde B_a(0)\rho_B] \nonumber\\
&&\left.+\alpha^2\sum_{a,b}\int_0^t ds  {\widetilde A}_a^\dag(t) e^{-i\eta{\widetilde H}_A(t)} {\widetilde K}_A(t,\eta) {\widetilde A}_b(t-s)  {\rm Tr}_B[ \widetilde B_b(-s-\eta) \widetilde B_a(0)\rho_B]\right\}.
\end{eqnarray}
We immediately find that Eq.~(\ref{timeevolutionequationworkopensystemorginteractionpictureprojectedexplicit}) is simply the RHS of Eq.~(\ref{timeevolutionequationheatopensystemorginteractionpictureprojectedexplicit}) if ${\widetilde {\hat{\rho}}}_A(t)$ therein is changed to $e^{-i\eta{\widetilde H}_A(t)} {\widetilde K}_A(t,\eta)$ and if the whole formula is multiplied by the exponential phase factor,  $e^{i\eta{\widetilde H}_A(t)}$. Therefore, after combining Eqs.~(\ref{timeevolutionequationworkopensystemorginteractionpictureprojectedexplicitmiddle}) and~(\ref{timeevolutionequationworkopensystemorginteractionpictureprojectedexplicit}) and applying the detailed process with respect to the concrete Hamiltonian $H_A(t)$ as we did in the heat case, we arrive at the general Eq.~(\ref{inclusiveworkmasterequationgeneralform}) for the WCOs on these open quantum systems. A very analogous procedure can be applied to the case of the exclusive work as well. Because this is straightforward, we do not present it here.

Before concluding this section, let us give two comments on the general Eqs.~(\ref{inclusiveworkmasterequationgeneralform}) and~(\ref{exclusiveworkmasterequationgeneralform}). First, the valid conditions of these equations are exactly the same as those of Eq.~(\ref{Heatmasterequationgeneralform}). Second, for the adiabatically driven Hamiltonian, Eq.~(\ref{inclusiveworkmasterequationgeneralform}) has a concise form~\cite{Liu2014a}. According to Eq.~(\ref{eigenoperatorAdiabaticallydrivenHamiltonian}), we can move the exponential term, $e^{-i\eta H_A(t)}$ in Eq.~(\ref{inclusiveworkmasterequationgeneralform}), outside of the first bracket to cancel the front term, $e^{i\eta H_A(t)}$, and find that the original equation is simplified to
\begin{eqnarray}
\label{inclusiveworkmasterequationadiabaticallydriven}
\partial_t K_A(t,\eta)&=& {\cal L}(t) [K_A(t,\eta)] +
\partial_t \left[ e^{i\eta H_A(t)}\right ]e^{-i\eta H_A(t)}K_{A}(t,\eta).
\end{eqnarray}
We see that this structure is consistent with that of Eq.~(\ref{timeevolutionequationclosedsysteminclusivework}) for the inclusive work in closed quantum systems. On the other hand, Eq.~(\ref{exclusiveworkmasterequationgeneralform}) also has a simpler formula for the case of a weakly driven Hamiltonian~\cite{Liu2014}. In this situation, considering
\begin{eqnarray}
\label{eigenoperatorsofweaklydrivenHamiltonian}
[H_0, A_a^\dag(\omega)]=\omega A_a^\dag(\omega), \hspace{1cm}[H_0, A_a(\omega)]=-\omega A_a(\omega).
\end{eqnarray}
we can cancel the two exponential terms about $H_0$ in Eq.~(\ref{exclusiveworkmasterequationgeneralform}) and rewrite the original equation as
\begin{eqnarray}
\partial_t K_{A}^0(t,\eta)&=& {\cal L}(t) [ K_{A}^0(t,\eta)] -i[e^{i\eta H_0},H_1(t)]e^{-i\eta H_0} K_{A}^0(t,\eta).
\end{eqnarray}
Intriguingly, this structure is consistent with that of Eq.~(\ref{timeevolutionequationclosedsystemexclusivework}) for the exclusive work in a closed quantum system.

\subsection{Mean heat and work}
\label{section4D}
It is interesting to compare the thermodynamic quantities defined in ST to those in EQT~\cite{Spohn1978,Spohn1978a,Alicki1979,Kohn2001,Boukobza2006,Kosloff2013,Szczygielski2013,Langemeyer2014,Alicki2017}. We expect that the averaged stochastic quantities of the former would agree with those of the latter. According to the definitions of CFs, Eqs.~(\ref{CFheatopensystemorg}) and~(\ref{CFinclusiveworkopensystemorg}), the averages of the stochastic heat and work can be calculated by taking derivatives of these functions with respect to $\eta$ at $\eta=0$, e.g., the released average heat,
\begin{eqnarray}
\langle Q_{lk} \rangle=-i\left.\frac{d\Phi_h(\eta)}{d\eta}\right|_{\eta=0}.
\end{eqnarray}
Alternatively, we can also obtain the same result by applying a Taylor series expansion of these CFs about $\eta=0$. For instance, we first rewrite Eq.~(\ref{Heatmasterequationgeneralform}) as
\begin{eqnarray}
\label{HeatmasterequationgeneralformforTaylorexpansion}
\partial_t \hat{{\rho}}_A(t,\eta)&=&{\cal{L}}(t)\hat{\rho}_A(t,\eta)+\sum_{\omega_t,a,b}\left(e^{i\eta \omega_t}-1\right)r_{ab}(\omega_t)A_b(\omega_t,t) \hat{{\rho}}_A(t,\eta) A_a^\dag(\omega_t,t)\nonumber\\
&\equiv&{\cal{L}}(t)\hat{\rho}_A(t,\eta)+{\cal Q}_{\eta}(t)\hat{\rho}_A(t,\eta).
\end{eqnarray}
Note that the lowest order of the superoperator ${\cal Q}_\eta$  of $\eta$ is the first order. Based on the superpropagator,  $G(t,s)$ in Eq.~(\ref{propagatorsolutionofheatequation}), where time $0$ there is replaced by time $s$ ($\le t$), the solution of the above equation has an integral expression:
\begin{eqnarray}
\label{Dysonsolutiongeneralheatmasterequation}
\hat{{\rho}}_A(t,\eta)&=&G(t,0)\rho_A(0)+\int_0^t ds G(t,s){\cal Q}_{\eta}(s)\hat{{\rho}}_A(s,\eta)\nonumber\\
&=&G(t,0)\rho_A(0)+\int_0^t ds G(t,s){\cal Q}_{\eta}(s)G(s,0){\rho}_A(0)\nonumber\\
&&\hspace{2cm}+\int_0^tds\int_0^s du G(t,s){\cal Q}_{\eta}(s) G(s,u){\cal Q}_{\eta}(u) \hat{{\rho}}_A(u,\eta).
\end{eqnarray}
Obviously, the first term on the RHS is the density matrix of the open quantum system A at time $t$, $\rho_A(t)$. Substituting the solution into Eq.~(\ref{CFheatopensystemgeneral}), using a Taylor series about $\eta=0$, and collecting all coefficients to first order in $\eta$, we formally obtain the average heat:
\begin{eqnarray}
\label{meanstochasticheat}
\langle Q_{lk} \rangle&=&\int_{0}^{t_f}ds \sum_{\omega_s,a,b}\omega_s r_{ab}(\omega_s){\rm Tr}_A\left[G(t,s) \left(A_b(\omega_s,s)\rho_A(s)A_a^
\dag (\omega_s,s)\right)\right]\nonumber \\
&=&\int_{0}^{t_f}ds \sum_{\omega_s,a,b}\omega_sr_{ab}(\omega_s){\rm Tr}_A[A_b(\omega_s,s)\rho_A(s)A_a^\dag (\omega_s,s)].
\end{eqnarray}
To obtain the second equation, we have used an important identity:
\begin{eqnarray}
\label{dualdefinitionofquantummasterequationgenerator}
{\rm Tr}_A\left[O_1 T_{\leftarrow} e^{\int_0^t ds{{\cal L}}(s)}(O_2)\right]={\rm Tr}_A\left[ T_{\rightarrow} e^{\int_0^t ds{{\cal L}}^*(s)}(O_1) O_2 \right],
\end{eqnarray}
where ${\cal L}^*(t)$ is the dual of ${\cal L}(t)$ and is equal to
\begin{eqnarray}
{\cal L}^*(s)O=&&i[H_A(t)+H_{LS}(t), O] \nonumber\\
&&+\sum_{\omega_t,a,b} r_{ab}(\omega_t)\left[A_a^\dag(\omega_t,t)O A_b(\omega_t,t)-\frac{1}{2}\left\{A_a^\dag(\omega_t,t)A_b(\omega_t,t),O\right\}\right].
\end{eqnarray}
The dual superoperator ${\cal L}^*(t)$ has the trivial property
\begin{eqnarray}
\label{dualoperatorproperty}
{\cal L}^*(t)(I) =0,
\end{eqnarray}
where $I$ represents the identity matrix.

Applying the analogous arguments to Eqs.~(\ref{CFinclusiveworkopensystemgeneralbasedheat}) and~(\ref{exclusiveworkcharacteristicoperatoropensystemorg}), we find that the average inclusive and exclusive works are
\begin{eqnarray}
\label{meanstochasticworkinclusive}
&&\left\langle W_{lnkm}\right \rangle={\rm Tr}_A[H_A(t_f)\rho_A(t_f)]-{\rm Tr}_A[H_A(0)\rho_A(0)] + \langle Q_{lk} \rangle,\\
\label{meanstochasticworkexclusive}
&&\left\langle W^0_{lnkm}\right \rangle={\rm Tr}_A[H_0\rho_A(t_f)]-{\rm Tr}_A[H_0\rho_A(0)] + \langle Q_{lk} \rangle,
\end{eqnarray}
respectively. Note that the second terms on the RHSes are the same because we have imposed the condition $H_1(0)=0$ in the case of the exclusive work. Obviously, the differences between the first two terms on the RHSes are the changes in the systems' mean inner energies within a finite time interval, $(0,t_f)$. Hence, Eqs.~(\ref{meanstochasticworkinclusive}) and~(\ref{meanstochasticworkexclusive}) are simply the first law of thermodynamics realized in the various MQMEs.

Let us check whether the above mean stochastic heat and work are the same as the definitions in the literature~\cite{Pusz1978,Alicki1979,Boukobza2006,Langemeyer2014,Szczygielski2013,Kosloff2013}. In adiabatically driven MQMEs, Alicki has split the change in the mean inner energy of the open quantum system A into two parts:
\begin{eqnarray}
\label{old1stlawinclusivework}
{\rm Tr}_A[H_A(t_f)\rho_A(t_f)]-{\rm Tr_A}[H_A(0)\rho_A(0)]=\int_0^{t_f} ds {\rm Tr}_A[\partial_s H(s) \rho(s)]+\int_0^{t_f} ds {\rm Tr}_A[H_A(s) \partial_s\rho_A(s)].
\end{eqnarray}
This formula is  a simple integral expression for the time derivative of the mean inner energy within the finite time interval, $(0,t_f)$, which  can always be done for any MQME and is not essentially confined to the adiabatically driven case. Since Pusz and Woronowicz~\cite{Pusz1978} interpreted the first term on the RHS as the work done by some external agent in the $C^*$-algebraic context, Alicki called the second integral the heat {\it supplied} to the system by its heat bath. The physical validity of the heat definition is supported by the second law of thermodynamics in adiabatically driven MQMEs. Alicki's heat and work definitions were broadly applied in EQT~\cite{Kosloff2013}. Superficially, Eq.~(\ref{meanstochasticheat}) is very distinct from Eq.~(\ref{old1stlawinclusivework}). However, when we insert the master equation~(\ref{masterequationadiabaticallydrivenHamiltonian}) into Alicki's heat definition, we find that
\begin{eqnarray}
\int_0^{t_f} ds {\rm Tr}[H_A(s) \partial_s\rho_A(s)]=-
\int_0^{t_f} ds\sum_{a,b,m,n} \omega_{mn}(s) r_{ab}(\omega_{mn}(s)){\rm Tr}_A[A_{b, {mn}}(s)\rho_A(s) A^\dag_{a,mn}(s)],
\end{eqnarray}
where we have used Eq.~(\ref{eigenoperatorAdiabaticallydrivenHamiltonian}).
The integral on the RHS is simply Eq.~(\ref{meanstochasticheat}) in these particular MQMEs.
Hence, the mean stochastic heat is indeed consistent with the heat proposed by Alicki. Because of the same first law of thermodynamics, the mean stochastic work is also equivalent to the work definition in Eq.~(\ref{old1stlawinclusivework}). Interestingly, this also reminds us that, in the adiabatically driven case, the mean stochastic work has an integral expression. Even so, we must emphasize that we cannot generalize these observations to the other types of MQMEs, that is, in general,
\begin{eqnarray}
&&\langle Q_{lk} \rangle\neq -\int_0^{t_f} ds {\rm Tr}[H_A(s) \partial_s\rho_A(s)],\\
&&\langle W_{lnkm}\rangle \neq\int_0^{t_f}ds {\rm Tr}[(\partial_sH_A(s))\rho_A(s)].
\end{eqnarray}
An obvious example is periodically driven MQMEs~\cite{Szczygielski2013,Langemeyer2014,Liu2016}. This finding is not very surprising. After all, the splitting of the LHS in Eq.~(\ref{old1stlawinclusivework}) is not unique in mathematics. Hence, there is no prior reason to treat these two terms precisely as the physical heat and heat. In contrast, Eq.~(\ref{meanstochasticheat}) is a measurement-based heat definition.

It is interesting to see that Eq.~(\ref{meanstochasticworkexclusive}) is consistent with other work definitions that are restricted to weakly driven MQMEs~\cite{Liu2014a}, where the Hamiltonian of the open quantum system A is Eq.~(\ref{weakdrivenHamiltonian}). In the spirit of Alick, the change in the average energy of the bare Hamiltonian, $H_0$, can also be split into two parts:
\begin{eqnarray}
\label{1stlawinclusiveworkweaklydrivenHamiltonian}
{\rm Tr}_A[H_0\rho_A(t_f)]-{\rm Tr_A}[H_0\rho_A(0)]=\int_0^{t_f} ds {\rm Tr}_A\left[-i [H_0,\gamma H_1(s)] \rho_A(s)\right]+\int_0^{t_f} ds {\rm Tr}_A[D^*(H_0)\rho_A(s)],
\end{eqnarray}
where the dual dissipator is
\begin{eqnarray}
D^*(O)=\sum_{\omega,a,b}r_{ab}(\omega)\left[A_a^\dag(\omega) OA_b(\omega)-\frac{1}{2}\left\{A_a^\dag(\omega) {A}_b(\omega),O\right \} \right].
\end{eqnarray}
We can prove that for weakly driven MQMEs, based on Eq.~(\ref{eigenoperatorsofweaklydrivenHamiltonian}), the second integral term in Eq.~(\ref{1stlawinclusiveworkweaklydrivenHamiltonian}) is equal to
\begin{eqnarray}
\int_0^{t_f} ds {\rm Tr}_A[D^*[H_0]\rho_A(s)]=-\int_0^{t_f}ds \sum_{\omega,a,b} \omega r_{ab}(\omega){\rm Tr}_A[A_b(\omega)\rho_A(s)A_a^\dag(\omega)].
\end{eqnarray}
The RHS is simply Eq.~(\ref{meanstochasticheat}) in these particular MQMEs. Hence, we also find that the mean stochastic exclusive work here has an integral expression; see the first integral in Eq.~(\ref{1stlawinclusiveworkweaklydrivenHamiltonian}). As is the inclusive work case, these formulas are only valid for these specific MQMEs. Boukobza and Tannor~\cite{Boukobza2006} have independently proposed very similar heat and work definitions. They thought that their definitions were also applicable to other MQMEs. Because of the non-uniqueness of the splitting of Eq.~(\ref{1stlawinclusiveworkweaklydrivenHamiltonian}), this statement shall be treated carefully. Finally, we want to note that, in periodically driven MQMEs, Eq.~(\ref{meanstochasticheat}) is also consistent with the heat definitions proposed by Langemeyer and Holthaus~\cite{Langemeyer2014} and Szczygielski et al.~\cite{Szczygielski2013}. We have illustrated this concretely in a two-level system~\cite{Liu2016a}. The interested reader is referred to our paper.

\section{Fluctuation theorems}
\label{section5}
As we mentioned in Sec.~(\ref{section1}), one of the primary motivations of studying  stochastic heat and work is to explore the validity of a variety of FTs in the quantum regime. In the classical situation, based on a theorem about average quantity or about the probability distribution of a stochastic thermodynamic quantity, we roughly divide these theorems into integral or detailed FT correspondingly. For instance, the JE belongs to the former, while Crooks' equality is an example of the latter. In this section, we also utilize these terms. Although a detailed FT can result in an integral FT, the proof of the latter is usually relatively easy because we do not need to resort to the time-reversal concept. In addition, the reader will see that the conditions of these two types of FTs are slightly different due to the involvement of quantum measurements.

\subsection{Integral FTs}
\label{section5A}
These theorems are implied in the mathematical structure of Eq.~(\ref{Heatmasterequationgeneralform}). Because the heat bath is in thermal equilibrium, the KMS condition on the Fourier transforms, Eq.~(\ref{DetailedBalanceCondition}), reminds us that the action of the superpropagator $\check{{\cal L}}(t,i\beta)$ in Eq.~(\ref{Heatmasterequationgeneralform}) on the identity operator $I$ is exactly zero,
\begin{eqnarray}
\label{checkLproperty}
\check{\cal L}(t,i\beta)(I)=0.
\end{eqnarray}
This almost trivial property can account for several integral fluctuation theorems. The first theorem is about the stochastic heat~(\ref{heatdefintionopensystem}). According to Eq.~(\ref{CFheatopensystemgeneral}) and the meaning of CF, we immediately find that
\begin{eqnarray}
\label{heatFT}
\left\langle e^{-\beta Q}\right\rangle_c=1,
\end{eqnarray}
where we have used the subscript $c$ to denote that the equality is valid for a completely random initial density matrix, which is denoted by $C$, e.g., in a $D$-level quantum open system, $C=I/D$.

Eq.~(\ref{checkLproperty}) also accounts for the remarkable equalities about the exclusive and inclusive work. According to Eq.~(\ref{CFinclusiveworkopensystemgeneralbasedheat}), if the system's initial density matrix $\rho_A(0)$ is the thermal equilibrium state with the heat bath inverse temperature $\beta$, we must have
\begin{eqnarray}
\label{inclusiveworkFT}
\left\langle e^{-\beta W}\right\rangle_{th} =\frac{Z(t_f)}{Z(0)},
\end{eqnarray}
where the subscript $th$ denotes the initial thermal  state and $Z(t)$ is the equilibrium partition function, with $Z(t)={\rm Tr}[e^{-\beta H_A(t)}]$. This theorem is simply the quantum JE in the general MQME. We can verify this intriguing result by solving the time evolution equation~(\ref{inclusiveworkmasterequationgeneralform}). Because $\eta=i\beta$, we easily see that it has a simple solution:
\begin{eqnarray}
K_A(t,i\beta)=\frac{e^{-\beta H_A(t)}}{Z(0)}.
\end{eqnarray}
Taking the trace over the degree of freedom of the quantum system A, we re-obtain the JE of the inclusive work. Very analogous results can be derived for the exclusive work, Eq.~(\ref{exclusiveworkdefinitionopensystem}):
\begin{eqnarray}
\label{exclusiveworkFT}
\left\langle e^{-\beta W_0}\right\rangle_{th} =1.
\end{eqnarray}
This is called the quantum BKE~\cite{Liu2014} in the general MQME, Eq.~(\ref{quantummasterequationgeneralformdensitymatrixcase}). Analogously, given $\eta=i\beta$, the solution of Eq.~(\ref{exclusiveworkmasterequationgeneralform}) is simply
\begin{eqnarray}
K_{0_A}(t,i\beta)=\frac{e^{-\beta H_0}}{Z(0)}.
\end{eqnarray}

The validity of the quantum JE is not trivial. We know that in the classical regime, the JE is always accompanied by a condition that the open classical system has an instantaneous thermal state solution~\cite{Jarzynski2007,Seifert2011}. This condition was also equivalent to the instantaneous detailed balance condition. Physically, this condition indicates that if the external protocol is fixed at a specific value, the system will relax to the thermal equilibrium state under this fixed protocol. If we extend this observation into the quantum regime, we might expect that only if the generator of Eq.~(\ref{quantummasterequationgeneralformdensitymatrixcase}) satisfies
\begin{eqnarray}
\label{instaneousthermalequilibriumcondition}
{\cal L}(t)[\rho_{th}(t)]=0,
\end{eqnarray}
would the quantum JE be valid. Indeed, for closed quantum systems and the adiabatically driven Hamiltonian case that we discussed before, this connection was well established~\cite{Crooks2008,Horowitz2012,Liu2012,Liu2014a}. However, the scope of validity of Eq.~(\ref{inclusiveworkFT}) is beyond this condition. For instance, there is no $\rho_{th}(t)$ in periodically driven MQMEs~\cite{Breuer2000,Alicki2006,Szczygielski2013} or in  weakly driven MQMEs. Therefore, our proof shows that the most fundamental aspect ensuring the validity of these finite-time FTs is the KMS condition, Eq.~(\ref{DetailedBalanceCondition}). It has been known for a long time that the KMS condition and the detailed balance condition have a deep connection in a time-homogenous dynamical semigroup possessing a stationary state~\cite{Agarwal1973,Alicki1976,Kossakowski1978,Spohn1978}. This connection is broken in the general time-dependent MQME, but the KMS condition remains because the heat bath is still in thermal equilibrium. Intriguingly, Spohn and Lebowitz~\cite{Spohn1978} noticed that the KMS condition was also essential for deriving some basic postulates of the thermodynamics of irreversible processes, e.g., the Onsager relations and the principle of minimal entropy production. They particularly noted that ``To state  it (the KMS condition denoted by us) in a somewhat oversimplified way: In our model irreversible thermodynamics follows from the fact that the reservoirs are at thermal equilibrium". The irreversible behaviors described by Eq.~(\ref{quantummasterequationgeneralformdensitymatrixcase})  of course occur for the same reason. Our conclusion also agrees with that drawn by Talker et al.~\cite{Talkner2009}. They proved that the JE is always true for very general open quantum systems with arbitrary time-dependent protocols, as long as the interaction between the quantum system and the heat bath is sufficiently weak. Contrary to our formulas, their results did not rely on Markovian or rotation wave approximations. The weak interaction condition  amounts to the condition that the heat bath is always in a thermal state and is unperturbed by the open quantum system.  

\subsection{Detailed FTs}
\label{section5B}
In classical stochastic systems, the integral FTs can always be traced to their detailed versions concerning the probability distributions of the stochastic thermodynamic quantities of forward and backward processes. Indeed, Talkner et al.~\cite{Talkner2007,Talkner2009,Campisi2011,Esposito2009,Talkner2009} showed that this is also true in closed quantum systems. In particular, they expressed the results using the symmetric property of CFs. In this section, for general MQMEs, Eq.~(\ref{quantummasterequationgeneralformdensitymatrixcase}), we present the detailed versions for the integral FTs, Eqs.~(\ref{heatFT}),~(\ref{inclusiveworkFT}), and~(\ref{exclusiveworkFT}). To this end, we need to define the backward non-equilibrium quantum process. We call a process forward if the protocol driving system A is $\lambda(t)$, whereas a process is backward if the same quantum system is driven by a time-reversed protocol,
\begin{eqnarray}
\overline{\lambda}(s)=\lambda(t),
\end{eqnarray}
where $t+s=t_f$. To distinguish the backward process from the forward process, we specifically use the symbols with overlines to represent the physical quantities of the backward process. We also use the notation $s$ to denote the time of the backward process. In addition, we further suppose that the Hamiltonian of the global system is time-reversal invariant (TRI). Specifically, the Hamiltonian of the composite system A and heat bath B of the backward process is
\begin{eqnarray}
\label{timereversalcompsiteHamiltonian}
{\overline H}(s)={\overline H}_A(s)+{\overline H}_B+{\overline V},
\end{eqnarray}
where the time-reversed Hamiltonian of system A, of the heat bath and of the interaction are
\begin{eqnarray}
\label{systemHamiltonianbackwardprocess}
&&{\overline H}_A(s)=\Theta H_A[{\overline \lambda}(s)]\Theta^{-1}=H_A[\overline {\lambda}(s)]=H_A(t),\\
\label{systemVHaimiltonianwardprocess}
&&{\overline H}_B=\Theta H_B\Theta^{-1}=H_B,\hspace{1cm}  {\overline V}=\Theta V\Theta^{-1}=V,
\end{eqnarray}
respectively, and $\Theta$ is the time reversal operator~\cite{Messiah1962}. The last equation in Eq.~(\ref{systemHamiltonianbackwardprocess}) is true because the time dependence of this Hamiltonian is realized only through the external protocol $\lambda(t)$. According to Eq.~(\ref{interactionHamiltonianVdetailedexpression}), the TRI interaction Hamiltonian $V$ indicates that the time-reversal types of the operators $A_a$ and $B_a$ must be consistent with each other. Specifically, if
\begin{eqnarray}
\overline A_a=\Theta A_a \Theta^{-1}=\delta_aA_a,
\end{eqnarray}
then
\begin{eqnarray}
\overline B_a=\Theta B_a\Theta^{-1}=\delta_a B_a,
\end{eqnarray}
where $\delta_a$ is equal to $+1$ or $-1$ depending on the time-reversal type of $A_a$ being even or odd. Comparing Eq.~(\ref{timereversalcompsiteHamiltonian}) with Eq.~(\ref{compositeHamiltonian}), we can see that the dynamics of the backward process is the same as that of the forward process except that the protocol is changed into the new process, $\overline\lambda (s)$. Therefore, the CF of the heat of the backward process is
\begin{eqnarray}
\label{backwardCFheat}
{\overline \Phi}_h(\eta)={\rm Tr}\left[ \overline {\hat{\rho}}_A (t_f,\eta)\right],
\end{eqnarray}
where the HCO, $\overline {\hat{\rho}}_A$, satisfies a time evolution equation given by
\begin{eqnarray}
\label{BackwardHeatmasterequationgeneralform}
\partial_s {\overline {\hat \rho}}
_A(s,\eta)&=&-i[{\overline H}_A(s)+{\overline H}_{LS}(s), {\overline {\hat \rho}}_A(s,\eta)]\nonumber\\
&&+\sum_{\overline\omega_s,a,b} { r}_{ab}(\overline \omega_s)\left[ e^{i\eta\overline \omega_s} {\overline A}_b(\overline \omega_s,s){\overline {\hat \rho}}_A(s,\eta){\overline A}_a^\dag(\overline \omega_s,s)-\frac{1}{2}\left\{ {\overline A}_a^\dag(\overline \omega_s,s){\overline A}_b(\overline \omega_s,s),{\overline {\hat \rho}}_A(s,\eta)\right\}\right]\nonumber\\
&\equiv&{\overline {\check{ \cal L}}}(s,\eta) {\overline {\hat \rho}}_A(s,\eta).
\end{eqnarray}
We do not specify the uncorrelated initial condition of the backward process yet. This general equation does not provide us the explicit expressions of $\overline\omega_s$, ${\overline A}_l(\overline \omega_s,s)$, and ${\overline A}_l(\overline \omega_s,s)^\dag$ or their concrete connections with those quantities of the forward process; we have to obtain them by investigating concrete Hamiltonian systems. Note that we did not add an overline over $r_{ab}$ because $V$ is TRI.

Given the above notations, we are in a position to discuss the relation between the CFs of the forward and backward processes. Let us start with the case of heat. According to the formal solution, Eq.~(\ref{propagatorsolutionofheatequation1}), we write Eq.~(\ref{CFheatopensystemgeneral}) with a completely random initial density matrix as
\begin{eqnarray}
\label{heatCFforwardandbackward}
\Phi_h(\eta)&=&{\rm Tr}_A \left[T_{\leftarrow} e^{\int_0^{t_f} du{\check{\cal L}}(u,\eta) }\left( C\right)\right] \nonumber\\
&=&{\rm Tr}_A \left[T_{\rightarrow} e^{\int_0^{t_f} du{\check{\cal L}}^*(u,\eta) }\left(C\right)\right] \nonumber\\
&=&{\rm Tr}_A \left[\Theta \left( T_{\rightarrow} e^{\int_0^{t_f} du{\check{\cal L}}^*(u,\eta) }\left( C\right)\right)^\dag \Theta^{-1}\right]  \nonumber\\
&=&{\rm Tr}_A \left[\Theta  T_{\rightarrow} e^{\int_0^{t_f} du{\check{\cal L}}^*(u,-\eta) }\left(C\right)\Theta^{-1}\right].
\end{eqnarray}
The second equation results from the fact that $C$ is proportional to the identity matrix and the dual definition
\begin{eqnarray}
\label{dualdefinitionofgeneralheatequationgenerator}
{\rm Tr}_A\left[O_1 T_{\leftarrow} e^{\int_0^t ds{\check{\cal L}}(s,\eta)}(O_2)\right]={\rm Tr}_A\left[ T_{\rightarrow} e^{\int_0^t ds{\check{\cal L}}^*(s,\eta)}(O_1) O_2\right],
\end{eqnarray}
where
\begin{eqnarray}
\label{definitiondualheatmasterequationgenerator}
\check{\cal L}^*(s,\eta)O=i[H_A(t)+H_{LS}(t), O ]+\sum_{\omega_t,a,b} r_{ab}(\omega_t)\left[ e^{i\eta\omega_t} A_a^\dag(\omega_t,t)O A_b(\omega_t,t)-\frac{1}{2}\left\{A_a^\dag(\omega_t,t)A_b(\omega_t,t),O\right\}\right].
\end{eqnarray}
In the third equation, we applied an identity for the time reversal operator $\Theta$~\cite{Messiah1962}:
\begin{eqnarray}
\label{timereversalproperty}
{\rm Tr}_A [\Theta O \Theta^{-1}]={\rm Tr}_A[O^\dag].
\end{eqnarray}
The last equation is obvious if one writes the time-ordered exponential term explicitly. Now, we want to show that the whole term in the trace of the last equation is the formal solution of Eq.~(\ref{BackwardHeatmasterequationgeneralform}) but with $\overline\eta=i\beta-\eta$ at time $t_f$, that is,
\begin{eqnarray}
\label{keyTRformula}
{\overline {\hat\rho}}_A(s,\overline\eta)=\Theta  T_{\rightarrow} e^{\int_{t}^{t_f} du{\check{\cal L}}^*(u,-\eta) }\left( C\right)\Theta^{-1}.
\end{eqnarray}
We again emphasize $s+t=t_f$. To prove this conjecture, we temporally denote the whole term on the RHS of Eq.~(\ref{keyTRformula}) ${{\hat\rho}}'_A(s,\overline\eta)$. Taking its derivative with respect to time $s$, we obtain
\begin{eqnarray}
\label{BackwardHeatmasterequationgeneralformneededtobeproved}
\partial_s {{\hat\rho}}'_A(s,\overline\eta)=&&-i\left[\Theta (H_A(t)+H_{LS}(t))\Theta^{-1}, { {\hat\rho}}'_A(s,\overline\eta)\right]\nonumber\\
&&+\sum_{\omega_{t},a,b}r_{ab}^*(\omega_{t})\left[ e^{i\eta\omega_{t}} \Theta A_a^\dag(\omega_t,t)\Theta^{-1}  { {\hat\rho}}'_A(s,\overline\eta) \Theta A_b(\omega_t,t)\Theta^{-1}\right.\nonumber\\
&&\hspace{3cm}\left.-\frac{1}{2}\left\{\Theta A_a^\dag(\omega_t,t)\Theta^{-1} \Theta A_b(\omega_t,t)\Theta^{-1},{ {\hat\rho}}'_A(s,\overline\eta) \right\}\right].
\end{eqnarray}
It is worth noting that the condition of the TRI interaction Hamiltonian $V$ implies that the Fourier transforms of the two-time correlation functions, Eq.~(\ref{FouriertransformCorrlationfunctions}), satisfy
\begin{eqnarray}
\label{TimereversalCorrelationfuncts}
r_{ab}^*(\omega)=\delta_a \delta_b r_{ab}(\omega).
\end{eqnarray}
This relation also indicates that the matrix $\{r_{ab}(\omega)\}$ is not only Hermitian~\cite{Breuer2002} but also composed of  real or purely imaginary numbers. A brief explanation of Eq~(\ref{TimereversalCorrelationfuncts}) is presented in Appendix D. Therefore, if Eq.~(\ref{BackwardHeatmasterequationgeneralformneededtobeproved}) is really equivalent to Eq.~(\ref{BackwardHeatmasterequationgeneralform}), we need to check the following two equations:
\begin{eqnarray}
\label{forwardbackwardidentity1}
\overline\omega_s&=&\omega_t,\\
\label{forwardbackwardidentity2}
{\overline A}_a^\dag(\overline \omega_s,s)&=&\delta_a\Theta A_a^\dag(\omega_t,t)\Theta^{-1}.
\end{eqnarray}
In Appendix C, we will show that Eqs.~(\ref{forwardbackwardidentity1}) and~(\ref{forwardbackwardidentity2}) are exactly true for all the MQMEs that we discussed so far. In the general situations, we can simply regard a process as backward if its ${\overline A}_a^\dag$ and Bohr frequency $\overline\omega$ are defined by the above two equations.

Substituting Eq.~(\ref{TimereversalCorrelationfuncts}) and these two identities into Eq.~(\ref{BackwardHeatmasterequationgeneralformneededtobeproved}), and noticing the KMS condition, Eq.~(\ref{DetailedBalanceCondition}), we find that
\begin{eqnarray}
\partial_s {{\hat\rho}}'_A(s,\overline\eta)=&&-i[{\overline H}_A(s), { {\hat\rho}}'_A(s,\overline\eta)]\nonumber\\
&&+\sum_{\overline\omega_{s},a,b}r_{ab} ({\overline \omega}_{s})\left[ e^{i{\overline\eta}{\overline \omega}_{s}} {\overline A}_b(\overline\omega_s,s)  { {\hat\rho}}'_A(s,\overline\eta){\overline A}_a^\dag({\overline\omega}_s,s)  -\frac{1}{2}\left\{\overline A_a^\dag(\overline\omega_s,s){\overline A}_b(\overline\omega_s,s) ,{ {\hat\rho}}'_A(s,\overline\eta) \right\}\right],
\end{eqnarray}
where its initial condition is $C$. Comparing the time evolution equation with Eq.~(\ref{BackwardHeatmasterequationgeneralform}), we conclude that  Eq.~(\ref{keyTRformula}) is indeed true.  Therefore, we can continue Eq.~(\ref{heatCFforwardandbackward}) and obtain
\begin{eqnarray}
\label{detailedFTforheatCFversion}
\Phi_h(\eta)={\rm Tr}_A\left[{\overline {\hat\rho}}_A(t_f,\overline\eta) \right]={\overline \Phi}_h(i\beta-\eta).
\end{eqnarray}
According to the relation between the CF and probability distributions, e.g., Eq.~(\ref{CFinclusivework}), we have
\begin{eqnarray}
\label{detailedFTforheatprobabilityversion}
P(Q)={\overline P}(-Q) e^{\beta Q}.
\end{eqnarray}

What we have done can be generalized to the cases concerning the inclusive and exclusive work. In the former, according to Eq.~(\ref{CFinclusiveworkopensystemgeneralbasedheat}), we have
\begin{eqnarray}
\label{inclusiveworkCFforwardandbackward}
\Phi_w(\eta)Z(0)&=&{\rm Tr}_A \left[e^{-i\eta H_A(0)}\rho_A(0) T_{\rightarrow} e^{\int_0^{t_f}ds{\check{\cal L}}^*(s,\eta) }\left( e^{i\eta H_A({t_f})} \right)\right] \nonumber\\
&=&{\rm Tr}_A \left[\Theta \left( e^{-i\eta H_A(0)}\rho_A(0) T_{\rightarrow} e^{\int_0^{t_f} ds{\check{\cal L}}^*(s,\eta) }\left( e^{i\eta H_A(t)} \right) \right)^\dag \Theta^{-1}\right]  \nonumber\\
&=& {\rm Tr}_A \left[\Theta T_{\rightarrow} e^{\int_0^{t_f} ds{\check{\cal L}}^*(s,-\eta) }\left( e^{-i\eta H_A({t_f})} \right) \Theta^{-1} \rho_A(0)e^{-i\eta H_A(0)}\right] \nonumber\\
&=&{\rm Tr}_A \left[e^{i\overline\eta H_A(0)} \Theta T_{\rightarrow} e^{\int_0^{t_f} ds{\check{\cal L}}^*(s,-\eta) }\left( e^{-i\eta H_A({t_f})}/Z({t_f}) \right) \Theta^{-1}  \right] Z({t_f}).
\end{eqnarray}
Based on the previous discussion, in the last equation, the whole term except for the first exponential operator  is simply the solution of the backward equation~(\ref{BackwardHeatmasterequationgeneralform}) with $\overline\eta=i\beta-\eta $. In particular, here, its initial condition is
\begin{eqnarray}
{\overline {\hat\rho}}_A(0,\overline\eta)=e^{-i{\overline \eta}H(t_f)}\rho_{th}(t_f).
\end{eqnarray}
Hence, the whole term, including the term $e^{i\overline\eta H_A(0)} $, is simply the WCO of the inclusive work of the backward process; see the definition in Eq.~(\ref{CFinclusiveworkopensystemgeneralbasedheat}). Therefore,
\begin{eqnarray}
\label{detailedFTforinclusiveworkCFversion}
\Phi_w(\eta)Z(0)=\overline{\Phi}_w(i\beta-\eta )Z(t_f).
\end{eqnarray}
If we represent this relation using probability distributions, we obtain the remarkable Crooks' equality about the inclusive work:
\begin{eqnarray}
\label{detailedFTforinclusiveworkprobabilityversion}
P(W) ={\overline P}(-W) e^{\beta (W-\Delta F)}.
\end{eqnarray}
where the free energy difference is $\beta\Delta F=\ln Z(t_f)-\ln Z(0)$. In the exclusive work case, we can utilize a very analogous discussion and ultimately obtain
\begin{eqnarray}
\label{exclusiveworkCrooksequality}
&&\Phi_{w_0}(\eta)=\overline{\Phi}_{w_0}(i\beta-\eta ),\\
\label{detailedFTforexclusiveworkprobabilityversion}
&&P(W_0)={\overline P}(-W_0) e^{\beta W_0}.
\end{eqnarray}
Note that in this equality, the initial condition of the backward process must be the thermal state $e^{-\beta H_0}/{\rm Tr}[e^{-\beta H_0}]$.
On the other hand, to ensure that the work definition based on the TMS scheme is reasonable, we also require that the initial density matrix is commutative with the Hamiltonian of the system at the initial time. This requirement is essential in both the forward and backward processes. Hence, in this case, Eq.~(\ref{exclusiveworkCrooksequality}) additionally imposes a condition that $H_1(t)$ must be zero at both the beginning and end time. Interestingly, this condition is absent in the quantum BKE, Eq.~(\ref{exclusiveworkFT}). In classical stochastic processes, this condition is not needed either. Finally, let us conclude this section by noting that, if we integrate these detailed FTs, Eqs.~(\ref{detailedFTforheatprobabilityversion}),~(\ref{detailedFTforinclusiveworkprobabilityversion}), and~(\ref{detailedFTforexclusiveworkprobabilityversion}), with respect to the stochastic heat and work, we will again obtain the previous integral FTs.

\section{Quantum jump trajectory }
\label{section6}
The density matrix and MQMEs are essentially an ensemble description of open quantum systems. Nevertheless, in practice, one usually addresses individual quantum systems. This situation is very similar to classical Brownian motion: we can mathematically describe the evolution of an ensemble of Brownian particles using probability distributions and the Fokker-Planck equation~\cite{Risken1984,Gardiner1983}, but in real experiments, e.g., single-molecule experiments~\cite{Ciliberto2017}, one often measures the motion of individual Brownian particles, and the equation describing the motion is the Langevin equation. Therefore, we may naturally ask whether there are formulas that can describe the dynamics of single open quantum systems. The QJT was shown to be one such formula~\cite{Breuer2002,Gardiner2004,Wiseman2010}. The idea behind the QJT started in the sixties and was intimately connected with theoretical efforts on fully quantum mechanical treatments of photon counting experiments in quantum optics, where a detector monitors a quantum system over a time interval and records the  arrival times of the photons~\cite{Kelley1964,Davies1969,Mollow1968,Mollow1975,Srinivas1981,Zoller1987,Carmichael1989,Ueda1990,Wiseman1993}. This concept became increasingly important after observing or manipulating single quantum system became possible~\cite{Nagourney1986,Sauter1986,Bergquist1986,Basch1995,Peil1999,Gleyzes2007,Murch2013,Sun2013,Vool2014,Campagne-Ibarcq2016}. In the developing history, the QJT has been given different names, e.g., the Monte-Carlo wave function method~\cite{Molmer93}, the stochastic wave-function method~\cite{Breuer2002}, the unravelling of the quantum master equations, and quantum trajectories~\cite{Carmichael1993}. We refer the reader interested in the historic overview of QJTs to the review papers~\cite{Plenio1998} or the relevant monographs~\cite{Carmichael1993,Breuer2002,Gardiner2004,Barchielli2009}. Although it was known very early that the QJT can be used to re-obtain all the results given by  MQMEs~\cite{Breuer2002} and that the latter are very fundamental in investigating irreversible thermodynamics~\cite{Spohn1978}, its potential applications in thermodynamics were almost completely unrecognized for almost four decades~\footnote{An exception might be the entropy production study conducted by Breuer~\cite{Breuer2003}.}. In this section, we give a self-contained explanation about QJTs. Our discussion is based on basic quantum mechanics and simple probability knowledge and does not rely on other advanced contents, e.g., quantum stochastic calculus~\cite{Parthasarathy1992}. In particular, we attempt to present our theory under the framework of the TEM scheme. In previous studies, this energetic perspective of the origin of QJTs was usually introduced directly without explicit explanation. We think that this might be an obstruction hindering the appreciation of the values of QJTs in SQT.

\begin{figure}
\includegraphics[width=1.\columnwidth]{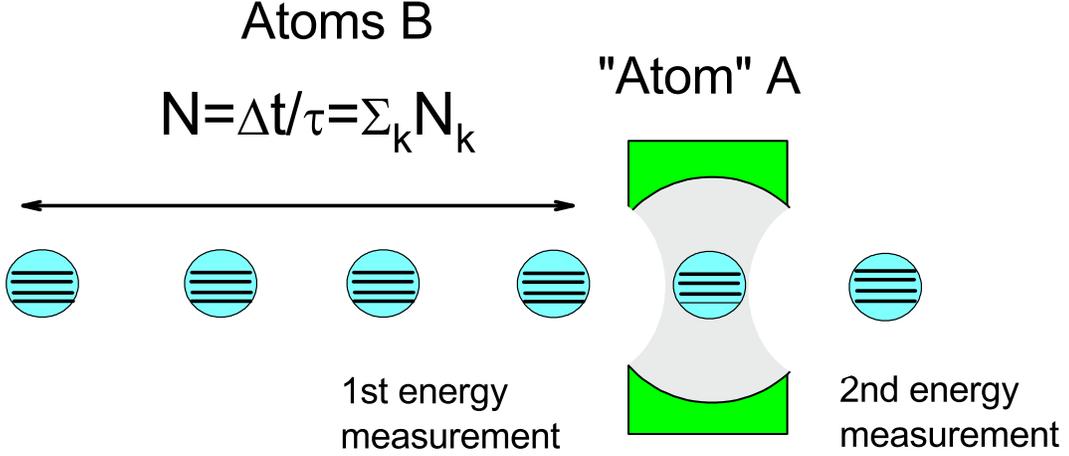}
\caption{A schematic diagram of the repeated interaction quantum model. The system or ``atom" A is represented by a mode in a cavity. A series of  heat bath B atoms in the thermal state pass through the cavity and interact with  atom A within brief intervals $\tau$. The TEM is applied to each atom B before it enters and after it leaves the cavity. Note that we only record the change in energy and ignore the exact energy values. If this conceptual experiment is conducted over a finite time $\Delta t$, the number of B atoms is $N=\Delta t/\tau$. These B atoms are composed of $N_k$ atoms at the energy eigenvector $|\chi_k\rangle$.  }
\label{figure2}
\end{figure}

We use a repeated interaction model~\cite{Kist1999,Brun2001,Attal2006,Horowitz2013} to explain QJTs: a quantum system A continuously interacts with a series of atoms B, and these B atoms play the role of the ``heat bath". Fig.~(\ref{figure2}) is a schematic diagram of the model. Compared with previous models, the model that we are considering is general, and we do not assume that these B atoms are two-level systems (q-bits)~\cite{Kist1999,Brun2001}. Before these B atoms interact with the quantum system A, we assume that they are in thermal equilibrium. These atoms then pass through a region wherein they interact with the system or the ``atom'' A successively for finite intervals. The global Hamiltonian of the composite A and B atoms is still Eq.~(\ref{compositeHamiltonian}). We further assume that there are no correlations or interactions among these B atoms. When a B atom enters and leaves the interaction region, its energy is measured, and we record the change in energy. This is simply the TEM scheme applied to the bath atoms. This model is not abstract because many experimental systems can approximately realize it~\cite{Brun2001}.

\subsection{Static Hamiltonian }
\label{subsection6A}
\subsubsection{Simple interaction}
We first consider the case of the static Hamiltonian with a simple interaction, $V=A\otimes B$, or Eq.~(\ref{interactionHamiltonianVdetailedexpression}) with $a=1$. We suppose that before the interaction with atom B occurs, system $A$ is initially in a pure state. These two restrictions are based on the following consideration: under this situation, a QJT can be described by the wave vector of the system; otherwise, the density matrix formulation has to be used. Although this is a specific case of the latter, the wave-vector description of QJTs seems to be more accessible to non-expert readers. In addition, it is easier to realize in numerical simulations as well. We have seen that a major portion of the literature studying SQT using QJTs has favored this formulation~\cite{DeRoeck2004,Horowitz2012, Horowitz2013,Hekking2013,Liu2014a,Liu2014,Gong2016,Suomela2014,Suomela2015}.

In analogy to the derivations of MQMEs, our discussion is presented in the interaction picture with respect to the free Hamiltonian, $H_A+H_B$. As we have mentioned, before a bath atom B enters the cavity, we measure its energy. Because these bath atoms are in a thermal equilibrium state, the measured atom shall remain at one of the eigenvectors of the Hamiltonian $H_B$. Suppose that the wave vector is $|\chi_k\rangle$ with eigenvalue $\chi_k$. After a time interval $\tau$, due to the interaction between the A and B atoms, the global wave vector of the composite A and B system evolves into
\begin{eqnarray}
\label{wavevectorofcompositeAandB}
|\widetilde\Psi(t+\tau)\rangle &=& T_{\leftarrow} e^{ -i \alpha\int_{t}^{t+\tau}ds  {\widetilde V}(s)}|\widetilde\psi(t)\rangle_A\otimes|\chi_k\rangle\nonumber\\
&=&\left[ 1-i\alpha\int_{t}^{t+\tau}ds {\widetilde V}(s)-\alpha^2\int_{t}^{t+\tau} ds\int_t^s du {\widetilde V}(s){\widetilde V}(u) \right] |\widetilde\psi(t)\rangle_A\otimes|k\rangle +{\cal O}(\alpha^3),
\end{eqnarray}
where $|\widetilde\psi(t)\rangle_A$ is the wave vector of atom A at time $t$ before entry of atom B.
Because we are using a perturbation approach, we redefine $V$ as $\alpha V$ with the small parameter $\alpha$. In the second equation, we collect all terms up to second order in $\alpha$; all higher order terms are neglected. According to Eq.~(\ref{decompositionofAoperatorstaticHamiltonian}), the interaction picture operator $V$ is
\begin{eqnarray}
\label{simpleinteractionstaticHamiltonian}
{\widetilde V}(s)=U_0^\dag (s)V U_0(s)=\sum_\omega {A}(\omega)e^{-i\omega s}\otimes {\widetilde B}(s).
\end{eqnarray}

When atom B passes by after a time interval $\tau$, we measure its energy again. According to the projective measurement assumption in quantum mechanics~\cite{Braginsky1992}, there are two types of outputs. One output is whereby the measured energy of atom B is still $\chi_k$. In this case, the wave vector of atom A after the measurement is
\begin{eqnarray}
\label{wavevectorAsystemoutputcontinue}
|\widetilde\psi(t+\tau)\rangle_A&=&\frac{\langle\chi_k|\widetilde\Psi(t+\tau)\rangle}{\sqrt{p_{kk}}}
=\frac{1}{{\sqrt{p_{kk}}}}\left[ 1+\alpha^2\sum_{\omega,\omega'}{A}^\dag(\omega){A}(\omega') e^{i(\omega-\omega')t} g_k(\tau,\omega,\omega') \right]|\widetilde\psi(t)\rangle_A.
\end{eqnarray}
where the coefficient $g_k$ is
\begin{eqnarray}\label{gfunction}
g_{k}(\tau,\omega,\omega')=-\int_0^\tau ds e^{i(\omega-\omega')s} \int_0^s du e^{i\omega'u}\langle \chi_k|{\widetilde B}(s){\widetilde B}(s-u)|\chi_k\rangle.
\end{eqnarray}
Noting the presence of the exponential term $e^{i(\omega-\omega')t}$ in Eq.~(\ref{wavevectorAsystemoutputcontinue}), which is analogous to that in Eq.~(\ref{timeevolutionequationheatopensystemorginteractionpictureprojectedexplicitstaticHamiltonian}), we may simplify Eq.~(\ref{wavevectorAsystemoutputcontinue}) by applying the rotation wave approximation:
\begin{eqnarray}
\label{wavevectorAsystemoutputcontinuesecularapproximation}
|\widetilde\psi(t+\tau)\rangle_A=\frac{1}{{\sqrt{p_{kk}}}}\left[ 1+\alpha^2\sum_{\omega}{A}^\dag(\omega){A}(\omega) g_k(\tau,\omega) \right]|\widetilde\psi(t)\rangle_A,
\end{eqnarray}
where $g_k(\tau,\omega)$ is the same as Eq.~(\ref{gfunction}), but $\omega$ and $\omega'$ are set equal here. The other possible output is whereby the measured energy value is $\chi_l$, and especially, $l\neq k$. In this situation, the wave vector of atom A after this measurement is
\begin{eqnarray}
\label{wavevectorAsystemoutputjump}
|\widetilde\psi(t+\tau)\rangle_A&=&\frac{\langle \chi_l|\widetilde\Psi(t+\tau)\rangle}{\sqrt{p_{lk}}}=
\frac{\alpha}{\sqrt{p_{lk}}}\sum_{\omega}  e^{-i(\omega-\chi_l+\chi_k)t}f_{lk}(\tau,\omega){A}(\omega)|\widetilde\psi(t)\rangle_A.
\end{eqnarray}
where the coefficient $f_{lk}$ is
\begin{eqnarray}
\label{fmn}
f_{lk}(\tau,\omega)=-i\int_{0}^{\tau}ds e^{-i\omega s}\langle \chi_l| {\widetilde B}(s)  |\chi_k\rangle.
\end{eqnarray}
Here, we keep the term of lowest order in $\alpha$, ${\cal O}(\alpha^1)$. This approximation convention is used throughout this paper, e.g., Eq.~(\ref{wavevectorofcompositeAandB}). In Eqs.~(\ref{wavevectorAsystemoutputcontinue}) and~(\ref{wavevectorAsystemoutputjump}), the coefficient $p_{lk}$ ($l=k$ or $l\neq k$) is used to ensure the normalization of the wave vector $|\widetilde\psi(t+\tau)\rangle_A$, which is
\begin{eqnarray}
\label{probtimetau}
p_{lk}=\left |\langle \chi_l|\widetilde\Psi(t+\tau)\rangle \right |^2.
\end{eqnarray}
Its concrete expression is temporarily unimportant.

Let us consider a new time scale $\Delta t$. We require that it is far larger than $\tau$ but still far smaller than the relaxation time of system A. During this interval, there are many B atoms passing by. Obviously, the number of B atoms is $N=\Delta t/\tau$ ($\gg 1$). We have assumed that the bath atoms are in a thermal equilibrium state. Hence, these atoms are composed of various atoms with energy eigenvalue $\chi_k$, and in particular, their numbers are
\begin{eqnarray}
N_k= Np_k=N\frac{e^{-\beta \chi_k}}{Z}.
\end{eqnarray}
When $N$ B atoms pass by, we want to count the energy changes in these atoms. One possibility is that none of the energies change even though they have interacted with system A. In this situation, according to Eq.~(\ref{wavevectorAsystemoutputcontinuesecularapproximation}), we find that the wave vector of system A after the longer time interval $\Delta t$ is
\begin{eqnarray}
\label{continuewavequationunnormalized}
|\widetilde\psi(t+\Delta t)\rangle_A\propto \left[ 1 + \alpha^2N\sum_{\omega}{A}^\dag(\omega){A}(\omega) \sum_k p_kg_k(\tau,\omega) \right]|\widetilde\psi(t)\rangle_A.
\end{eqnarray}
Notice that we did not yet normalize the vector. Breuer and Petruccione~\cite{Breuer1995} proved that the summation term with respect to the energy quantum number $k$ is simply
\begin{eqnarray}
\label{BreuerPetruccioneformula1}
-\tau\left[ \frac{1}{2}r(\omega) +iS(\omega)\right],
\end{eqnarray}
where the functions $r$ and $S$ have been defined in Eq.~(\ref{onesideFouriertransformcorrelationfunc}). To arrive at this result, we have considered that the decay time of the correlation function of the B atoms is far shorter than the interaction time so that $\tau$ in $g(\tau,\omega)$ can be replaced by $+\infty$. This is simply the Markovian approximation. Substituting the above result into Eq.~(\ref{continuewavequationunnormalized}) and imposing the normalization requirement, we obtain
\begin{eqnarray}
\label{continuewavequation}
|\widetilde\psi(t+\Delta t)\rangle_A =\left \{ 1- i H_{LS} \Delta t +\frac{\Delta t }{2} \sum_{ \omega}r(\omega)\left[\langle{A}^\dag(\omega){A}(\omega)\rangle -{A}^\dag(\omega){A}(\omega)\right]\right\}|\widetilde \psi(t)\rangle_A +{\cal O}(\Delta t^2),
\end{eqnarray}
where $H_{LS}$ has been given in Eq.~(\ref{LambshiftH}), and the average $\langle\cdots\rangle$ is with respect to the initial wave vector, $|\widetilde \psi(t)\rangle_A$. We have eliminated the parameter $\alpha$ here.

On the other hand, we may find that some B atoms have indeed changed their energies after the time interval $\Delta t$. According to Eq.~(\ref{probtimetau}), compared with the case of no atoms changing their energies, however, such a probability is very small. Hence, we suppose that, within the time interval $\Delta t$, only one B atom changed its energy. Obviously, the choice of $\Delta t$ becomes very important to the validity of this assumption. Consider that the energy change is from $\chi_k$ to $\chi_l$. The explicit expression of Eq.~(\ref{fmn}) is
\begin{eqnarray}
\label{explicitfmn}
f_{lk}(\tau,\omega)=\frac{e^{-i[\omega-(\chi_l-\chi_k)]\tau}-1}{\omega-(\chi_l-\chi_k)}\langle \chi_l|B|\chi_k\rangle.
\end{eqnarray}
We see that the ratio term on the RHS is a rapidly oscillating and decaying function of $\tau$; it is significantly non-zero only if
\begin{eqnarray}
\label{energyconservationcondition}
\omega=\chi_l-\chi_k.
\end{eqnarray}
This condition, which we call the energy conservation condition throughout this paper, implies that the major contributions in the sum, Eq.~(\ref{wavevectorAsystemoutputjump}), are the $A(\omega)$ terms, which exactly satisfy Eq.~(\ref{energyconservationcondition}).
Conversely, this also indicates that if we really observed an energy change in one B atom within the time interval $\Delta t$, its value must be equal to one of  the Bohr frequencies. These discussions lead us to the conclusion that the wave vector of system A after an observation of an energy change $\omega$ is
\begin{eqnarray}
\label{jumpwavequation}
|\widetilde\psi(t+\Delta t)\rangle_A=\frac{A(\omega)|\widetilde \psi(t)\rangle_A}{\sqrt{\langle{A}^\dag(\omega){A}(\omega)\rangle} } .
\end{eqnarray}
The normalization condition has been considered here. We call such a change in the wave vector a jump of type $\omega$.
The reader is reminded that as long as one $\omega$ is given, although different quantum numbers, $l$ and $k$, (or different wave vectors $|\chi_l\rangle$) correspond to different coefficients, $f_{lk}(\tau,\omega)$, they cannot alter the above result due to the normalization. However, if the interaction Hamiltonian $V$ is complex, namely, $a\ge 2$ in Eq.~(\ref{interactionHamiltonianVdetailedexpression}), we cannot obtain the simple Eq.~(\ref{jumpwavequation}). We will discuss this in the next subsection. Because we are only concerned with the energy change $\omega$ of the bath atoms and ignore the information of the exact quantum numbers $k$, our follow-up question concerns the probability of the jump of type $\omega$. After a simple probability argument, we find that it is
\begin{eqnarray}
\label{probjumpduringtimescaleDeltatoriginalform}
P_\omega(t)=\alpha^2N\sum_{k} p_k f_{lk}(\tau,\omega)^*f_{lk}(\tau,\omega) \langle  {A}^\dag (\omega){A}(\omega) \rangle.
\end{eqnarray}
Here, the quantum number $l$ is restricted by the energy conservation condition, Eq.~(\ref{energyconservationcondition}). The sum over the quantum number $k$ is because the jump might occur with every B atom interacting with  system A within the time interval $\Delta t$. Of course, this is only possible if the energy eigenvalues of the bath atoms are sufficiently dense to satisfy the energy conservation condition. Breuer and Petruccione proved that the above sum term is equal to $\tau r(\omega)$~\cite{Breuer1995}. Hence, Eq.~(\ref{probjumpduringtimescaleDeltatoriginalform}) is simplified as
\begin{eqnarray}
\label{onejumpprobabilitygivenpsi}
P_\omega(t)=r(\omega)\langle  {A}^\dag (\omega){A}(\omega) \rangle \Delta t.
\end{eqnarray}
Here, we eliminated $\alpha$ as well. Although we obtain the probability in the interaction picture, it is the same in the Schr$\ddot{o}$dinger picture.

Based on the above discussion, we have a physical picture of the evolution of system A and the energy changes in these B atoms. The B atoms successively pass through the interaction region, and we continue recording their energy changes. At each time interval $t\sim t+\Delta t$, we  have the probability
\begin{eqnarray}
\label{alljumpsprobabilitygivenpsi}
\Delta t\Gamma(t)=\Delta t\sum_\omega r(\omega)\langle  {A}^\dag (\omega){A}(\omega) \rangle
\end{eqnarray}
of finding that an  energy change has occurred. Obviously, $\Gamma(t)$ has an interpretation of being a total instantaneous jump rate. If a jump of type $\omega$ is indeed recorded, the wave vector of system A is updated to the new wave vector, Eq.~(\ref{jumpwavequation}). Because the probability is very small, we more often see that there is no change in the energy. In this situation, system A deterministically evolves according to Eq.~(\ref{continuewavequation}). These two types of evolutions alternatingly proceed until the process is terminated at the end time $t_f$. We call the full evolution of $|\widetilde\psi\rangle_A$ a QJT in the system's Hilbert space and denote it as $\{|\widetilde\psi\rangle_A \}$.

Eqs.~(\ref{continuewavequation}) and~(\ref{jumpwavequation}) can be written in a concise form:
\begin{eqnarray}
\label{stochasticdifferentialequationvector}
\Delta |\widetilde\psi\rangle_A &=&|\widetilde\psi(t+\Delta t)\rangle_A-|\widetilde\psi(t)\rangle_A  \nonumber\\
&=& \prod_{\omega}[1-\Delta N_\omega(t)]\Delta t \left\{- i H_{LS}+\frac{1 }{2} \sum_{\omega}r (\omega)\left[{A}^\dag(\omega){A}(\omega)-\langle{A}^\dag(\omega){A}(\omega)\rangle \right]\right\}|\widetilde \psi(t)\rangle_A\nonumber\\
&&\hspace{1cm}+\sum_{\omega }\Delta N_\omega(t) \left(\frac{A(\omega) }{\sqrt{\langle {A}^\dag(\omega){A}(\omega)\rangle} }-1\right)|\widetilde \psi(t)\rangle_A.
\end{eqnarray}
The stochastic variable $\Delta N_\omega(t)$ is equal to $1$ or $0$ with corresponding probabilities $P_\omega(t)$ or $1-P_\omega(t)$. Obviously, its average is
\begin{eqnarray}
\label{onejumpaverage}
E[\Delta N_\omega(t)]=r(\omega)\langle  {A}^\dag (\omega){A}(\omega) \rangle \Delta t.
\end{eqnarray}
The reader is reminded that the above average is performed over a given $|\widetilde\psi(t)\rangle_A$. Hence, it is a conditional average. On the other hand, we may re-express Eq.~(\ref{stochasticdifferentialequationvector}) using the density matrix formula
\begin{eqnarray}
\widetilde\sigma_A(t)=|\widetilde\psi(t)\rangle_A\langle\widetilde\psi(t)|.
\end{eqnarray}
It is straightforward to prove that
\begin{eqnarray}
\label{stochasticdifferentialequationdensitymatrix}
\Delta \widetilde\sigma_A &=&\widetilde\sigma_A (t+\Delta t)-\widetilde\sigma_A(t)  \nonumber\\
&=&\prod_{\omega}[1-\Delta N_\omega(t)]\Delta t \left\{- i [H_{LS},
\widetilde \sigma_A] -\frac{1}{2} \sum_{\omega}r(\omega)\left\{{A}^\dag(\omega){A}(\omega),\widetilde\sigma_A \right\} + \sum_{\omega}r(\omega){\rm Tr}_A[ {A}^\dag(\omega){A}(\omega)\widetilde\sigma_A]\widetilde\sigma_A \right\} \nonumber\\
&&\hspace{1cm}+\sum_{\omega }\Delta N_\omega(t) \left(\frac{A(\omega)\widetilde\sigma_A A^\dag(\omega)}{{\rm Tr}_A[  {A}(\omega)\widetilde\sigma_A{A}^\dag(\omega) ] }-\widetilde\sigma_A\right),
\end{eqnarray}
We also call the time evolution of $\widetilde\sigma_A$ a QJT in the system's density matrix space and denote it as $\{\widetilde\sigma_A\}$. The reader is reminded that Eq.~(\ref{stochasticdifferentialequationdensitymatrix}) can be obtained by applying the density matrix formula at the beginning rather than being indirectly derived from the wave vector. Although Eqs.~(\ref{stochasticdifferentialequationdensitymatrix}) and Eq.~(\ref{stochasticdifferentialequationvector}) are equivalent to each other, we will see that only the former form can be extended to situations with the general system's initial conditions and/or with a complex interaction Hamiltonian.

Let us discuss several consequences of the QJT concept. First, a QJT is combined with alternating deterministic evolution and stochastic jumps. Hence, QJTs  are typical piecewise deterministic processes (PDPs)~\cite{Breuer2002} concerning the wave vector or density matrix of  system A. One can study the dynamic equations of their distribution functions based on the standard theory of PDPs. We refer the interested reader to the textbook by Breuer and Petruccione~\cite{Breuer2002}. Second, a QJT is a random process. For each realized process, the density matrix $\widetilde\sigma_A(t)$ at a given time $t$ is random. One may prove that its mean over all processes,
\begin{eqnarray}
\label{Mdefintion}
\widetilde \rho_A(t)=M[\widetilde\sigma_A],
\end{eqnarray}
satisfies a time evolution equation:
\begin{eqnarray}
\label{quantummasterequation}
\Delta \widetilde\rho_A &=&\widetilde\rho_A (t+\Delta t)-\widetilde\rho_A(t)  \nonumber\\
&=&\Delta t \left\{- i [H_{LS},
\widetilde \rho_A] + \sum_{\omega}r(\omega) \left[A(\omega)\widetilde\sigma_A A^\dag(\omega)- \frac{1}{2} \left\{{A}^\dag(\omega){A}(\omega),\widetilde\sigma_A \right\}\right]\right\}.
\end{eqnarray}
As $\Delta t$ goes to zero, the above equation is simply the MQME for the static Hamiltonian in the interaction picture, Eq.~(\ref{timeevolutionequationopensystemorginteractionpictureprojectedexplicitstaticHamiltonianfinaldensitymatrixcase}). Notice that the mean~(\ref{Mdefintion}) is not the same as the conditional average~(\ref{onejumpaverage}). To obtain Eq.~(\ref{quantummasterequation}), we have used a key identity~\cite{Wiseman2010},
\begin{eqnarray}
\label{averageidentity}
M[f[\widetilde\sigma_A]\Delta N_\omega(t)]=M[f(\widetilde\sigma_A)r(\omega){\rm Tr}_A[A^\dag(\omega)A(\omega)\widetilde\sigma_A]]\Delta t,
\end{eqnarray}
where $f$ is an arbitrary function of $\widetilde\sigma_A$. The reason for Eq.~(\ref{averageidentity}) is simple: the probability of finding a specific $\widetilde\sigma_A$ and $\Delta N_\omega(t)$ equals a multiplication of the probability of finding the density matrix and the conditional probability of finding $\Delta N_\omega(t)$ given $\widetilde\sigma_A$, where the latter  is simply $P_\omega(t)$, Eq.~(\ref{onejumpprobabilitygivenpsi}). Finally, we can simulate Eqs.~(\ref{stochasticdifferentialequationvector}) or~(\ref{stochasticdifferentialequationdensitymatrix}) using a Monte-Carlo method. According to Eqs.~(\ref{onejumpprobabilitygivenpsi}) and~(\ref{alljumpsprobabilitygivenpsi}), the probability of observing a deterministic evolution from time $t'$ until $t$ and then being interrupted by a jump of type $\omega$ during the time interval $t\sim t+\Delta t$ is
\begin{eqnarray}
P_\omega(t) e^{-\int_{t'}^{t} \Gamma(s)ds}.
\end{eqnarray}
Hence, we can construct the probability of finding a QJT that starts at time 0, jumps $N$ times with types $\omega_i$ at time intervals $t_i\sim t_i+\Delta t_i$ ($i=1,\cdots,N$), and ends at time $t$:
\begin{eqnarray}
\label{probQJvector}
P\{|\psi\rangle_A\}= e^{-\int_{t_{N}}^{t} \Gamma(s_{N})ds_{N} } P_{\omega_N}(t_N) e^{-\int_{t_{N-1}}^{t_N} \Gamma(s_{N-1})ds_{N-1}} \cdots P_{\omega_1}(t_1) e^{-\int_0^{t_1} \Gamma(s_1)ds_1} .
\end{eqnarray}
This formula is very analogous to the probability of observing a classical jump trajectory in discrete state space~\cite{Seifert2011}. We can also write the above probability alternatively using the density matrix of the QJT:
\begin{eqnarray}
\label{probQJdensitymatrix}
P\{\sigma_A\}&=&{\rm Tr}_A\left[ G_0(t,t_{N}) J(\omega_N)G_0(t_N,t_{N-1})\cdots J(\omega_1)G_0(t_1,0)\sigma_A(0)\right]\prod_{i=1}^N\Delta t_i,
\end{eqnarray}
where 
the superoperators $J(\omega)$ are defined as
\begin{eqnarray}
\label{jumppartsimplecase}
J(\omega)O=r(\omega)A(\omega) O A^\dag(\omega),
\end{eqnarray}
and $G_0(t_i,t_{i-1})$ ($i=1,\cdots,N$ and $t_0=0$) is the superpropagator of the time evolution equation,
\begin{eqnarray}
\label{linearquantummastersimplecase}
\partial_t \pi_A(t)&=&{L}_{0}\pi_A(t)\nonumber\\
&=&-i[H_A+H_{LS},\pi_A(t)]-\frac{1}{2}\sum_{\omega}r(\omega)\left\{A^\dag(\omega)A(\omega),\pi_A(t)\right\},
\end{eqnarray}
namely,
\begin{eqnarray}
\label{solutionlineardensitymatrixequationsimplecase}
G_0(t,t')=T_\leftarrow e^{ \int_{t'}^t ds{L}_0 }.
\end{eqnarray}
We have transformed back to the Schr${\ddot{o}}$dinger picture. Both superoperators act on the symbols on their RHSes. Note that $\pi_A(t)$ is not yet normalized. An explanation for the equivalence of Eqs.~(\ref{probQJvector}) and ~(\ref{probQJdensitymatrix}) is given in Appendix E.  We will see that these two equations, especially the second equation, play central roles in deriving various FTs at the QJT level.

\subsubsection{Complex interaction}
Here, we want to extend the previous results to the case of the general interaction Hamiltonian, Eq.~(\ref{interactionHamiltonianVdetailedexpression}) ($a\ge 2$). We first concretely explain why the wave-vector description of QJTs fails even if the initial state of  system A is a pure state. If the initial state is a general mixed state, the failure is apparent. Assume that at time $t$ the wave vector of system A is $|\widetilde\psi(t)\rangle_A$. We can perform an analogous analysis to that performed previously and find that, after the time interval $\Delta t$, if an energy change $\omega$ occurred at one B atom, the system state is updated to
\begin{eqnarray}
\label{jumpedstatevectortimetaugeneral}
|\widetilde\psi(t+\tau)\rangle_A=
\frac{\sum_{a} {A}_a(\omega) f_{lk}^a(\tau,\omega)|\widetilde\psi(t)\rangle_A}{(\sum_{a,b}f_{lk}^a(\tau,\omega)^*f_{lk}^b(\tau,\omega)\langle A_a^\dag(\omega)A_b(\omega)\rangle)^{1/2}}.
\end{eqnarray}
where the energy conservation condition, Eq.~(\ref{energyconservationcondition}), has been considered due to the coefficient
\begin{eqnarray}
\label{fmnk}
&&f_{lk}^a(\tau,\omega)=-i\int_{0}^{\tau}ds e^{-i\omega s}\langle\chi_l| {\widetilde B}_a(s)  |\chi_k\rangle.
\end{eqnarray}
The presence of the sum over the index $a$ is due to the fact that, given the same Bohr frequency $\omega$,  there are several different corresponding operators $A_a$; also see Sec.~(\ref{section2B1}). Obviously, for different quantum numbers $l$, the wave vector Eq.~(\ref{jumpedstatevectortimetaugeneral}) is distinct even if the types of jumps are exactly the same. Because we do not record the concrete quantum number that is in charge of the jump, the density matrix formula is naturally needed. To this end, consider the density matrices of a B atom and of system A to be $|\chi_k\rangle \langle \chi_k|$ and $\widetilde\sigma_A(t)$, respectively. The global density matrix of the combined A and B systems after a brief interaction time $\tau$ is
\begin{eqnarray}
\widetilde\rho(t+\tau) &=&\left[{T}_{\leftarrow} e^{ -i \alpha\int_{t}^{t+\tau}ds  {\widetilde V}(s)}\right]\widetilde\sigma_A(t) \otimes|\chi_k\rangle\langle \chi_k|\left[{T}_{\leftarrow} e^{ -i \alpha\int_{t}^{t+\tau}ds  {\widetilde V}(s)}\right]^\dag\nonumber\\
&=&\widetilde\sigma_A(t)\otimes|\chi_k\rangle\langle \chi_k|-i\alpha\int_{t}^{t+\tau}ds {\widetilde V}(s)|\chi_k\rangle\langle \chi_k|\otimes \widetilde\sigma_A(t)
+i\alpha\widetilde\sigma_A(t)\otimes|\chi_k\rangle\langle \chi_k| \int_{t}^{t+\tau}ds {\widetilde V}(s)\nonumber\\
&&-\alpha^2\int_t^{t+\tau} ds\int_t^s du {\widetilde V}(s){\widetilde V}(u)|\chi_k\rangle\langle \chi_k|\otimes\widetilde\sigma_A(t)-\alpha^2\widetilde\sigma_A(t)\otimes|\chi_k\rangle\langle \chi_k|\int_t^{t+\tau} ds\int_t^s du {\widetilde V}(u){\widetilde V}(s) \nonumber\\
&&+\alpha^2\int_t^{t+\tau} ds\widetilde V(s) \widetilde\sigma_A(t)\otimes|\chi_k\rangle\langle \chi_k|\int_t^{t+\tau} du\widetilde V(u) + {\cal O}(\alpha^3) .
\end{eqnarray}
Then, we measure the energy of the output B atom. If the atom remains at the same eigenvector with eigenvalue $\chi_k$, the density matrix of the system A is then
\begin{eqnarray}
\label{continuousdensitymatrixunnormalizedorig}
\widetilde\sigma_A(t+\tau)&=&\frac{{\rm Tr}_B [|\chi_k\rangle\langle \chi_k|\widetilde\rho(t+\tau)]\chi_k\rangle\langle \chi_k|]}{{\rm Tr}\left[|\chi_k\rangle\langle \chi_k|\widetilde\rho(t+\tau)\right] } \nonumber\\
&\propto& \widetilde\sigma_A(t)+\alpha^2\sum_{a,b,\omega,\omega'} e^{i(\omega-\omega')t}g_{ab}^k(\tau,\omega,\omega'){A}_a^\dag(\omega){A}_b(\omega') \widetilde\sigma_A(t)\nonumber\\
&& \hspace{2cm}+\alpha^2\sum_{a,b,\omega,\omega'} e^{-i(\omega-\omega')t} g_{ab}^k(\tau,\omega,\omega')^*\widetilde\sigma_A(t){A}_b^\dag(\omega'){A}_a(\omega).
\end{eqnarray}
where
\begin{eqnarray}
\label{functiongkln}
g_{ab}^{k}(\tau,\omega,\omega')=-\int_0^\tau ds\int_0^s du e^{i(\omega-\omega')s +i\omega'u}\langle \chi_k|{\widetilde B}_a(s){\widetilde B}_b(s-u)|\chi_k\rangle.
\end{eqnarray}
To be similar with the previous argument, according to the rotating wave approximation, we may simplify Eq.~(\ref{continuousdensitymatrixunnormalizedorig}) by keeping all terms with $\omega=\omega'$. In particular,
within a longer time interval $\Delta t$, in which there are $N$ ($\gg 1$) B atoms interacting with system A, if we do not observe a change in energy of these B atoms, we find that at time $t+\Delta t$ the system's density matrix evolves to \begin{eqnarray}
\label{continuousdensitymatrixunnormalized}
\widetilde\sigma_A(t+\Delta t)&\propto& \widetilde\sigma_A(t)+\alpha^2N\sum_{a,b,\omega}\left[\sum_n p_ng_{ab}^k(\tau,\omega)\right]{A}_a^\dag(\omega){A}_b(\omega) \widetilde\sigma_A(t)\nonumber\\
&& \hspace{2cm}+\alpha^2N\sum_{a,b,\omega}\left[\sum_n p_ng_{ab}^k(\tau,\omega)\right]^*\widetilde\sigma_A(t){A}_b^\dag(\omega){A}_a(\omega),
\end{eqnarray}
where $g^k_{ab}(\tau,\omega)$ is Eq.~(\ref{functiongkln}) except that $\omega=\omega'$ here.
Using the same argument in proving Eq.~(\ref{BreuerPetruccioneformula1})~\cite{Breuer1995}, we find that the above sum over the quantum number $k$ is equal to
\begin{eqnarray}
-\tau\left[ \frac{1}{2}r_{ab}(\omega) +iS_{ab}(\omega)\right].
\end{eqnarray}
Substituting this into Eq.~(\ref{continuousdensitymatrixunnormalized}) and normalizing, we obtain
\begin{eqnarray}
\label{continuousnormalizeddensitymatrix}
\widetilde\sigma_A(t+\Delta t)=\widetilde\sigma_A(t) &&-i\Delta t[H_{LS},\widetilde\sigma_A(t)]-\Delta t\sum_{\omega,k,l}\frac{1}{2}r_{ab}(\omega)\left\{A_a^\dag(\omega)A_b(\omega),\widetilde\sigma_A(t)\right\}  \nonumber\\
&&+\Delta t\sum_{\omega,a,b }r_{ab}(\omega){\rm Tr}_A[ A_a^\dag(\omega)A_b(\omega)\widetilde\sigma_A(t)]\widetilde\sigma_A(t) +{\cal O}(\Delta t^2).
\end{eqnarray}

The other possible output is that there is an energy change at one of these B atoms. Let us again suppose that the change in the energy of the B atom is $\omega$ and that this has occurred from an eigenvalue $\chi_k$ to $\chi_l$. In this situation, the density matrix of  atom A jumps to
\begin{eqnarray}
\widetilde\sigma_A(t+\tau)&=&\frac{{\rm Tr}_B[|\chi_l\rangle \langle \chi_l|\widetilde\rho(t+\tau)|\chi_l\rangle \langle \chi_l|]}{{\rm Tr}\left[|\chi_l\rangle\langle \chi_l|\widetilde\rho(t+\tau)\right] } \propto  \alpha^2\sum_{a,b} f^{a}_{lk}(\tau,\omega)f^{b}_{lk}(\tau,\omega)^*  {A}_a(\omega)\widetilde\sigma_A(t){A}_b^\dag(\omega).
\end{eqnarray}
We have considered the energy conservation condition Eq.~(\ref{energyconservationcondition}). Obviously, the density matrix after the jump depends on the concrete quantum number $k$. Because we only record the change in energy $\omega$ instead of the exact quantum number $k$, the genuine density matrix at time $t+\Delta t$ shall be a sum of the weighted density matrix with a specific $k$ value, namely,
\begin{eqnarray}
\label{jumpednormalizeddensitymatrix}
\widetilde\sigma_A(t+\Delta t)&=&\frac{  \sum_{a,b} \left[\sum_k p_k f^{a}_{lk}(\tau,\omega)f^{b}_{lk}(\tau,\omega)^* \right] {A}_a(\omega)\widetilde\sigma_A(t){A}_b^\dag(\omega)}{\sum_{a,b} \left[\sum_k p_k f^{a}_{lk}(\tau,\omega)f^{b}_{lk}(\tau,\omega)^* \right]  {\rm Tr}_A[{A}_a(\omega)\widetilde\sigma_A(t){A}_b^\dag(\omega)]}\nonumber\\
&=&\frac{\sum_{a,b} r_{ab}(\omega) {A}_a(\omega)\widetilde\sigma_A(t){A}_b^\dag(\omega)}{\sum_{a,b} r_{ab}(\omega)  {\rm Tr}_A[{A}_a(\omega)\widetilde\sigma_A(t){A}_b^\dag(\omega)]}.
\end{eqnarray}
This density matrix has been normalized. To derive the second equation, the following identity was used~\cite{Breuer1995}:
\begin{eqnarray}
 \sum_k p_k f^{a}_{lk}(\tau,\omega)f^{b}_{lk}(\tau,\omega)^*  \approx \tau r_{ab}(\omega).
\end{eqnarray}
According to Eq.~(\ref{jumpednormalizeddensitymatrix}), we see that the possibility of finding such a change in energy $\omega$ is
\begin{eqnarray}
\label{onejumpprobabilitygivenpsi2}
P_\omega(t)= \Delta t \sum_{a,b}  r_{ab}(\omega)  {\rm Tr}_A[{A}_a(\omega)\widetilde\sigma_A(t){A}_b^\dag(\omega)].
\end{eqnarray}
Hence, the probability of observing an energy change during a time interval $t\sim t+\Delta t$ is
\begin{eqnarray}
\label{alljumpsprobabilitygivensigma}
\Delta t\Gamma(t)=\Delta t \sum_{\omega,a,b}  r_{ab}(\omega) {\rm  Tr}_A[{A}_a(\omega)\widetilde\sigma_A(t){A}_b^\dag(\omega)].
\end{eqnarray}

We can combine Eqs.~(\ref{continuousnormalizeddensitymatrix}) and~(\ref{jumpednormalizeddensitymatrix}) into a compact form:
\begin{eqnarray}
\label{stochasticdifferentialequationdensitymatrixcomplexcase}
\Delta \widetilde \sigma_A &=&  \widetilde \sigma_A (t+\Delta t) - \widetilde \sigma_A (t) \nonumber\\
&=& \prod_{\omega}[1-\Delta N_\omega(t)]\Delta t \left\{- i [H_{LS},\widetilde \sigma_A (t)]-\frac{1 }{2} \sum_{\omega,a,b}r_{ab}(\omega)\left\{A_a^\dag(\omega)A_b(\omega),\widetilde\sigma_A(t)\right\}  \right.\nonumber\\
&&\left.+\sum_{\omega,a,b }r_{ab}(\omega){\rm Tr}_A[ A_a^\dag(\omega)A_b(\omega)\widetilde\sigma_A(t)] \widetilde\sigma_A(t)\right\} \nonumber\\
&&+\sum_{\omega }\Delta N_\omega(t) \left(\frac{\sum_{a,b} r_{ab}(\omega) {A}_a(\omega)\widetilde\sigma_A(t){A}_b^\dag(\omega)}{\sum_{a,b}  r_{ab}(\omega) {\rm Tr}_A[{A}_a(\omega)\widetilde\sigma_A(t){A}_b^\dag(\omega)]}-\widetilde\sigma_A(t)\right ),
\end{eqnarray}
where $\Delta N_\omega(t)$ is also a stochastic variable, equal to $0$ or $1$, and
\begin{eqnarray}
E[\Delta N_\omega(t)]= \Delta t \sum_{a,b} r_{ab}(\omega) {\rm  Tr}_A[{A}_a(\omega)\widetilde\sigma_A(t){A}_b^\dag(\omega)].
\end{eqnarray}
We call the density matrix evolution described by Eq.~(\ref{stochasticdifferentialequationdensitymatrixcomplexcase}) a QJT in the system's density matrix space. Analogous to the case of the simple interaction Hamiltonian, if we perform an average of Eq.~(\ref{stochasticdifferentialequationdensitymatrixcomplexcase}) over all QJTs, we can find that the mean,
\begin{eqnarray}
\label{averageddensitymatrixstaticHamiltoniancomplexcase}
\widetilde \rho_A(t)=M[\widetilde\sigma_A(t)],
\end{eqnarray}
satisfies the time evolution equation
\begin{eqnarray}
\label{timeevolutionequation}
\partial_t {\widetilde {\rho}_A(t)} =&-&i[{H}_{LS},{\widetilde {\rho}_A(t)} ]\nonumber \\
&+& \sum_{\omega,a,b}r_{ab}(\omega)\left[{ A}_b(\omega){\widetilde {\hat{\rho}}_A(t)} A_a^\dag(\omega)-\frac{1}{2}\left\{A_a^\dag(\omega) {A}_b(\omega),{\widetilde {\rho}_A(t)}\right \} \right],
\end{eqnarray}
where we have used an extension of Eq.~(\ref{averageidentity}):
\begin{eqnarray}
\label{averageidentity2}
M\left[f[\widetilde\sigma_A]\Delta N_\omega(t)\right]=M [f(\widetilde\sigma_A)  \sum_{a,b} r_{ab}(\omega)  {\rm Tr}_A[{A}_a(\omega)\widetilde\sigma_A(t){A}_b^\dag(\omega)] ]\Delta t.
\end{eqnarray}
Eq.~(\ref{timeevolutionequation}) is simply the MQME for the static Hamiltonian in the interaction picture or Eq.~(\ref{timeevolutionequationheatopensystemorginteractionpictureprojectedexplicitstaticHamiltonianfinal}) with $\eta=0$.

Eq.~(\ref{stochasticdifferentialequationdensitymatrixcomplexcase}) can also be simulated by the Monte-Carlo method, which is almost the same as that in the preceding subsection. In particular, the probability of observing a trajectory is also described by Eq.~(\ref{probQJvector}) or ~(\ref{probQJdensitymatrix}) except that $P_\omega(t)$ is equal to Eq.~(\ref{onejumpprobabilitygivenpsi2}) instead of~(\ref{onejumpprobabilitygivenpsi}), where the superoperator
\begin{eqnarray}
\label{jumppartcomplexcase}
J(\omega)O=\sum_{ab}r_{ab}(\omega)A_a(\omega) O A_b^\dag(\omega),
\end{eqnarray}
and $G_0(t_i,t_{i-1})$ is the superpropagator of the following equation:
\begin{eqnarray}
\label{linearquantummastercomplexcase}
\partial_t \pi_A(t)=&-&i[H_A+{H}_{LS},\pi_A]-\frac{1}{2}\sum_{\omega,a,b}r_{ab}(\omega)\left\{A_a^\dag(\omega) {A}_b(\omega),\pi_A\right \}.
\end{eqnarray}
Note that we have transformed back to the Schr$\ddot{o}$dinger picture here. Obviously, if both the subscripts $a$ and $b$ are equal to $1$, all the results in this subsection  are reduced to those in the case of the simple interaction Hamiltonian. Finally, we want to emphasize that our treatment of the repeated interaction quantum model and the explanation of the QJT concept is not very rigorous. For instance, we did not clarify how choosing $\Delta t$ makes it far larger than the brief interaction time interval $\tau$ while ensuring an energy change has  only occurred at one B atom. We refer the interested reader to a rigorous theory of the quantum model given by Attal and Yan~\cite{Attal2006}.

\subsection{Other time-dependent Hamiltonians}
\label{subsection6B}
In this section, we want to generalize the previous results in the static Hamiltonian case to the cases of time-dependent Hamiltonians. From the above discussion, we noted that there is no fundamental differences in the spirit of obtaining the QJT concept in the cases of simple or complex interaction Hamiltonians, although the latter are indeed more complicated notationally. Hence, for the sake of simplicity, the results for the cases of time-dependent Hamiltonians that we are presenting are only for the case of simple interactions.

\subsubsection{Weakly driven Hamiltonian}
The Hamiltonian of the quantum system A is Eq.~(\ref{weakdrivenHamiltonian}). Because $\gamma$ is a small parameter, we may take a Taylor series for Eq.~(\ref{simpleinteractionstaticHamiltonian}) at $\gamma=0$ up to zeroth order:
\begin{eqnarray}
{\widetilde V}(s)=\sum_\omega {A}(\omega)e^{-i\omega s}\otimes {\widetilde B}(s)+{\cal O}(\gamma).
\end{eqnarray}
The reader is reminded again that here $A(\omega)$ is the decomposition with respect to the eigenvectors and eigenvalues of the bare Hamiltonian $H_0$. Because we have assumed that the magnitude of $\gamma$ is the same as that of $\alpha$ and because we only consider the lowest order approximation, most of the formulas in this case are retained except that, in Eqs.~(\ref{linearquantummastersimplecase}), $H_A$ is changed into the weakly driven Hamiltonian, Eq.~(\ref{weakdrivenHamiltonian}).

\subsubsection{Periodically driven Hamiltonian cases}
For the periodic Hamiltonian, Eq.~(\ref{perodicallydrivenHamiltonian}), based on Eq.~(\ref{decompositionofAoperatorperiodicallydrivenHamiltonian}), the key interaction picture operator for $V$ is \begin{eqnarray}
\label{simpleinteractionperiodicallydrivenHamiltonian}
{\widetilde V}(s)=\sum_\omega {A}(\omega,0)e^{-i\omega s}\otimes {\widetilde B}(s).
\end{eqnarray}
When we compare Eqs.~(\ref{simpleinteractionperiodicallydrivenHamiltonian}) and~(\ref{simpleinteractionstaticHamiltonian}), we find that they are almost identical in form. Hence, without further derivations, in this case, we conclude that the QJT concept and its physical interpretation can be well established, as we did in the case of the static Hamiltonian, and previous formulas are also retained without significant revisions. The reader is reminded that the parameter $0$ in $A(\omega,0)$ is essential if one wants to obtain formulas in the Sch$\ddot{o}$dinger picture; see Eq.~(\ref{AtransformationbetweenSandHperiodicallydrivenHamiltonian}).

\subsubsection{Adiabatically driven Hamiltonian case}
This case is slightly complicated. According to Eq.~(\ref{decompositionofAoperatoradiabaticallydrivenHamiltonian1}), the interaction picture operator of $V$ is
\begin{eqnarray}
\label{simpleinteractionadiabticallydrivenHamiltonian}
{\widetilde V}(s)=\sum_{m,n} {A}_{mn}(t,0)e^{-i\mu_{mn}(t)}\otimes {\widetilde B}(s),
\end{eqnarray}
where the decomposition $A_{nm}(t,0)$ depends on time $t$, which is significantly distinct from the previous three cases. Substituting it into Eq.~(\ref{wavevectorofcompositeAandB}), if  no energy change has occurred at a B atom staying at the state $|\chi_k\rangle$, the wave vector of the quantum system A after the second energy measurement is then
\begin{eqnarray}
\label{wavevectorAsystemoutputcontinueadibaticallydrivenHamiltonian}
|\widetilde\psi(t+\tau)\rangle_A&\propto& \left[1-\int_0^\tau ds \int_0^s du \langle\chi_k|{\widetilde V}(t+s) {\widetilde V}(t+s-u)|\chi_k\rangle\right]\widetilde\psi(t)\rangle_A \nonumber \\
&\propto& \left[ 1+\alpha^2\sum_{m_1,n_1,m_2,n_2}{A}_{m_1n_1}^\dag(t,0){A}_{m_2n_2}(t,0)e^{i\left[\mu_{m_1n_1}(t) -\mu_{m_2n_2}(t)\right]} g_k\left(\tau,\omega_{m_1n_1}(t),\omega_{m_2n_2}(t)\right) \right]|\widetilde\psi(t)\rangle_A.\nonumber\\
&\propto& \left[ 1+\alpha^2\sum_{m_1n_1}{A}_{m_1n_1}^\dag(t,0){A}_{m_1n_1}(t,0) g_k\left(\tau,\omega_{m_1n_1}(t)\right) \right]|\widetilde\psi(t)\rangle_A.\nonumber\\
\end{eqnarray}
To obtain the second equation, we have regarded $s$ and $s-u$ as small parameters and used the approximation in Eq.~(\ref{decompositionofAoperatoradiabaticallydrivenHamiltonian2}). The last step is due to the application of the rotating wave approximation. If we further consider the influence of  $N$  successive interactions between the system A and B atoms within the time interval $\Delta t$, because the external protocol varies slowly (adiabatic approximation), we may reasonably suppose that the time parameter $t$ in the last equation of Eq.~(\ref{wavevectorAsystemoutputcontinueadibaticallydrivenHamiltonian}) is fixed. Hence, in this case, we still obtain an evolution analogous to Eq.~(\ref{continuewavequation}) except that $A(\omega)$ therein is replaced by $A_{m_1,n_1}(t,0)$ and that the sum is about $m_1$ and $n_1$ rather than $\omega$. On the other hand, if a B atom changes its energy from eigenvalues $\chi_k$ to $\chi_l$ after a brief interaction with system A, the wave function of the system jumps to
\begin{eqnarray}
\label{wavevectorAsystemoutputjumpadibaticallydrivenHamiltonian}
|\widetilde\psi(t+\tau)\rangle_A&\propto& \sum_{mn}e^{-i\left[\mu_{mn}(t)-(\chi_l -\chi_k)t\right]}f_{lk}(\tau,\omega_{mn}(t)) {A}_{mn}(t,0) |\widetilde\psi(t)\rangle_A.
\end{eqnarray}
We have noted that the coefficient $f_{lk}$ is significantly non-zero only when the condition of energy conservation is satisfied. In the current case, the condition is
\begin{eqnarray}
\label{energyconservationconditionadiabaticallydrivenHamiltonian}
\omega_{mn}(t)=\chi_l -\chi_k.
\end{eqnarray}
Because we have assumed that $\omega_{mn}(t)$ corresponds to a unique set of  quantum numbers $(m,n)$ (see Sec.\ref{section2B4}), the above energy conservation condition indicates that there is only one term remaining in the sum of Eq.~(\ref{wavevectorAsystemoutputjumpadibaticallydrivenHamiltonian}). Hence, for the case of an adiabatically driven Hamiltonian, we still obtain a jump in the wave vector of the system as Eq.~(\ref{jumpwavequation}) except that $A(\omega)$ therein is replaced by $A_{mn}(t,0)$.

\subsection{QJT for the general MQME}
\label{subsection6C}
We have explained the QJT concept for several open quantum systems. Therefore, could we have a unified approach to obtain these results in a more efficient manner? We hope that this approach could be directly performed on  general MQMEs, Eq.~(\ref{quantummasterequationgeneralformdensitymatrixcase}). This may then provide convenience when we discuss the QST of QJTs from a general perspective. Such a type of approach indeed exists and was mainly developed by Srinivas and Davies~\cite{Srinivas1981}, Carmichael~\cite{Carmichael1993}, Wiseman and Milburn~\cite{Wiseman1993}, and Kist et al.~\cite{Kist1999}. However, their discussions were limited to time-homogenous MQMEs. Here, we will show that their ideas can be easily extended to the time-dependent situation.

We first rewrite the general equation~(\ref{quantummasterequationgeneralformdensitymatrixcase}) in the following form:
\begin{eqnarray}
\label{decomposedgeneralquantummasterequation}
\partial_t \rho_A(t)={\cal L}_{0}(t)\rho_A(t)+\sum_{\omega_t}J(\omega_t,t)\rho_A(t),
\end{eqnarray}
where these two superoperators are defined as
\begin{eqnarray}
\label{decompositioncontinuouspart}
{\cal L}_{0}(t)O=-i[H_A(t)+H_{LS}(t),O]-\frac{1}{2}\sum_{\omega_t,a,b}r_{ab}(\omega_t)\left\{A_a^\dag(\omega_t,t)A_b(\omega_t,t),O\right\},
\end{eqnarray}
and
\begin{eqnarray}
\label{decompositionjumppart}
J({\omega_t},t)O=\sum_{a,b}r_{ab}(\omega_t)A_b(\omega_t,t)O A_a^\dag(\omega_t,t),
\end{eqnarray}
respectively. Now, treating these $J$-terms as ``perturbations", we can obtain its Dyson series solution,
\begin{eqnarray}
\label{Dysonsolution}
\rho_A(t)&=&G_0(t,0)\left[\rho_A(0)\right]\nonumber \\
&+& \sum_{N=1}^\infty\sum_{\{\omega_{t_i}\}} \left(\prod_{i=N}^1 \int_{0}^{t_{i+1}}\right)   \left(\prod_{i=N}^1dt_i\right) G_0(t,t_N)J(\omega_{t_N},t_N)G_0(t_N,t_{N-1})\cdots J(\omega_{t_1},t_1)G_0(t_1,0)\rho_A(0)\nonumber\\
&\equiv&\int_C {\cal D}(t)\hspace{0.1cm} 
G_0(t,t_N)J(\omega_{t_N},t_N)G_0(t_N,t_{N-1})\cdots J(\omega_{t_1},t_1)G_0(t_1,0)\rho_A(0),
\end{eqnarray}
where $\{\omega_{t_i}\}$$=$$\{\omega_{t_N},\cdots,\omega_{t_1}\}$, the sums are over all possible $\omega_{t_i}$ at times $t_i$  ordered by $t=t_{N+1}\ge t_N\ge t_1\ge 0$, and the propagator is
\begin{eqnarray}
\label{propagatorG0generalQME}
G_0(t,t')=T_{\leftarrow}e^{\int_{t'}^{t}du{\cal L}_0(u)}.
\end{eqnarray}
The reader is reminded that these superoperators act on all terms on their right-hand side. For simplicity, we used the abbreviation ${\cal D}(t)$ and the subscript {\it C} to denote that these integrals and sums are with respect to all possible arrangements.

At first sight, the physical relevance of this integral formal solution is ambiguous. After all, the decomposition of Eq.~(\ref{decomposedgeneralquantummasterequation}) is non-unique. However, we see that the integrand in Eq.~(\ref{Dysonsolution}) is very similar to the whole term in the trace of Eq.~(\ref{probQJdensitymatrix}). Indeed, for the case of static Hamiltonians with a simple interaction Hamiltonian, the superoperators ${\cal L}_0$ and $J$ in Eqs.~(\ref{decompositioncontinuouspart}) and (\ref{decompositionjumppart}) are reduced to those in Eqs.~(\ref{jumppartsimplecase}) and (\ref{linearquantummastersimplecase}). In particular, according to Eq.~(\ref{probQJdensitymatrixttoQJvector}) in Appendix E, we find that
\begin{eqnarray}
\label{probabilityinterpretationofstaticQJT}
\sigma_A(t) P\{\sigma_A\}=G_0(t,t_N)J(\omega_{t_N},t_N)G_0(t_N,t_{N-1})\cdots J(\omega_{t_1},t_1)G_0(t_1,t_0)\rho_A(0)\prod_{i=1}^N\Delta t_i ,
\end{eqnarray}
where $\sigma_A(t)$ is the density matrix of the system along the trajectory $\{\sigma_A\}$ at time $t$, and the probability of observing a QJT, $P\{\sigma_A\}$, has been given by Eq.~(\ref{probQJdensitymatrix}).  Hence, the RHS of Eq.~(\ref{Dysonsolution}) is  a weighted sum over all possible QJTs of the open quantum system A, and the Dyson series solution is simply
\begin{eqnarray}
\label{densitymatrixaveragerepresentation}
\rho_A(t)
=M[\sigma_A(t)],
\end{eqnarray}
or the Schr$\ddot{o}$dinger picture of Eq.~(\ref{averageddensitymatrixstaticHamiltoniancomplexcase}). Although our current discussion is on the static Hamiltonian case, the results are relevant to other MQMEs as well. The interested reader may check them individually. Finally, we shall note that the trace of the integrand in Eq.~~(\ref{Dysonsolution}),
\begin{eqnarray}
\label{probQJgeneralcase}
p\{\sigma_A\}={\rm Tr}_A[G_0(t,t_N)J(\omega_{t_N},t_N)G_0(t_N,t_{N-1})\cdots J(\omega_{t_1},t_1)G_0(t_1,t_0)\rho_A(0)],
\end{eqnarray}
has the interpretation of a probability {\it density} of observing a particular QJT in the density matrix space of the system: it has an initial density matrix $\rho_A(0)$, which undergoes $N$ jumps at increasing times $t_i$ ($i$$=$$1$, $\cdots$, $N$), with an order of jump types $\{\omega_{t_i}\}$.

\section{Trajectory heat and work }
\label{section7}
The previous discussion has clearly shown that the presence of QJT is due to the fact that some external detector is continuously monitoring the changes in energy of the bath atoms: within a time interval $t\sim t+\Delta t$, if the system A evolves deterministically, there are no  energy changes characterizing these passing B atoms; on the contrary, if the state of system A jumps, an energy change, $\omega_t$, essentially occurs at one of these B atoms; see the energy conservation condition, Eq.~(\ref{energyconservationcondition}). Hence, given a QJT $\{\sigma_A\}$ within a finite time interval, $(0,t_f$), if we record all these energy changes (not the exact energy values of these B atoms) and sum up these numbers, we will obtain the total energy exchanged between the single open quantum system A and the bath B atoms. From the perspective of thermodynamics, we naturally interpret this exchanged energy as the heat released along the QJT. Hence, given that there are $N$ jumps of type ${\omega_{t_i}}$ ($i=1,\cdots,N$), the trajectory heat is formally defined as~\cite{Breuer2003,DeRoeck2004,DeRoeck2006,Derezinski2008,Horowitz2012,Leggio2013,Hekking2013,Liu2014,Liu2014a,Suomela2015,Gong2016}
\begin{eqnarray}
\label{heatdefinitionQJ}
Q\{\sigma_A\}=\sum_{i=1}^N \omega_{t_i}.
\end{eqnarray}
Note that we do not impose any restrictions on the initial density matrix of the system, $\sigma_A(0)$.

We may also define the trajectory work. In contrast to the trajectory heat definition, we have to apply the TEM scheme~\cite{Kurchan2000,Campisi2011} to the single system A in addition to continuously monitoring the energy changes in these B atoms. Because the energy measurement is performed on system A at the beginning, its initial density matrix is no longer arbitrary. This is the same as what we have considered in Sec.~\ref{subsection4A}. Assuming that the system's Hamiltonian $H_A(t)$ has instantaneous eigenvectors $|\varepsilon_n(t)\rangle$ with discrete eigenvalues $\varepsilon_n(t)$, for a QJT $\{\sigma_A\}$ with $N$ jump types $\{\omega_{t_i}\}$, starting from the density matrix $|\varepsilon_m(0)\rangle\langle \varepsilon_m(0)|$ and ending at the density matrix $|\varepsilon_n(t_f)\rangle\langle \varepsilon_n(t_f)|$, because of the last energy measurement at the end time $t_f$, the trajectory work is defined as
\begin{eqnarray}
\label{inclusiveworkdefinitionQJ}
W_{nm}\{\sigma_A\}=\varepsilon_n(t_f)-\varepsilon_m(0)+ Q\{\sigma_A\}.
\end{eqnarray}
Obviously, this is the inclusive work at the trajectory level and is the first law of thermodynamics under the concept of QJTs. In addition, if the open quantum system has the Hamiltonian in Eq.~(\ref{freeopenquantumsystemHamiltonian}), we may alternatively define the trajectory exclusive work if the TEM is applied to the bare Hamiltonian, $H_0$. In this case, the exclusive work is
\begin{eqnarray}
\label{exclusiveworkdefinitionQJ}
W^0_{nm}\{\sigma_A\}=\varepsilon_n -\varepsilon_m+ Q\{\sigma_A\}.
\end{eqnarray}
where $\varepsilon_n$ and $\varepsilon_m$ are the energy eigenvalues of $H_0$.

\section{Trajectory characteristic operators}
\label{section8}
Obviously, the trajectory heat and work, Eqs.~(\ref{heatdefinitionQJ})-(\ref{exclusiveworkdefinitionQJ}), are random due to the random characteristics of QJTs. Hence, it shall be interesting to explore their statistical features. In particular, because QJTs have very strong connections with the MQMEs, one may naturally speculate that their statistics should consist of those of the stochastic heat and work defined by applying the TEM scheme to the combined system and heat bath; see Eqs.~(\ref{heatdefintionopensystem}),~(\ref{inclusiveworkdefinitionopensystem}) and~(\ref{exclusiveworkdefinitionopensystem}). However, there are no ad hoc reasons to support this conjecture. After all, very different measurement schemes are involved in these two types of stochastic thermodynamic quantities. Because the QJT has a classical probability measure, we can also define the CFs for the random trajectory heat and work in analogy to what we did in Sec.~\ref{section4}. For the heat case, its CF is apparently
\begin{eqnarray}
\label{trajectoryCFheatdef}
\Psi_h(\eta)=\int_C {\cal D}(t)p\{\sigma_A\}  e^{i\eta Q\{\sigma_A\}}.
\end{eqnarray}
Substituting Eqs.~(\ref{probQJgeneralcase}) and~(\ref{heatdefinitionQJ}), the RHS of the above equation becomes
\begin{eqnarray}
\label{CFheatmiddle}
{\rm Tr}_A\left[\int_C {\cal D}(t) \hspace{0.1cm}
 G_0(t,t_N)e^{i\eta\omega_{t_N}}J(\omega_{t_N},t_N)G_0(t_N,t_{N-1}) \cdots e^{i\eta\omega_{t_1}}J(\omega_{t_1},t_1)G_0(t_1,t_0)\rho_A(0) \right].
\end{eqnarray}
We immediately find that the entire term in the square brackets is almost the same as the Dyson series solution, Eq.~(\ref{Dysonsolution}), for the general MQME; the only difference is that each superoperator $J(\omega_{t},t)$ in the latter is multiplied by an exponential ``phase'' factor,  $e^{i\eta\omega_t}$ here. Hence, without further derivations, this analogy leads us to conclude that, if we define the whole integral in Eq.~(\ref{CFheatmiddle}) as the trajectory HCO, it must satisfy a time evolution that is exactly the same as that of $\hat\rho_A(t,\eta)$; see Eq.~(\ref{Heatmasterequationgeneralform}). Therefore, we do not need to introduce new notation for the trajectory HCO. In a word, we have confirmed the conjecture in the heat case. The reader is reminded that during the construction of the QJT concept, we  implicitly applied the weak coupling limit. For instance, Eqs.~(\ref{wavevectorAsystemoutputcontinuesecularapproximation}) and~(\ref{BreuerPetruccioneformula1}) achieve rigorous mathematical meaning only in this limit. Hence, the statistics of the stochastic heat and work are defined by either the TEM scheme on the composite system or  the QJTs having the same mathematical foundation. It would be interesting to say some words about the history of Eq.~(\ref{Heatmasterequationgeneralform})~\cite{Gardiner2004,Garrahan2010}. Although this result was usually attributed to Espositor~\cite{Esposito2009}, the earliest version shall be credited to Mollow~\cite{Mollow1975}, who investigated the probability distribution of the number of photons of a two-level atom that is driven by a weak classical field and simultaneously interacts with a bath of modes of a radiation field. Zoller et al.~\cite{Zoller1987} extended Mollow's result to three-level systems. At that time, the concept of QJTs was still in its infancy, and the emission and absorption of photons were not interpreted as heat.

The next thermodynamic quantity is the trajectory work, Eq.~(\ref{inclusiveworkdefinitionQJ}). Its CF is
\begin{eqnarray}
\label{trajectoryCFworkdef}
\Psi_w(\eta)=\sum_{n,m} \int_C {\cal D}(t)  p_{nm}\{\sigma_A\}P_m(0)  e^{i\eta W_{nm}\{\sigma_A\}}.
\end{eqnarray}
where $P_m(0)$ is the probability of observing the quantum system A at the eigenvector $|\varepsilon_m(0)\rangle$, and the probability density of observing a QJT that has an initial density matrix $|\varepsilon_m(0)\rangle \langle \varepsilon_m(0)|$, that undergoes $N$ jumps at increasing times $t_i$ ($i$$=$$1$, $\cdots$, $N$) with an order of jump types $\{\omega_{t_i}\}$ and that is measured to be at the energy eigenvector $|\varepsilon_n(t_f)\rangle\langle\varepsilon_n(t_f)|$ at time $t_f$ is equal to
\begin{eqnarray}
\label{trajectoryprobabilityforwork}
p_{nm}\{\sigma_A\}&=&{\rm Tr}_A\left[|\varepsilon_n(t_f)\rangle\langle \varepsilon_n(t_f)|\sigma_A(t_f)\right]p\{\sigma_A\}\nonumber\\
&=&{\rm Tr}_A\left[|\varepsilon_n(t_f)\rangle\langle \varepsilon_n(t_f)G_0(t,t_N)J(\omega_{t_N},t_N)G_0(t_N,t_{N-1})\cdots  J(\omega_{t_1},t_1)G_0(t_1,0)|\varepsilon_m(0)\rangle\langle \varepsilon_m(0)|\right].
\end{eqnarray}
To obtain the second equation, we have used Eq.~(\ref{probabilityinterpretationofstaticQJT}). Substituting Eqs.~(\ref{inclusiveworkdefinitionQJ}) and~(\ref{trajectoryprobabilityforwork}) into Eq.~(\ref{trajectoryCFworkdef}), we immediately find that its RHS is equal to
\begin{eqnarray}
\label{CFworkmiddle}
{\rm Tr}_A\left[e^{i\eta H_A(t)}\int_C {\cal D}(t)
G_0(t,t_N)e^{i\eta\omega_{t_N}}J(\omega_{t_N},t_N)G_0(t_N,t_{N-1})\cdots e^{i\eta \omega_{t_1}}J(\omega_{t_1},t_1)G_0(t_1,t_0)e^{-i\eta H_A(0)}\rho_A(0) \right],
\end{eqnarray}
where the initial density matrix is
\begin{eqnarray}
\rho_A(0)=\sum_m P_m(0)|\varepsilon_m(0)\rangle \langle \varepsilon_m(0)|.
\end{eqnarray}
We have assumed that this is indeed the initial density matrix of the quantum system A and that it is not affected by the first energy projection measurement. In comparison with Eq.~(\ref{CFheatmiddle}), in addition to the fact that the initial density matrix is replaced by $e^{-i\eta H_A(0)}\rho_A(0)$, the other difference is the presence of an operator $e^{i\eta H_A(t_f)}$. Therefore, we find that the whole term in the trace of Eq.~(\ref{CFworkmiddle}) is simply the whole term in the trace of Eq.~(\ref{CFinclusiveworkopensystemgeneralbasedheat}). Specifically, it is exactly the same as the WCO,  $K_A(\eta,t_f)$. Therefore, analogous to the heat case, we do not need to define the trajectory WCO at all. Note that all these discussions are also relevant to the case of the trajectory exclusive work, Eq.~(\ref{exclusiveworkdefinitionQJ}).
We do not write them here.

Based on the above discussion, we arrive at the following important conclusion: for all MQMEs that we discussed so far, the stochastic heat and work, which are   defined either by the TEM scheme applied to the combined system and heat bath or at the QJTs, their statistics are indeed exactly equivalent. This conclusion has two important consequences. On the one hand, this equivalence reminds us that we may measure these thermodynamic quantities experimentally by these two seemly very distinct measurement approaches. So far, we have not been very clear on how to realize the TEM measurement for the heat and work experimentally, even though their definitions are very simple and straightforward. On the contrary, QJT experiments have become a reality~\cite{Nagourney1986,Sauter1986,Bergquist1986,Basch1995,Peil1999,Gleyzes2007,Murch2013,Sun2013,Vool2014,Campagne-Ibarcq2016}. On the other hand, from a computational perspective, we can calculate the distributions of these thermodynamic quantities either by  numerically solving the time evolution equation, e.g., Eqs.~(\ref{Heatmasterequationgeneralform}) or~(\ref{inclusiveworkmasterequationgeneralform}), and then performing a Fourier inverse transformation, or by performing a Monte-Carlo simulation. In many cases, the former is analogous to solving the evolution equation of the density matrix Eq.~(\ref{quantummasterequationgeneralformdensitymatrixcase}) of an open quantum system. Suppose that the Hilbert space of the system has dimensions of $D$. The number of variables of the matrices, $\hat\rho_A(t,\eta)$ or $K_A(t,\eta)$, is $D^2$. In contrast, the number of variables involved in the Monte-Carlo method when solving the wave function evolution, Eq.~(\ref{stochasticdifferentialequationvector}), is $D$. Hence, if the dimension $D$ is very large, the required computational resources of the Monte-Carlo method increase as $\sim D$, which is substantially smaller than the requirement of solving the evolution equations, i.e., $\sim D^2$~\cite{Breuer2002}.

\section{Trajectory fluctuation theorems}
\label{section9}
\begin{figure}\label{figure3}
\includegraphics[width=1\columnwidth]{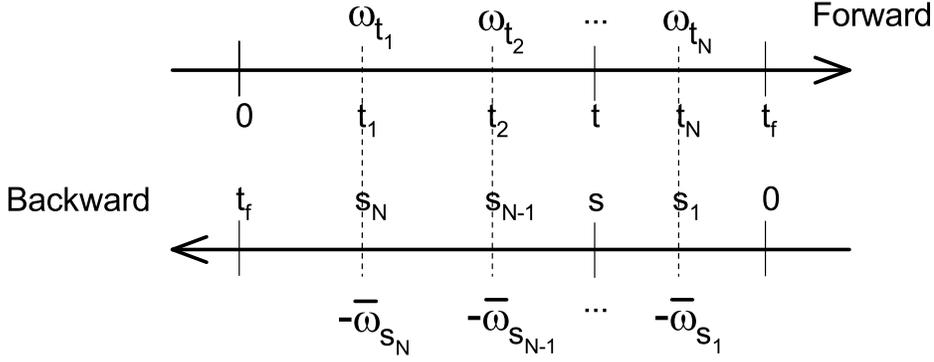}
\caption{A QJT of the forward MQME (the upper line) and its backward trajectory (the down line). The arrows indicate the time directions of the evolutions. The jump types of the forward QJT are ordered by increasing times as $\omega_{t_1},\cdots,\omega_{t_N}$. Then, the corresponding jump types of the backward QJT are ordered by increasing time as  $-\overline\omega_{ s_1},\cdots,-\overline\omega_{s_N}$, where $\overline \omega_{s_i}=\omega_{t_j}$ with $s_i+t_j=t_f$ and $i+j=N+1$. }
\end{figure}
Because the time evolution equations of the trajectory HCO and WCO are still Eqs.~(\ref{Heatmasterequationgeneralform}) and ~(\ref{inclusiveworkmasterequationgeneralform}), respectively, their CFs of course satisfy the same symmetries as those in Sec.~\ref{section5B}. Furthermore, the establishment of the QJT concept implies that there exist new FTs at the trajectory level. In classical ST, trajectory FTs are well known, e.g., the total entropy production FT~\cite{Seifert2005}. Horowitz and Vaikuntanathan~\cite{Horowitz2010} even generalized these FTs to the case in the presence of feedback. In the quantum regime, very analogous efforts have been made by Crooks~\cite{Crooks2008}, Horowitz and his colleagues~\cite{Horowitz2012,Horowitz2013,Horowitz2014,Manzano2015}. In this section, we want to present the trajectory versions of several previous FTs. The first version concerns the ration of the probability densities of observing a forward QJT to a backward QJT:
\begin{eqnarray}
\label{trajectoryFTforheat}
\frac{p\{ \sigma_A \}} {\overline p \{ \overline{ \sigma}_A\}}=
e^{\beta\sum_{i=1}^N\omega_{t_i}}.
\end{eqnarray}
where the forward trajectory $\{\sigma_A\}$ comes from the forward QME~(\ref{quantummasterequationgeneralformdensitymatrixcase}) with an assigned initial density matrix, the completely random matrix $C$. The backward trajectory comes from the backward MQME, i.e., Eq.~(\ref{BackwardHeatmasterequationgeneralform}) with vanishing $\eta$. Assume that the QJT $\{\sigma_A\}$ has $N$ jumps with increasing time orders,  $\omega_{t_1},\cdots,\omega_{t_N}$. Its backward trajectory $\{\overline\sigma_A\}$ is then also started from the completely random matrix $C$ but consists of $N$ jumps with increasing time order $-\overline\omega_{ s_1},\cdots,-\overline\omega_{s_N}$, where $\overline\omega_{s_i}= \omega_{t_j}$ (see Eq.~(\ref{forwardbackwardidentity1})) and particularly
\begin{eqnarray}
\label{siandtj}
s_i+t_j=t_f
\end{eqnarray}
with $i+j=N+1$ $(i=1,\cdots,N)$. According to our previous convention, the parameter $s$ was used to denote the time of the backward process.

To prove Eq.~(\ref{trajectoryFTforheat}), let us define the duals for the superoperator $J$ and the propagator $G_0$ in Eqs.~(\ref{decompositionjumppart}) and~(\ref{propagatorG0generalQME}):
\begin{eqnarray}
&&G^*_0(t,t')=T_\rightarrow e^{\int_{t'}^{t}du{\cal L}^*_0(u)},\\
&&J^*({\omega_t},t)O=\sum_{a,b}r_{ab}(\omega_t)A_a^\dag(\omega_t,t)O A_b(\omega_t,t),
\end{eqnarray}
where
\begin{eqnarray}
\label{continuouspartdual}
{\cal L}_{0}^*(t)O=i[H_A(t)+H_{LS}(t),O]-\frac{1}{2}\sum_{\omega_t,a,b}r_{ab}(\omega_t)\left\{A_a^\dag(\omega_t,t)A_b(\omega_t,t),O\right\}.
\end{eqnarray}
The definition of the dual was given in Eq.~(\ref{dualdefinitionofgeneralheatequationgenerator}). On the other hand, we note that, if we apply adjoint transforms to these superoperators, we have
\begin{eqnarray}
&&[G^*_0(t,t')(O)]^\dag=G^*_0(t,t')(O^\dag), \\
&&[J^*({\omega_t},t)(O)]^\dag=J^*({\omega_t},t)(O^\dag).
\end{eqnarray}
Under these notations and formulas, we can rewrite the probability density of the forward QJT, Eq.~(\ref{probQJgeneralcase}), as
\begin{eqnarray}
\label{forwardprobabilitydensityQJTforheatmiddle}
p\{\sigma_A\}&=&{\rm Tr}_A[G_0(t_f,t_N)(J(\omega_{t_N},t_N)(G_0(t_N,t_{N-1})\cdots (J(\omega_{t_1},t_1)(G_0(t_1,0)(C)))\cdots))] \nonumber\\
&=&{\rm Tr}_A[G^*_0(t_1,0) (J^*(\omega_{t_1},t_1)(G^*_0(t_2,t_1)\cdots  (J^*(\omega_{t_N},t_N)(G^*_0(t_f,t_N)(C)))\cdots))]\nonumber\\
&=&{\rm Tr}_A[\Theta(  G^*_0(t_1,0) (J^*(\omega_{t_1},t_1)(G^*_0(t_2,t_1)\cdots  (J^*(\omega_{t_N},t_N)(G^*_0(t_f,t_N)(C)))\cdots)))^\dag\Theta^{-1}]\nonumber\\
&=&{\rm Tr}_A[\Theta G^*_0(t_1,0) (J^*(\omega_{t_1},t_1)(G^*_0(t_2,t_1)\cdots  (J^*(\omega_{t_N},t_N)(G^*_0(t_f,t_N)(C)))\cdots))\Theta^{-1}].
\end{eqnarray}
We have added circle brackets to clarify the acting regime of each superoperators.

To connect this result with the backward QJT, we still need two additional formulas. The first formula concerns the superoperator of  $J$:
\begin{eqnarray}
\label{forwardbackwardjumps}
e^{-\beta \omega_t} \Theta J^*(\omega_t,t)(O)\Theta^{-1}=\overline J(-\overline\omega_s,s)(\Theta O\Theta^{-1}),
\end{eqnarray}
where $\overline J$ is the $J$-superoperator but of the backward MQME, that is,
\begin{eqnarray}
\overline J(\overline\omega_s,s)O=\sum_{a,b}r_{ab}(\overline \omega_s){\overline A}_b(\overline\omega_s,s)O {\overline A}^\dag_a(\overline\omega_s,s).
\end{eqnarray}
The proof is straightforward if we apply the KMS condition, Eq.~(\ref{DetailedBalanceCondition}), and Eqs.~(\ref{TimereversalCorrelationfuncts})-(\ref{forwardbackwardidentity2}). The other formula is
\begin{eqnarray}
\label{forwardbackwardcontinuous}
\Theta G_0^*(t_j,t)(O)\Theta^{-1}=\overline G_0(s,s_i)(\Theta O\Theta^{-1}),
\end{eqnarray}
where $t_{j-1}\le t\le t_{j}$, $s_{i}\le s\le s_{i+1}$, $t+s=t_f$, and $\overline G_0(s,s_i)$ ($s_i \le s $) is the propagator of the deterministic evolution of the backward process. The proof of Eq.~(\ref{forwardbackwardcontinuous}) is very similar to that of Eq.~(\ref{keyTRformula}). Let the LHS of Eq.~(\ref{forwardbackwardcontinuous}) be $\pi_A'(s)$. We then take a derivative with respect to time $s$ and obtain
\begin{eqnarray}
&&\partial_s \pi_A'(s)=\Theta {\cal L}^*_0(t_f-s)\left(\Theta^{-1} \pi_A'(s) \Theta \right)\nonumber\\
&=&-i[{\overline H}_A(s)+{\overline H}_{LS}(s), \pi_A'(s) ]-\frac{1}{2}\sum_{\overline\omega_s,a,b} r_{ab}(\overline \omega_s) \left\{ {\overline A}_a^\dag(\overline \omega_s,s){\overline A}_b(\overline \omega_s,s),\pi_A'(s)\right\}.
\end{eqnarray}
The last equation was obtained by inserting the explicit formula for ${\cal L}^*_0$, Eq.~(\ref{continuouspartdual}), and applying Eqs.~(\ref{TimereversalCorrelationfuncts})-(\ref{forwardbackwardidentity2}). This is clearly the continuous evolution equation of the backward MQME. Considering the initial condition, $\pi_A'(s_i)=\Theta O\Theta^{-1}$, we have proved Eq.~(\ref{forwardbackwardcontinuous}).

Now, we are in a position to prove the trajectory FT for the heat. By repeatedly applying Eqs.~(\ref{forwardbackwardjumps}) and~(\ref{forwardbackwardcontinuous}) to Eq.~(\ref{forwardprobabilitydensityQJTforheatmiddle}), we immediately find that the original one is  equal to
\begin{eqnarray}
p(\sigma_A)&=&e^{\beta \sum_{i=1}^N \omega_{t_i}}{\rm Tr}_A[ \overline G_0(t_f,s_N) (\overline J(-\overline \omega_{s_N},s_N)(\overline G_0(s_N,s_{N-1})\cdots \nonumber\\
&&\hspace{5cm} (\overline J(-\overline\omega_{s_1},s_1)(\overline G_0(s_1,0)(C)))\cdots)) ].
\end{eqnarray}
The trace term on the RHS is simply the probability density $\overline p(\overline\sigma_A)$ of observing the backward QJT $\{\overline\sigma_A\}$ of the backward process. Therefore, we have proved Eq.~(\ref{trajectoryFTforheat}).

We can consider analogous discussions for the cases of trajectory inclusive and exclusive work. We first notice that the trajectory probability density, Eq.~(\ref{trajectoryprobabilityforwork}), can be rewritten as
\begin{eqnarray}
\label{forwardprobabilitydensityQJTforworkmiddle}
p_{nm}\{\sigma_A\}&=&{\rm Tr}_A[|\varepsilon_n(t_f)\rangle\langle \varepsilon_n(t_f)|G_0(t_f,t_N)(J(\omega_{t_N},t_N)(G_0(t_N,t_{N-1})\cdots\nonumber\\ &&\hspace{4cm}(J(\omega_{t_1},t_1)(G_0(t_1,0)(|\varepsilon_m(0)\rangle\langle \varepsilon_m(0)|)))\cdots))] \nonumber\\
&=&e^{\beta \sum \omega_{t_i}}{\rm  Tr}_A[\Theta |\varepsilon_m(0)\rangle\langle \varepsilon_m(0)|\Theta^{-1}  \overline G_0(t_f,s_N) (\overline J(-\overline \omega_{s_N},s_N)(\overline G_0(s_N,s_{N-1})\cdots \nonumber\\
&&\hspace{4cm} (\overline J(-\overline\omega_{s_1},s_1)(\overline G_0(s_1,0)(\Theta|\varepsilon_n(t_f)\rangle\langle \varepsilon_n(t_f)|\Theta^{-1})))\cdots)) ]\nonumber\\
&=&e^{\beta \sum_{i=1}^N \omega_{t_i}} \overline p_{mn}\{\overline\sigma_A\}.
\end{eqnarray}
On the RHS,  $\overline p_{mn}$ is the probability density of observing the backward trajectory that initially starts from the energy eigenvector $|\varepsilon_n(t_f)\rangle$, with $N$ jumps with time order $-\overline\omega_{ s_1},\cdots,-\overline\omega_{s_N}$, and is measured to be at the energy eigenvector $|\varepsilon_m(0)\rangle$ at the end time $t_f$. Note that we have assumed that $H_A(t)$ is TRI; see Eq.~(\ref{systemHamiltonianbackwardprocess}). Based on this equality, we further have
\begin{eqnarray}
\label{trajectoryFTforinclusivework}
\frac{p_{nm}\{ \sigma_A \}P_m(0)} {\overline p_{mn} \{ \overline{ \sigma}_A\}P_n(t_f)}=\frac{Z(0)}{Z(t_f)}e^{\beta W_{nm}\{\sigma_A\}},
\end{eqnarray}
where $P_m(t)=e^{-\beta\varepsilon_m(t)}/Z(t)$ and $t=0$, $t_f$. Obviously, the numerator on the LHS is the joint probability of measuring system A at the eigenvector $|\varepsilon_m(0)\rangle$ at time 0 and the QJT being $\{\sigma_A\}$. The denominator has the same meaning but is initially measured at the eigenvector $|\varepsilon_n(t_f)\rangle$  and concerns the backward process. Because $P_m(0)$ and $P_n(t_f)$ have canonical distributions, the initial states of these two processes are in thermal equilibrium with the heat bath at the inverse temperature $\beta$, and in particular, their Hamiltonians are $H_A(0)$ and $H_A(t_f)$, respectively. An analogous formula can also be derived for the trajectory exclusive work:
\begin{eqnarray}
\label{trajectoryFTforexclusivework}
\frac{p_{nm}\{ \sigma_A \}P_m^0} {\overline p_{mn} \{ \overline{ \sigma}_A\}P_n^0}= e^{\beta W^0_{nm}\{\sigma_A\}},
\end{eqnarray}
where $P_m^0=e^{-\beta\varepsilon_m}/Z_0$. The meanings of $p_{nm}$ and $\overline p_{mn}$ here are similar to those in the case of inclusive work. For instance, $p_{nm}$ is the probability density of the QJT $\{\sigma_A\}$ starting from $|\varepsilon_m\rangle$ and being measured at  $|\varepsilon_n\rangle$; both of these wave vectors are the eigenvectors of the bare Hamiltonian $H_0$ in Eq.~(\ref{freeopenquantumsystemHamiltonian}). Note that we must impose the condition that $H_1$ is zero at times $0$ and $t_f$. The reason has been provided in Sec.~\ref{section5B}.

\section{Total entropy production}
\label{section10}
In addition to the stochastic heat and work, the total entropy production (TEP) is also an important thermodynamic quantity in irreversible thermodynamics~\cite{Spohn1978}. In the classical situation, it has been well established that the TEP can be defined at the classical trajectory level, and in particular, it also satisfies intriguing FTs~\cite{Lebowitz1999,Seifert2005}. It shall be interesting to investigate their correspondence to MQMEs. Analogous to the cases of heat and work, we can realize this aim through two methods: one method is based on an analogous TEM scheme but one not acting on the system's energy, and the other method is to define a TEP along a QJT. Because the whole consideration is very similar to the cases of heat and work, our discussion about the TEP is brief.

The density matrix of the open quantum system is Hermitian. It possess a diagonal form in its instantaneous orthogonal basis $|\rho_\mu(t)\rangle $. Specifically, we suppose that the spectral decomposition of the density matrix is
\begin{eqnarray}
\rho_A(t)=\sum_{\mu} \rho_\mu (t)|\rho_{\mu}(t)\rangle\langle \rho_{\mu}(t)|.
\end{eqnarray}
The eigenvalue $\rho_\mu(t)$ has an interpretation that it is the probability of finding the system at the eigenvector $|\rho_\mu(t)\rangle$.
To define a stochastic TEP, we measure the instantaneous eigenvector $|\rho_\mu(t)\rangle$ and energy eigenvector $|\chi_l\rangle $ of the open quantum system A and heat bath B, respectively, at the beginning and end of the non-equilibrium process. Then, the stochastic quantity is
\begin{eqnarray}
\label{TEPtwomeasurementscheme}
S_{\nu l\mu k}=k_B[-\ln \rho_\nu(t_f)+\ln \rho_\mu(0)]+\frac{1}{T}(\chi_l-\chi_k).
\end{eqnarray}
Here, we have assumed that these algorithm functions are well-defined, or $\rho_\mu(t)>0$. Note that Eq.~(\ref{TEPtwomeasurementscheme}) is usually distinct from the classical Shannon entropy~\cite{Seifert2005}: here, $-k_B\ln\rho_\mu(t)$ is related to the von Neumann entropy, and these two types of entropy are not essentially the same~\cite{Breuer2002}. The above definition is very analogous to the definition of the inclusive work, Eq.~(\ref{inclusiveworkdefinitionopensystem}). Hence, this similarity leads us to obtain the CF of the TEP of the open quantum system A:
\begin{eqnarray}
\label{CFinclusiveworkQJT}
\Phi_s(\eta)&=&{\rm Tr}[e^{i\eta [-k_B\ln \rho_A(t_f)+H_B] }U(t_f) e^{-i\eta [-k_B\ln\rho_A(0)+H_B]}\rho_A(0)\otimes\rho_B U(t_f)^\dag]\nonumber\\
&=&{\rm Tr}_A\left[\rho_A(t_f)^{-i\eta k_B} T_{\leftarrow} e^{\int_0^t ds{\check{\cal L}}(s,\eta)} \left(\rho_A(0)^{i\eta k_B}\rho_A(0)\right)\right].
\end{eqnarray}
Obviously, choosing $\eta=i/k_B$ and considering Eq.~(\ref{checkLproperty}), this CF is simply equal to 1. Therefore, the integral FT of the TEP is established as follows:
\begin{eqnarray}
\label{entropyprodequality}
\left\langle e^{- S/k_B}\right\rangle=1.
\end{eqnarray}
In contrast to the previous FTs for heat production and work, the current FT is true for very general initial conditions. Hence, we did not add any subscripts in this equality. According to Jensen's inequality, Eq.~(\ref{entropyprodequality}) implies that $\langle S\rangle\ge 0$, or the second law of thermodynamics for the general MQME, Eq.~(\ref{quantummasterequationgeneralformdensitymatrixcase}). Spohn~\cite{Spohn1978a} and Spohn and Lebowitz~\cite{Spohn1978}  presented a general formula of the TEP for an arbitrary time-homogeneous  quantum dynamical semigroup having a stationary state. In this specific situation, our results are consistent with their formula.

On the other hand, for a QJT that starts from an initial density matrix $|\rho_\mu(0)\rangle\langle\rho_\mu(0) |$, jumps $N$ times with types $\omega_{t_i}$ ($i=1,\cdots,N$), and ends at $|\rho_\nu(t_f)\rangle$ due to the second measurement of the system at time $t_f$, we define its trajectory TEP as
\begin{eqnarray}
\label{entropyproddefinitionQJT}
S_{\nu\mu}\{\sigma_A\}=k_B\left([-\ln \rho_\nu(t_f)+\ln \rho_\mu(0)]+\frac{1}{\beta}Q\{\sigma_A\} \right).
\end{eqnarray}
Eq.~(\ref{heatdefinitionQJ}) presents the explicit expression for $Q\{\sigma_A\}$. We can write the probability density of observing such a QJT as
\begin{eqnarray}
\label{forwardprobabilitydensityQJTforworkmiddle}
p_{\nu\mu}\{\sigma_A\}&=&{\rm Tr}_A[|\rho_\nu(t_f)\rangle\langle \rho_\nu(t_f)|G_0(t_f,t_N) J(\omega_{t_N},t_N)G_0(t_N,t_{N-1})\cdots  J(\omega_{t_1},t_1)G_0(t_1,0)|\rho_\mu(0)\rangle\langle\rho_\mu(0)|].
\end{eqnarray}
Furthermore, we have the following trajectory FT for the TEP:
\begin{eqnarray}
\label{trajectoryprobabilityratiorforTEP}
\frac{p_{\nu\mu}\{ \sigma_A \}\rho_\mu(0)} {\overline p_{\mu\nu} \{ \overline{ \sigma}_A\} \rho_\nu(t_f)}=e^{{S_{\nu\mu}\{\sigma_A\}}/{ k_B }}.
\end{eqnarray}
Note that here, $\overline p_{\mu\nu}$ is the probability density of observing a backward trajectory that starts from $\Theta |\rho_\nu(t_f)\rangle $ with $N$ jumps with time order $-\overline\omega_{ s_1},\cdots,-\overline\omega_{s_N}$ and is measured at the eigenvector $\Theta|\rho_\mu(0)\rangle$. The proof of the above equation is the same as the proof of Eq. (\ref{trajectoryFTforinclusivework}). Based on Eq.~(\ref{trajectoryprobabilityratiorforTEP}), we can easily obtain the detailed FT of the TEP:
\begin{eqnarray}
P(S)=\overline P(-S) e^{S/k_B}.
\end{eqnarray}
The reader is reminded that the initial density matrix for the backward MQME is $\Theta\rho_A(t_f)\Theta^{-1}$.

\section{Example: Periodically driven two-level system}
\label{section11}
Several simple models have been used to illustrate the computations of distributions of the stochastic heat and work and their FTs~\cite{Horowitz2012,Liu2014,Liu2014a,Liu2016,Suomela2014,Suomela2015,Gong2016,Gasparinetti2014,Cuetara2015}. A major portion of them are two-level systems (TLS), including cases of being driven by a weak external field, adiabatical external field, and periodical external field. In this section, we want to use the last model to show the computation of the inclusive work distribution. On the one hand, the thermodynamics of this model at the ensemble level has recently attracted substantial interest~\cite{Breuer2000,Kohn2001,Szczygielski2013,Langemeyer2014}. On the other hand, to the best of our knowledge, whether there is the JE  in this system  has not been definitively answered in the literature. Although we have formally proved that the equality is true, it shall be more desirable to obtain numerical verifications. Finally, using this non-trivial model, we also want to show the reader the consistency of the two methods in computing work distributions: one method is a direct simulation of QJTs, and the other method is to solve the time evolution equations for the HCO and then to perform a Fourier inverse transform of the CF.

The TLS Hamiltonian is
\begin{eqnarray}
\label{FloquetHamiltonian}
H_A(t)=\frac{\hbar\omega_0}{2} \sigma_z +\frac{\hbar\Omega}{2}\left(\sigma_+ e^{-i\omega_L t}+\sigma_- e^{i\omega_L t}\right),
\end{eqnarray}
where $\omega_0$ is the frequency of transitions between these two levels, $\Omega$ is the Rabi frequency, and $\omega_L$ is the frequency of the periodic external field. We can analytically solve for the Floquet basis of this system~\cite{Breuer1997,Szczygielski2013,Langemeyer2014}:
\begin{eqnarray}
\label{FloquetbasisTLS}
|u_{\pm}(t)\rangle =\frac{1}{\sqrt{2\Omega'}}
\left(\begin{array}{c}
 \pm \sqrt{\Omega'\pm\delta}\\
 e^{i\omega_L t}\sqrt{\Omega'\mp\delta},
\end{array}\right),
\end{eqnarray}
where $\Omega'=\sqrt{\delta ^2+\Omega^2}$ and the detuning parameter $\delta=\omega_0-\omega_L$. The corresponding quasi-energies of these two bases are
\begin{eqnarray}
\epsilon_\pm=\frac{\hbar}{2}(\omega_L\pm \Omega'),
\end{eqnarray}
respectively. Assume that $\omega_L-\Omega'>0$ and that the coupling between the TLS and heat reservoir is transverse~\cite{Breuer1997,Szczygielski2013,Langemeyer2014}, namely, $A=\sigma_x$ in the interaction Hamiltonian $V$. In this case, there are a total of six $A(\omega_t,t)$: Three of them are
\begin{eqnarray}
\label{Lindbladoperators}
&&A(\omega_L,t)=\frac{\Omega}{2\Omega'}\left(|u_+(t)\rangle\langle u_+(t)|-|u_-(t)\rangle\langle u_-(t)| \right)e^{-i\omega_L t},\nonumber\\
&&A(\omega_L-\Omega',t)=\left(\frac{\delta-\Omega'}{2\Omega'} \right)|u_+(t)\rangle\langle u_-(t)|e^{-i\omega_L t},\\
&&A(\omega_L+\Omega',t)=\left(\frac{\delta+\Omega'}{2\Omega'} \right) |u_-(t)\rangle\langle u_+(t)| e^{-i\omega_L t},\nonumber
\end{eqnarray}
and the other three operators are $A(\omega_t,t)$, where $\omega_t=-\omega_L$, $-(\omega_L-\Omega')$, and $-(\omega_L+\Omega')$ are the adjoint operators of Eq.~(\ref{Lindbladoperators}). Note that these $\omega_t$ are time independent. Assuming that the bath is composed of an electromagnetic field in thermal equilibrium at a certain temperature $T$ and that the coupling between the TLS and the reservoir is a dipole interaction, the Fourier transforms of the bath correlation function have the following standard form~\cite{Breuer2002,Szczygielski2013}: if $\omega<0$,
\begin{eqnarray}
r(\omega)=A|{\omega}|^3\frac{1}{e^{\hbar |\omega|/k_BT}-1};
\end{eqnarray}
otherwise, $\omega>0$, and using the KMS condition, $r(\omega)=e^{\hbar \omega/k_BT}r(-\omega)$, where the coefficient $A$ depends on the dipole strength.  At the initial time, we assume that the TLS is in a thermal state with temperature $T$.

Using the Monte-Carlo technique~\cite{Breuer2002}, we may easily realize the QJT simulation and construct histograms for the inclusive work after collecting adequate data. The simulation details are presented in Appendix F. Fig.~(\ref{figure4}) illustrates a distribution at a specific time under a given set of parameters.
\begin{figure}
\includegraphics[width=1.\columnwidth]{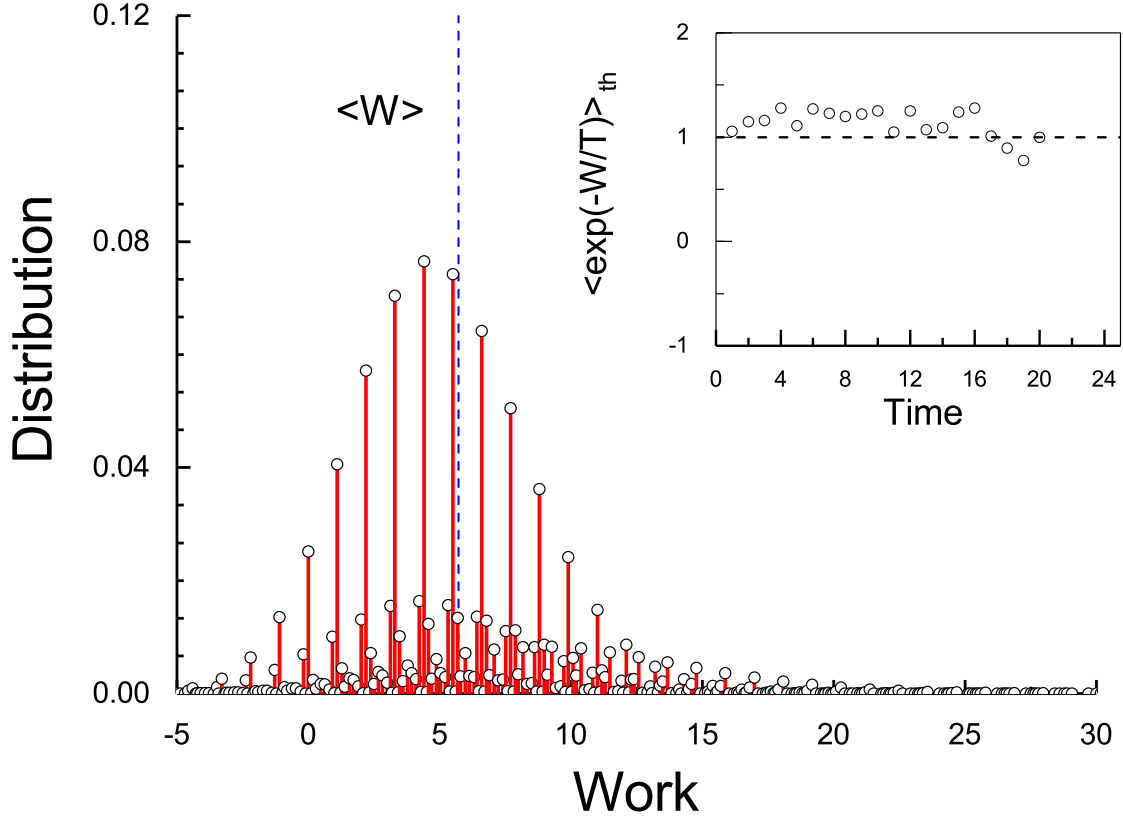}
\caption{The distribution of the inclusive work (in units of $\hbar\omega_0$) in a periodically driven TLS at time $t=10\omega_0^{-1}$. We calculated it by simulating the QJTs. The initial density matrix is set to be the thermal equilibrium state at temperature $T$. The utilized parameters are $\omega_L=1.1\omega_0$, $\Omega=0.8\omega_0$, ${\cal A}=1.0\omega_0^{-2}$, and $T=1.0 \hbar\omega_0/k_B$. For convenience, we have let $k_B=1$, $\hbar=1$, and $\omega_0=1$. The inset shows the results of the RHS of Eq.~(\ref{inclusiveworkFT}) calculated at different times through simulation. The dashed line therein is provided as visual guidance. We have used the same model and the same set of parameters to investigate the heat distribution and the heat equality, Eq.~(\ref{heatFT})~\cite{Liu2016}. }\label{figure4}
\end{figure}
We see that the distribution is complex, although the model seems to be a simple TLS. To verify the JE, we need to simulate a series of work distributions at different times and then calculate the LHS of Eq.~(\ref{inclusiveworkFT}); see the inset of the same figure. Note that the ratio of the partition functions of the RHS of the same equation is 1 and is independent of time.  We see that Eq.~(\ref{inclusiveworkFT}) is satisfactory, although there are apparent deviations at certain times. We attribute these deviations to statistical errors due to the finite number of QJTs. To show the consistency of the two methods in computing the inclusive work distributions, we calculate Eq.~(\ref{CFinclusiveworkopensystemgeneralbasedheat}) by numerically solving the time evolution equation~(\ref{Heatmasterequationgeneralform}), performing an inverse Fourier transform and comparing the obtained probability distribution to that obtained by direct simulation~\footnote{We may as well solve Eq.~(\ref{inclusiveworkmasterequationgeneralform}) for $K_A(t,\eta)$ and then obtain the CF, $\Phi_w(\eta)={\rm Tr}_A[K_A(t_f,\eta)]$. This is true in principle. However, we have to face an additional difficulty in the numerical realization of the term $
\partial_t[ e^{i\eta H_A(t)}]e^{-i\eta H_A(t)}$. We have met an analogous problem in the quantum piston model; see Fig.~(\ref{figure0}) }. However, considering that we had the probability distribution, it is obviously easier to make it a Fourier transform and to compare the result to the CF obtained by solving Eq.~(\ref{Heatmasterequationgeneralform}). We did this when we considered the inclusive work distribution in the quantum piston model; see Sec.~\ref{section3}. Fig.~(\ref{figure5}) shows the CFs obtained by these two routines, where the work distribution is that shown in Fig.~(\ref{figure4}).
\begin{figure}
\includegraphics[width=1.\columnwidth]{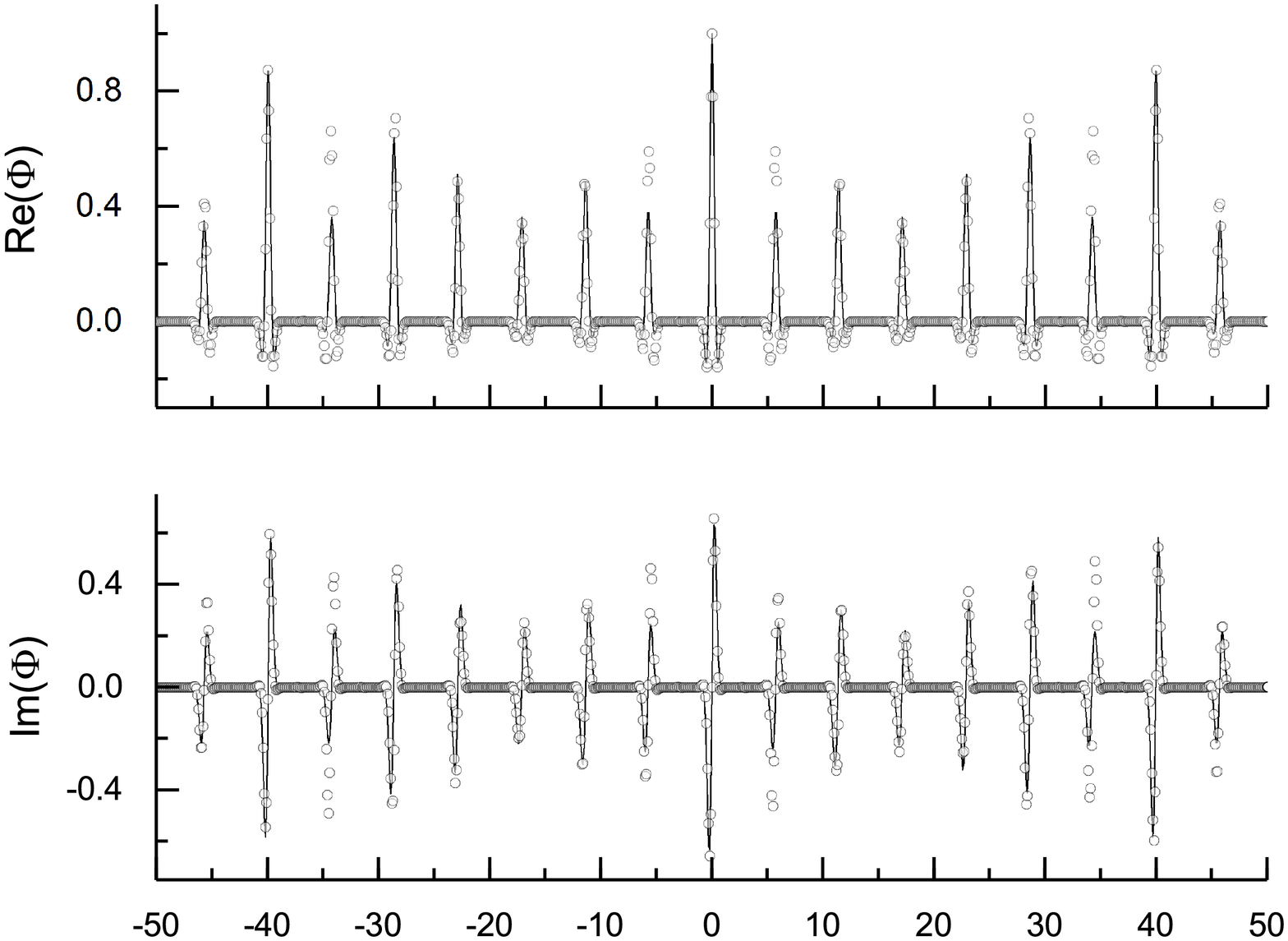}
\caption{The CFs calculated by simulating the QJT method (circles) and by solving the time evolution Eq.~(\ref{Heatmasterequationgeneralform}) (curves). The former is obtained by performing the Fourier transform on the work distribution shown in Fig.~(\ref{figure4}). The computational parameters are the same as those previously used. }\label{figure5}
\end{figure}
The computational details of solving the time evolution equation are also presented in Appendix F. We see that these two CFs are in excellent agreement with each other. Because in the current case the time evolution equation has an exact solution, the small differences between the two results shown in Fig.~(\ref{figure5}) are from the statistical errors of the QJT simulations.

\section{Discussion and conclusion}
\label{section12}

In this paper, we reviewed the stochastic heat and work in MQMEs and two strategies of studying them. One strategy is to treat the system and its surrounding heat bath as a closed quantum system; the evolution of the composite system is unitary under the control of a time-dependent total Hamiltonian, and the heat and work are defined as the changes in energy by the TEM scheme applied to the composite system. The other strategy is to unravel these MQMEs into QJTs, and the stochastic heat and work are defined along the individual trajectories. Our results clearly show that these two strategies lead to the same statistics for these thermodynamic quantities, although their definitions are based on these seemingly very distinct measurement approaches. It is useful to summarize the advantages of these two strategies. Although the first strategy is based on the idea of quantum master equations, because the physical picture of the composite system is always meaningful, this strategy can be extended in a straightforward manner to more complex situations, e.g., strong interaction between quantum systems and their environments and non-Markovian behaviors. The second strategy would be more attractive if we were given an MQME in advance. We have observed that a QJT has an intuitive physical explanation; its mathematics is  simple and easily simulated by computers, and in particular, it is more realistic experimentally.

Let us finish this paper by pointing out several extensions. The first extension is to study scenarios containing multiple reservoirs or particle transport. So far, we have only been concerned with one heat bath with a thermal temperature, and particle exchange is not allowed. Some studies have already considered this issue~\cite{Esposito2009,Szczygielski2013,Zinidarifmmodeheckclseci2014,Pigeon2015,Cuetara2015}. However, few studies have applied the concept of QJTs. The second possible extension is to account for non-Markovian effects~\cite{Breuer2004,Hush2015,Vega2017}. Non-Markovian properties usually lead to a failure of the MQME~(\ref{quantummasterequationgeneralformdensitymatrixcase}), on which we heavily rely. However, the extension of the state space of an open system may retrieve this key form~\cite{Breuer2004,Hush2015}. It would be interesting to determine what type of role the KMS condition plays in QST in this situation. Finally, the role of the many-body interactions of the quantum system in ST is a very intriguing issue. Some studies have shown that QJT could be very useful in studying this issue~\cite{Daley2014}.\\
\\
\\
{\noindent \it Acknowledgment.} This work was supported by the National Science Foundation of China under Grant Nos. 11174025 and 11575016. We also appreciate the support of the CAS Interdisciplinary Innovation Team, No. 2060299.


\section*{Appendix A: Feynman-Kac formulas}
In 1947, Kac~\cite{Kac1949} proposed the following question: Given a functional
\begin{eqnarray}
\label{functional}
Q\{z\}=\int_0^t V(z(\tau))d\tau
\end{eqnarray}
of a classical Brownian process $\{z\}$, because $Q$ is a random variable, how can one calculate its distribution $p(Q)$? Rather than on the distribution itself, Kac focused on its moment-generating
function (MGF)~\footnote{The moment-generating function is the Laplace transform of probability distributions, rather than the Fourier transform used in CF. However, if all  the moments of a stochastic quantity exist and are finite, there are no essential differences between two functions; a moment-generating function may simply be regarded as a CF evaluated on the imaginary axis~\cite{Gardiner1983}. If one first calculates the MGF, the distribution can be obtained by taking an inverse Laplace transform.}, or
\begin{eqnarray}
\label{Momentgeneratingfunction}
\Phi(\eta)=\int_0^\infty e^{-\eta Q} p(Q)dQ=\left \langle e^{-\eta\int_0^t V(z(\tau))d\tau}\right\rangle .
\end{eqnarray}
The last equation indicates that the average is taken over all Brownian stochastic trajectories, which possess probability weights.
He found that this MGF can be calculated by an integral of a function $G(z,t)$, namely,
\begin{eqnarray}
\Phi(\eta)=\int_{-\infty}^{+\infty} dz G(z,t,\eta)dz,
\end{eqnarray}
while the function $G$ satisfies a partial differential equation given by
\begin{eqnarray}
\label{FKevolutionequation}
\partial_t G(z,t,\mu)=D\partial_z^2 G(z,t,\mu)  - \mu V(z) G(z,t,\mu),
\end{eqnarray}
where $D$ is the diffusion constant. The above three equations are collectively known as the celebrated FK formula~\cite{Stroock2005}. The formula indicates that if we could solve Eq.~(\ref{FKevolutionequation}), after integrating the solution over $z$ and inverting the Laplace transform Eq.~(\ref{Momentgeneratingfunction}), we can obtain the distribution $p(Q)$.

FK has numerous applications in diverse fields~\cite{Majumdar2005}; it also has many extensions. One such extension that is relevant to us is that the Brownian diffusion can be replaced by a deterministic dynamics controlled by a Hamiltonian $H(z,t)$. In this situation, Eqs.~(\ref{functional})-(\ref{FKevolutionequation}) remain valid except that in the last equation, the diffusion operator, $D\partial_z^2$, is changed to the Poisson bracket, $\{H(z,t),$ $ \}_{PB}$. This is not unexpected because the Hamiltonian dynamics can be regarded as a special stochastic process. Hence, in Eq.~(\ref{timeevolutionequationclosedclassicalsystem}), according to the FK formula, we immediately find that the integral of its solution is equal to
\begin{eqnarray}
\label{CFclassicalinclusivework}
\int_{-\infty}^{+\infty} dz K(z,t,\eta)dz,=\left \langle e^{i\eta \int_0^t \partial_\tau H(z(\tau),\tau)d\tau}\right\rangle.
\end{eqnarray}
The average here is taken over all deterministic trajectories starting from a given initial distribution. In classical mechanics, the integral
\begin{eqnarray}
\label{classicalinclusivework}
W\{z\}=\int_0^t \partial_\tau H(z(\tau),\tau)d\tau
\end{eqnarray}
has an interpretation of the inclusive work along a deterministic trajectory in the phase space of the closed classical system~\cite{Jarzynski2007}. Therefore, we arrive at a conclusion that Eq.~(\ref{CFclassicalinclusivework}) is the CF for the classical inclusive work.

On the other hand, because Eq.~(\ref{timeevolutionequationclosedclassicalsystem}) can be thought of as a result of the quantum-classical correspondence of the operational Eq.~(\ref{timeevolutionequationclosedsysteminclusivework}), it is natural to call a collection of Eqs.~(\ref{CFinclusivework}),~(\ref{CFworkclosedsystem}), and~(\ref{timeevolutionequationclosedsysteminclusivework}) the quantum Fk formula~\cite{Liu2012,Liu2014b} of the inclusive work. Note that this name is not simply a literal meaning; it is also fully consistent with the spirit of Kac's original question: The quantum inclusive work definition, Eq.~(\ref{inclusiveworkdefinition}), is a functional of the evolution and measurements of the wave vector of the closed quantum system; the Fourier transform of its probability distribution $P(W)$ is equal to the trace of the WCO $K(t,\eta)$, i.e., Eq.~(\ref{CFworkclosedsystem}). Furthermore, $K(t,\eta)$ satisfies the time evolution equation~(\ref{timeevolutionequationclosedsysteminclusivework}). The reader is reminded that the LHS of Eq.~(\ref{CFclassicalinclusivework}) can also be thought of as the classical correspondence of Eq.~(\ref{CFworkclosedsystem})~\footnote{The quantum-classical correspondence issue is very complex and subtle~\cite{Schlosshauer2007,Polkovnikov2010}. Recently, the correspondence between quantum and classical work  attracted the attention of various authors~\cite{Jarzynski2015,Quan2012,Zhu2016,Wang2017a}. We have simply thought of Eq.~(\ref{timeevolutionequationclosedclassicalsystem}) as a  result of Eq.~(\ref{timeevolutionequationclosedsysteminclusivework}) when the Planck constant $\hbar$ tends to zero. However, even without the term $\cal W_\eta$, the correctness of this statement is model dependent~\cite{Ballentine2014}. Hence, the quantum-classical work correspondence from the perspective of the time evolution equation is an interesting topic worth serious study.}.

The quantum FK formula is of course not essentially limited to closed quantum systems. We easily see that what we discussed about the inclusive work in  open quantum systems or defined for QJTs is  still following the spirit of  Kac. Therefore, it is also appropriate to call a collection of Eqs.~(\ref{inclusiveworkdefinitionopensystem}) and~(\ref{inclusiveworkmasterequationgeneralform}) or a collection of Eqs.~(\ref{inclusiveworkdefinitionQJ}) and~(\ref{inclusiveworkmasterequationgeneralform}) the quantum Feynman-Kac formulas for the inclusive work. How they are different is that the ``trajectories" of the former are present in the Hilbert space of the composite system and heat bath, whereas for the latter, the QJTs are present in the Hilbert space of the quantum system alone.

\section*{Appendix B: Fourier transforms of correlations functions in Eq.~(\ref{timeevolutionequationheatopensystemorginteractionpictureprojectedexplicitstaticHamiltonian})}
The one-sided Fourier transform, the LHS of Eq.~(\ref{onesideFouriertransformcorrelationfunc}), can be replaced by the double-sided Fourier transform:
\begin{eqnarray}
&&\int_0^\infty ds e^{i\omega s} {\rm Tr}_B[{\widetilde B}_a(s)B_b\rho_B]={\cal F}\left [ {\rm Tr}_B[{\widetilde B}_a(s)B_b\rho_B]\Theta(s) \right]=\frac{1}{2\pi}{\cal F}[{\rm Tr}_B\left [{\widetilde B}_a(s)B_b\rho_B]\right]\star{\cal F}[\Theta(s)]]
\end{eqnarray}
where $\Theta(s)$ is the step function, ${\cal F}$ represents the Fourier transform, and the star $\star$ denotes the convolution operation. Noting that the first transform term is $r_{ab}(\omega)$, while the second transform term is
\begin{eqnarray}
{\cal F}[\Theta(s)]=P.V. \frac{1}{i\omega}+ \pi \delta(\omega).
\end{eqnarray}
We then obtain the RHS of Eq.~(\ref{onesideFouriertransformcorrelationfunc}). Using the same procedure, the other three integrals in Eq.~(\ref{timeevolutionequationheatopensystemorginteractionpictureprojectedexplicitstaticHamiltonian}) are as follows:
\begin{eqnarray}
\int_0^\infty ds e^{i\omega s}{\rm Tr}_B[\widetilde B_a(s-\eta)B_b\rho_B]&=&\frac{1}{2}r_{ab}(\omega) e^{\i\eta\omega} +i S_{ab}^{\eta}(\omega),\\
\int_0^\infty ds e^{i\omega s}{\rm Tr}_B[\widetilde B_b(-s)B_a\rho_B]&=&\frac{1}{2}r_{ab}(-\omega)^* - i S_{ab}(-\omega)^*, \\
\int_0^\infty ds e^{i\omega s}{\rm Tr}_B[\widetilde B_b(-s-\eta)B_a\rho_B]&=&\frac{1}{2}r_{ab}(-\omega)^*e^{-i\eta\omega} - i {S_{ab}^{-\eta}}(-\omega)^*.
\end{eqnarray}
For simplicity, here, we define
\begin{eqnarray}
S_{ab}^\eta(\omega)=\frac{1}{2\pi i}P.V. \int_{-\infty}^\infty d\omega' \frac{r_{ab}(\omega')}{\omega-\omega'}e^{i\omega'\lambda},
\end{eqnarray}
It is interesting to see that these functions $S_{ab}^\eta(\omega)$ will be exactly canceled in these time evolution equations due to the following identity:
\begin{eqnarray}
S_{ab}^{\eta}(\omega)={S_{ba}^{-\eta}(\omega)}^*.
\end{eqnarray}

\section*{Appendix C: Proof of Eq.~(\ref{TimereversalCorrelationfuncts})}
According to the definition, Eq.~(\ref{FouriertransformCorrlationfunctions}), the conjugation of $r_{ab}(\omega)$ is
\begin{eqnarray}
r_{ab}^*(\omega)&=&\int_{-\infty}^{\infty} du e^{-i\omega u}({\rm Tr_B}[\widetilde B_a(u)B_b\rho_B])^*
=\int_{-\infty}^{\infty}du  e^{-i\omega u} {\rm Tr_B}[\Theta \widetilde B_a(u)B_b\rho_B\Theta^{-1}]\nonumber\\
&=&\delta_a\delta_b\int_{-\infty}^{\infty} du e^{-i\omega u} {\rm Tr_B}[{\widetilde {B}}_a(-u){B}_b\rho_B ]
=\delta_a\delta_b\int_{-\infty}^{\infty} e^{i\omega u} {\rm Tr_B}[{\widetilde {B}}_a(u){B}_b\rho_B]\nonumber\\
&=&\delta_a\delta_b r_{ab}(\omega).
\end{eqnarray}
It was known early on that the matrix $[r_{ab}(\omega)]$ is Hermitian~\cite{Breuer2002}. Obviously, if $B_a$ and $B_b$ are of different time-reversal types, $r_{ab}(\omega)$ is purely imaginary; otherwise, it is real. Using the same argument, we can also prove that
\begin{eqnarray}
S_{ab}^*(\omega)=\delta_a\delta_b S_{ab}(\omega).
\end{eqnarray}

To illustrate Eq.~(\ref{TimereversalCorrelationfuncts}), we present a simple example about a two-level system damped by a bath of harmonic oscillators~\cite{Rivas2012,Carmichael1993}. The total Hamiltonian is
\begin{eqnarray}
H=H_A+H_B+V=\frac{\omega_0}{2}\sigma_z +\int_0^{\omega_{max}}d\omega a_\omega^\dag a_\omega + \int_0^{\omega_{max}}d\omega'h(\omega')(\sigma_+ a_{\omega'}+\sigma_- a^\dag_{\omega'}).
\end{eqnarray}
where $a_{\omega'}$ is the continuous bosonic operator and $\omega'$ is the frequency of each mode. The last integral term, the interaction $V$, can be rewritten as
\begin{eqnarray}
V=A_1\otimes B_1 + A_2\otimes B_2=\sigma_x\otimes \int_0^{\omega_{max}}d\omega'\frac{(a_{\omega'}+a^\dag_{\omega'})}{2} + \sigma_y\otimes \int_0^{\omega_{max}}d\omega'\frac{i(a_{\omega'}-a^\dag_{\omega'})}{2}.
\end{eqnarray}
Obviously, the spectral decompositions of $A_1$ and $A_2$ are
\begin{eqnarray}
A_1&=&A_1(-\omega_0)+A_1(\omega_0)=\sigma_+ +\sigma_-,\\
A_2&=&A_2(-\omega_0)+A_2(\omega_0)=-i\sigma_+ + i \sigma_-.
\end{eqnarray}
The reader is reminded that these Pauli matrices are not actual spin operators. Hence, the time reversals of these $A_a$ and $B_a$ ($a=1,2$) operators are
\begin{eqnarray}
\label{timereversalconcreteAB}
\Theta A_1 \Theta^{-1}=A_1,\hspace{0.5cm}\Theta A_2 \Theta^{-1} =-A_2\\
\Theta B_1 \Theta^{-1}=B_1,\hspace{0.5cm}\Theta B_2 \Theta^{-1} =-B_2,
\end{eqnarray}
respectively. This also indicates that the interaction Hamiltonian $V$ is TRI. According to Eqs.~(\ref{TimereversalCorrelationfuncts}) and~(\ref{timereversalconcreteAB}), we can predict that $r_{11}$ and $r_{22}$ are real and that $r_{12}$ and $r_{21}$ are purely imaginary for the distinct Bohr frequencies $-\omega_0$ and $\omega_0$. One can verify this by explicitly calculating these coefficients:
\begin{eqnarray}
&&\begin{pmatrix}
  r_{11} & r_{12} \\
  r_{21} & r_{22} \\
\end{pmatrix}
(-\omega_0)=\frac{\pi}{2}J(\omega_0)\bar{n}(\omega_0)
\begin{pmatrix}
  1 & i \\
  -i & 1 \\
\end{pmatrix},\\
&&\begin{pmatrix}
  r_{11} & r_{12} \\
  r_{21}& r_{22} \\
\end{pmatrix}
(\omega_0) =\frac{\pi}{2}J(\omega_0)[1+\bar{n}(\omega_0)]
\begin{pmatrix}
  1 & i \\
  -i & 1 \\
\end{pmatrix},
\end{eqnarray}
where $J(\omega_0)=h(\omega_0)^2$ and $\bar{n}(\omega_0)=1/[\exp(\beta\omega_0)-1]$.

\section*{Appendix D: time reversal in the MQMEs}
For the static and weakly driven Hamiltonian, the Bohr frequency $\omega$ is constant. According to our definition of backward processes, we easily see that the Bohr frequency and operators $\overline A_a(\overline \omega)^\dag$ for the backward process are the same as those of the forward process, namely,
\begin{eqnarray}
\overline \omega&=&\omega.\\
\overline A_a^\dag (\overline \omega)&=&A_a^\dag(\omega).
\end{eqnarray}
On the other hand, according to the definition of $A^\dag_a(\omega)$, Eq.~(\ref{LoperatorsstaticHamiltoniancase}), its time reversal is
\begin{eqnarray}
\label{timereversalAkdag}
\Theta A_a^\dag(\omega) \Theta^{-1}&=&\Theta \sum_{n,m } \delta_{\omega,\varepsilon_n- \varepsilon_m} \langle \varepsilon_n|A_a|\varepsilon_m\rangle   |\varepsilon_n\rangle \langle  \varepsilon_m|\Theta^{-1}\nonumber \\
&=&\sum_{n,m } \delta_{\omega,\varepsilon_n- \varepsilon_m} \langle\varepsilon_n|A_a|\varepsilon_m\rangle^*   \Theta |\varepsilon_n\rangle \langle  \varepsilon_m|\Theta^{-1}\nonumber\\
&=&\delta_a\sum_{n,m } \delta_{\omega,\varepsilon_n-\varepsilon_m} \langle\varepsilon_n|\Theta^{-1}{A}_a \Theta|\varepsilon_m\rangle \Theta |\varepsilon_n\rangle \langle  \varepsilon_m|\Theta^{-1}.
\end{eqnarray}
Considering that the Hamiltonian $H_A$ is TRI and that its eigenvectors are non-degenerate~\cite{Sakurai1994}, we have
\begin{eqnarray}
\Theta|\varepsilon_m\rangle=|\varepsilon_m\rangle.
\end{eqnarray}
Hence, the last equation is simply equal to $\delta_a{A}_a^\dag(\omega)$. In  other words, we proved Eqs.~(\ref{forwardbackwardidentity1}) and.~(\ref{forwardbackwardidentity2}) for the static and weakly driven Hamiltonian cases.

Now, let us check the adiabatically driven Hamiltonian case. Because the Hamiltonian of the backward process is Eq.~(\ref{systemHamiltonianbackwardprocess}), according to Eq.~(\ref{LoperatorsAdiabaticallydrivencase}), the Bohr frequency and explicit expression of $\overline A_a^\dag(\overline \omega_s)$ are
\begin{eqnarray}
\overline \omega(s)&=&\omega(t_f-s)=\omega(t),\\
{\overline A}_a^\dag({\overline \omega}_s)&=&A_a^\dag(\omega_{t_f-s})=A_a^\dag(\omega_{t}),
\end{eqnarray}
respectively. This is because we have assumed that the Hamiltonian $H_A(t)$ depends on $t$ only through the protocols $\lambda(t)$; see Eq.~(\ref{closedsystemHamiltonian}). Analogous to the static Hamiltonian case, according to the definition of $A_a^\dag(\omega_t)$,  Eq.~(\ref{LoperatorsAdiabaticallydrivencase}), its time reversal is
\begin{eqnarray}
\Theta A_a^\dag (\omega_t)\Theta^{-1}&=&\Theta \sum_{n,m } \delta_{\omega,\varepsilon_n(t)- \varepsilon_m(t)} \langle \varepsilon_n(t)|A_a|\varepsilon_m(t)\rangle   |\varepsilon_n(t)\rangle \langle  \varepsilon_m(t)|\Theta^{-1},\nonumber\\
&=&\delta_aA_a^\dag(\omega_t).
\end{eqnarray}
The proof is obvious because it is completely analogous to Eq.~(\ref{timereversalAkdag}); the additional time parameter $t$ here does not affect the action of the time reversal. Hence, we confirmed the second equation, Eq.~(\ref{forwardbackwardidentity2}), for the adiabatically driven case.

The last case is about the periodically driven Hamiltonian~(\ref{perodicallydrivenHamiltonian}). This is slightly involved because the Floquet basis~(\ref{Floquetbasis}) is time dependent. We first note that the Hamiltonian of the backward process,
\begin{eqnarray}
\overline H_A(s)=H_A(t_f-s),
\end{eqnarray}
is also periodic,  with the same periodicity $\Omega$. Then, its Floquet basis can be constructed using the Floquet basis of the forward process:
\begin{eqnarray}
\label{connectionFloquetbasisFBprocesses}
|{\overline\epsilon}_n(s)\rangle =\Theta |\epsilon_n(t_f-s)\rangle=\Theta |\epsilon_n(t)\rangle .
\end{eqnarray}
The quasi-energy remains the same as that of the forward process, $\epsilon_n$. Let us emphasize that these quasi-energies are constant. The results can be directly confirmed using the definition of the Floquet basis, Eq.~(\ref{Floquetbasis}). Because the Floquet basis is degenerate, we cannot expect that the RHS of Eq.~(\ref{connectionFloquetbasisFBprocesses}) is simply $|\epsilon_n(t)\rangle $, as in previous cases. Because the Bohr frequency $\omega$ is constant, we of course have $\overline\omega=\omega$. According to Eq.~(\ref{Loperatorspreodicallydrivencase}), the explicit expression of ${\overline A}_a^\dag({\overline \omega}_s,s)$ is
\begin{eqnarray}
\label{BackwardprocessLoperatorperiodicallydrivencase}
{\overline A}_a^\dag({\overline \omega},s) =\sum_{n,m } \delta_{\omega,\epsilon_n- \epsilon_m+q\Omega} \langle\langle \overline\epsilon_{n,q}|A_a|\overline\epsilon_m\rangle \rangle  |\overline\epsilon_n(s)\rangle \langle  \overline\epsilon_m(s)| e^{iq\Omega s},
\end{eqnarray}
where
\begin{eqnarray}
\label{backwardcoeff}
\langle \langle\overline \epsilon_{n,q}|A_a|\overline\epsilon_m\rangle\rangle =\frac{\Omega}{2\pi}\int_0^{2\pi/\Omega} du e^{-iq\Omega u} \langle\overline\epsilon_n(u)|A_k|\overline\epsilon_m(u)\rangle.
\end{eqnarray}
The reader is reminded that the time parameter $s$ is present here because the Floquet basis is usually no longer only a function of the protocol, $\overline\lambda(s)$. Now, we are in position to check Eq.~(\ref{forwardbackwardidentity2}) for periodically driven Hamiltonians. Recalling $t+s=t_f$, we have
\begin{eqnarray}
\label{timereversalAkdagperiodicallydrivencase}
\Theta A_a^\dag (\omega,t)\Theta^{-1}&=&\sum_{n,m } \delta_{\omega,\epsilon_n- \epsilon_m+q\Omega} \langle\langle \epsilon_{n,q}|A_a|\epsilon_m\rangle \rangle^*\Theta  |\epsilon_n(t)\rangle \langle  \epsilon_m(t)|\Theta^{-1}e^{-iq\Omega t} \nonumber\\
&=&\sum_{n,m } \delta_{\omega,\epsilon_n- \epsilon_m+q\Omega} \langle\langle \epsilon_{n,q}|A_a|\epsilon_m\rangle \rangle^*e^{-iq\Omega t_f} |\overline\epsilon_n(s)\rangle \langle \overline \epsilon_m(s)|e^{iq\Omega s}.
\end{eqnarray}
The multiplication of the first two terms after the Kronecker delta function is
\begin{eqnarray}
&&\frac{\Omega}{2\pi}\int_0^{2\pi/\Omega} du e^{iq\Omega (u-t_f)} \langle\epsilon_n(u)|A_a|\epsilon_m(u)\rangle^*
=\delta_a\frac{\Omega}{2\pi}\int_0^{2\pi/\Omega } du e^{-iq\Omega u} \langle\overline\epsilon_n(u)|A_a|\overline\epsilon_m(u)\rangle.
\end{eqnarray}
To obtain the equation, we have performed a change of variables $u-t_f\rightarrow-u$ and used the periodic property of the Floquet basis. The integral term above is simply Eq.~(\ref{backwardcoeff}). Hence, we arrive at
\begin{eqnarray}
\Theta A_a^\dag (\omega,t)\Theta^{-1}=\delta_a{\overline A}_a^\dag({\overline \omega},s).
\end{eqnarray}

\section*{Appendix E: equivalence of Eqs.~(\ref{probQJvector}) and~(\ref{probQJdensitymatrix}) }
First, we write the deterministic part of Eq.~(\ref{stochasticdifferentialequationdensitymatrix}) in the Schr$\ddot{o}$dinger picture:
\begin{eqnarray}
\label{deterministicequationsimpleinteractionHamiltonian}
\partial_t\sigma_A=- i [H_A+H_{LS},
 \sigma_A] -\frac{1}{2} \sum_{\omega}r(\omega)\left\{{A}^\dag(\omega){A}(\omega),\sigma_A \right\}
 + \sigma_A  \sum_{\omega}r(\omega){\rm Tr}_A[ {A}^\dag(\omega){A}(\omega)\sigma_A].
\end{eqnarray}
The density matrix $\sigma_A(t)$ is obviously normalized, but the equation is non-linear. Noticing that its last sum term is simply the instantaneous total jump rate, Eq.~(\ref{alljumpsprobabilitygivenpsi}), we suppose that
\begin{eqnarray}
\sigma_A(t)= e^{\int_{t'}^tds\Gamma(s)}\pi_A(t).
\end{eqnarray}
Inserting this into Eq.~(\ref{deterministicequationsimpleinteractionHamiltonian}), we immediately obtain the linear evolution equation for $\pi_A(t)$, Eq.~(\ref{linearquantummastersimplecase}). On the other hand, if we substitute the above equation into Eq.~(\ref{probQJdensitymatrix}), we have
\begin{eqnarray}
\label{probQJdensitymatrixttoQJvector}
P\{\sigma_A\}&=&{\rm Tr}_A\left[ G_0(t,t_{N}) J(\omega_N)G_0(t_N,t_{N-1})\cdots J(\omega_1) e^{\int_0^{t_1} \Gamma(s_1)ds_1} G_0(t_1,0)\sigma_A(0)\right] e^{-\int_0^{t_1} \Gamma(s_1)ds_1} \prod_{i=1}^N\Delta t_i \nonumber\\
&=&{\rm Tr}_A\left[ G_0(t,t_{N}) J(\omega_N)G_0(t_N,t_{N-1})\cdots J(\omega_1)\sigma_A(t_1)\right] e^{-\int_0^{t_1} \Gamma(s_1)ds_1} \prod_{i=1}^N\Delta t_i \nonumber \\
&=&{\rm Tr}_A\left[ G_0(t,t_{N}) J(\omega_N)G_0(t_N,t_{N-1})\cdots \frac{J(\omega_1)}{{\rm Tr}_A[  {A}(\omega_1)\sigma_A(t_1){A}^\dag(\omega_1) ]}\sigma_A(t_1)\right] \\
&&{{\rm Tr}_A[  {A}(\omega_1)\sigma_A(t_1){A}^\dag(\omega_1) ]} e^{-\int_0^{t_1} \Gamma(s_1)ds_1} \prod_{i=1}^N\Delta t_i \nonumber\\
&=&{\rm Tr}_A\left[ G_0(t,t_{N}) J(\omega_N)G_0(t_N,t_{N-1})\cdots \sigma_A'(t_1)\right]P_{\omega_1}(t_1)e^{-\int_0^{t_1} \Gamma(s_1)ds_1} \prod_{i=2}^N\Delta t_i \nonumber\\
&=&\cdots\cdots\nonumber\\
&=&{\rm Tr}_A\left[ \sigma_A(t)\right]e^{-\int_{t_{N}}^{t} \Gamma(s_{N})ds_{N} } P_{\omega_N}(t_N) e^{-\int_{t_{N-1}}^{t_N} \Gamma(s_{N-1})ds_{N-1}} \cdots P_{\omega_1}(t_1) e^{-\int_0^{t_1} \Gamma(s_1)ds_1}. \nonumber
\end{eqnarray}
We note that $\sigma'_A(t_1)$ is the normalized density matrix at time $t_1$, which is from the normalized density matrix $\sigma_A(t_1)$ with the jump type $\omega_{1}$. The above analysis can be continued, and thus, we skip the intermediate steps up to the last step. Because at time $t$ the system's density matrix is normalized, we proved the equivalence between Eqs.~(\ref{probQJvector}) and ~(\ref{probQJdensitymatrix}).
An analogous argument is also applicable to the case of complex interaction Hamiltonians.

On the other hand, in the case of the simple interaction Hamiltonian, Eq.~(\ref{probQJvector}) has the other expression if the system's initial state is pure, e.g., $|\psi(0)\rangle \langle \psi(0)|$. Then, we have
\begin{eqnarray}
\label{probQJvector2}
P[\{\psi(t)\}]= \left | U_0(t,t_N) A(\omega_N)U_0(t_N,t_{N-1})\cdots A(\omega_1)U_0(t_1,0)|\psi(0)\rangle\right |^2 \prod_{i=1}^N r(\omega_i),
\end{eqnarray}
where $U_0(t,t')$ is the effective time-evolution operator of the following equation:
\begin{eqnarray}
\label{unnormalizedwavevectortimeevolutionequation}
\partial_t |\phi(t)\rangle=-i{\hat H}|\phi(t)\rangle=\left[-i (H_A+H_{LS})+\frac{1}{2} \sum_{\omega}r(\omega) {A}^\dag(\omega){A}(\omega) \right]|\phi(t)\rangle,
\end{eqnarray}
namely,
\begin{eqnarray}\label{solutionlinearwavevectorequationsimplecase}
U_0(t,t')=e^{-i{\hat H} (t-t')}.
\end{eqnarray}
The reader is reminded that $|\phi(t)\rangle$ is unnormalized. There are two ways to prove this equivalence. The first way is to note that the propagator $G_0$, Eq.~(\ref{solutionlineardensitymatrixequationsimplecase}), and the time evolution operator $U_0$, Eq.~(\ref{solutionlinearwavevectorequationsimplecase}) satisfy the following relation:
\begin{eqnarray}
G_0(t,t')O=U_0(t,t')O U_0(t,t')^\dag.
\end{eqnarray}
This structure is the same as that of $J(\omega)$, Eq.~(\ref{jumppartsimplecase}). Hence, Eq.~(\ref{probQJvector2}) is  an alternative expression of Eq.~(\ref{probQJdensitymatrix}). Therefore, it is not strange that Eq.~(\ref{probQJvector}) is equal to Eq.~(\ref{probQJvector2}). The other way is to note that the solution of the deterministic equation in Eq.~(\ref{stochasticdifferentialequationvector}) in the Schr$\ddot{o}$dinger picture is related to the solution of Eq.~(\ref{unnormalizedwavevectortimeevolutionequation}) by
\begin{eqnarray}
|\psi(t)\rangle_A=e^{\int_{t'}^t ds \Gamma(s)}|\phi(t)\rangle.
\end{eqnarray}
Substituting this result into Eq.~(\ref{probQJvector2}), applying the same procedure as Eq.~(\ref{probQJdensitymatrixttoQJvector}), and noting the definition in Eq.~(\ref{onejumpprobabilitygivenpsi}), we may directly prove the equivalence of Eqs.~(\ref{probQJvector}) and~(\ref{probQJvector2}) without relying on Eq.~(\ref{probQJdensitymatrix}). Let us again emphasize that Eq.~(\ref{probQJvector2}) is valid only under the restrictions of a pure initial state of the system and a simple interaction Hamiltonian.

\section*{Appendix F: Computational details of the periodically driven TLS}
For the reader's convenience, we give some computational details about the TLS model. Although most of the details have been given in our previous paper~\cite{Liu2016}, because we are concerned with the inclusive work here, some modifications are needed.

\subsection{Directly simulating QJTs}
Because the Floquet basis, Eq.~(\ref{FloquetbasisTLS}), is complete and orthogonal, it shall be convenient to write operators and wave vectors in this time-dependent basis. For instance, the $A(\omega,t)$ operators in Eq.~(\ref{Lindbladoperators}) are as follows:
\begin{eqnarray}
A(\omega_L,t)&\doteq&\frac{\Omega}{2\Omega'}\sigma_z(t),\nonumber\\
A(\omega_L-\Omega',t)&\doteq&\left(\frac{\delta-\Omega'}{2\Omega'}\right)\sigma_+(t),\nonumber\\
A(\omega_L+\Omega',t)&\doteq&\left(\frac{\delta+\Omega'}{2\Omega'}\right)\sigma_-(t).
\end{eqnarray}
The other three operators $A(\omega,t)$ with $\omega=-\omega_L$, $-(\omega_L-\Omega')$, and $-(\omega_L+\Omega')$ are their adjoint operators. Note that to indicate that these Pauli matrices are not the conventional Pauli matrices, we add time parameters after these symbols.

According to our previous discussion, a QJT is a time evolution of a wave vector $\Psi(t)$ in the Hilbert space of the TLS. This evolution is alternatively composed of a deterministic continuous evolution and stochastic jumps; see Eq.~(\ref{stochasticdifferentialequationvector}). Assuming that the continuous process starts from time $t$ and ends at time $t+\tau$, during this process, its deterministic equation is
\begin{eqnarray}
\label{wavevectorevolvingeq}
\frac{d}{ds}\Psi(t+s)=-\frac{i}{\hbar} \left[H(t+s)-\frac{i\hbar}{2}\sum_{\omega_{t+s}}r(\omega_{t+s}) A^\dag(\omega_{t+s},t+s)A(\omega_{t+s},t+s)\right]\Psi(t+s),
\end{eqnarray}
$0\le s\le \tau$. We expand the wave vector in the Floquet basis, that is,
\begin{eqnarray}
\label{wavevectorunnormalized}
\Psi(t+s)=\mu_+(s)e^{-i(t+s)\epsilon_+/\hbar}|u_+(t+s)\rangle +\mu_-(s)e^{-i(t+s)\epsilon_-/\hbar}|u_-(t+s)\rangle.
\end{eqnarray}
Substituting this into Eq.~(\ref{wavevectorevolvingeq}), we obtain
\begin{eqnarray}
\label{componentsequation1}
  \frac{d\mu_\pm}{ds}&=&-\frac{1}{\tau_\pm}\mu_\pm,
\end{eqnarray}
where the coefficients are
\begin{eqnarray}
\frac{1}{\tau_\pm}=\frac{1}{2}\left[\left(\frac{\Omega}{2\Omega'}\right)^2\left(
    r(\omega_L) +
    r(-\omega_L) \right)+r(\mp(\omega_L-\Omega'))\left(\frac{\delta-\Omega'}{2\Omega'}\right)^2
    +r(\pm(\omega_L+\Omega'))\left(\frac{\delta+\Omega'}{2\Omega'}\right)^2\right],
\end{eqnarray}
respectively. Eqs.~(\ref{componentsequation1}) have simple solutions:
\begin{eqnarray}
\mu_\pm(s)&=&\mu_\pm(0)e^{-s/\tau_\pm}.
\end{eqnarray}
$\Psi(t+s)$ is not normalized. We denote the one $\overline{\Psi}(t+s)$, which is the same as Eq.~(\ref{wavevectorunnormalized}) except that $\mu_\pm(s)$ therein are replaced by $\overline{\mu}_\pm(s)=\mu_\pm(s)/\sqrt{\|\mu_+(s)\|^2+\|\mu_-(s)\|^2} $. We can determine the time duration $\tau$ by solving the following equation:
\begin{eqnarray}
\eta =\|\Psi(t+\tau)\|^2= \|\mu_+(0)\|^2\exp\left(-\frac{2\tau}{\tau_+}\right)+\|\mu_-(0)\|^2\exp\left(-\frac{2\tau}{\tau_-}\right),
\end{eqnarray}
where $\eta\in(0,1)$ is a uniform random number.

This smooth evolution is interrupted by a jump at time $t+\tau$. The wave vector after the jump is
\begin{eqnarray}
A(\omega_{t+\tau},t+\tau)\Psi(t+\tau)/\| A(\omega_{t+\tau},t+\tau)\Psi(t+\tau)\|.
\end{eqnarray}
The probabilities of these jumps are proportional to
\begin{eqnarray}
r(\omega_{t+\tau})\|A(\omega_{t+\tau},t+\tau)\Psi(t+\tau)\|^2.
\end{eqnarray}
We list these six vectors in the following table:
\begin{eqnarray}
\begin{array}{ccccccccccccccccc}
\hline
  \hbox{State after a jump} && \hbox{Probabilities }\propto && \hbox{Heat produced }\\
  \hline
  \overline{\Psi}'(t+\tau)  && \gamma(\omega_L)({\Omega}/{2\Omega'})^2 && \hbar\omega_L\\
  |u_+(t+\tau)\rangle \hspace{0.5cm}({\rm if}\hspace{0.2cm} \mu_-\neq0)&& r(\omega_L-\Omega')(\|\mu_-(\tau)\|(\delta-\Omega')/{2\Omega'})^2&& \hbar(\omega_L-\Omega')\\
  |u_-(t+\tau)\rangle \hspace{0.5cm}({\rm if}\hspace{0.2cm}  \mu_+\neq0)&& r(\omega_L+\Omega')((\|\mu_+(\tau)\|\delta+\Omega')/{2\Omega'})^2 &&\hbar(\omega_L+\Omega')\\
  \overline{\Psi}'(t+\tau)  && r(-\omega_L)({\Omega}/{2\Omega'})^2 &&-\hbar\omega_L\\
  |u_-(t+\tau)\rangle \hspace{0.5cm}({\rm if}\hspace{0.2cm}  \mu_+\neq0)&& r(-(\omega_L-\Omega'))(\|\mu_+(\tau)\|(\delta-\Omega')/{2\Omega'})^2 && -\hbar(\omega_L-\Omega')\\
  |u_+(t+\tau)\rangle \hspace{0.5cm}({\rm if}\hspace{0.2cm}  \mu_-\neq0)&&  r(-(\omega_L+\Omega'))(\|\mu_-(\tau)\|(\delta+\Omega')/{2\Omega'})^2 &&-\hbar(\omega_L+\Omega')\\
  \hline
  \end{array}
   \end{eqnarray}
where
\begin{eqnarray}
\overline{\Psi}'(t+\tau)=\bar{\mu}_+(\tau)e^{-i(t+\tau)\epsilon_+/\hbar}|u_+(t+\tau)\rangle-\bar{\mu}_-(\tau)e^{-i(t+\tau)\epsilon_-/\hbar}|u_-(t+\tau)\rangle. \end{eqnarray}
After a vector is randomly chosen from the above, new rounds with continuous evolution and stochastic jump start until the end time is obtained. Finally, because we are concerned with the inclusive work, the initial wave vector is one of the energy eigenvectors for $H_A(0)$.

\subsection{Solving Eq.~(\ref{Heatmasterequationgeneralform}) }
In the Floquet basis, we may expand the HCO $\hat{{\rho}}_A$ as follows:
\begin{eqnarray}
\hat{\rho}_A(t,\eta)=\frac{p_+(t)+p_-(t)}{2}I+\frac{p_+(t)-p_-(t)}{2}\sigma_z(t)+p_1(t)\sigma_+(t)+p_2(t)\sigma_-(t),
\end{eqnarray}
where $p_{\pm}$ and $p_{n}$, $n=1,2$, are the diagonal and non-diagonal elements of this density matrix in this basis, respectively. Substituting the expansion into Eq.~(\ref{Heatmasterequationgeneralform}) and doing simple algebra, we obtain the time-evolution equations for $\hat{p}_\pm(t)$ and $p_{n}$:
\begin{eqnarray}
\frac{dp_\pm}{dt}&=&\left[(e^{i\xi\omega_L}-1)r(\omega_L)\left(\frac{\Omega}{2\Omega'}\right)^2 +    (e^{-i\xi\omega_L}-1)r(-\omega_L)\left(\frac{\Omega}{2\Omega'}\right)^2\right. \nonumber\\
&&-\left. r(\mp(\omega_L-\Omega'))\left(\frac{\delta-\Omega'}{2\Omega'}\right)^2-
r(\pm(\omega_L+\Omega'))\left(\frac{\delta+\Omega'}{2\Omega'}\right)^2\right]p_\pm\nonumber\\
&&+\left[e^{\pm i\xi(\omega_L-\Omega')}r(\pm(\omega_L-\Omega'))\left(\frac{\delta-\Omega'}{2\Omega'}\right)^2
    +e^{\mp i\xi(\omega_L+\Omega')}r(\mp(\omega_L+\Omega'))\left(\frac{\delta+\Omega'}{2\Omega'}\right)^2
\right]p_\mp,\\
\frac{dp_n}{dt}&=&-\left[(e^{i\eta\omega_L}+1)r(\omega_L)\left(\frac{\Omega}{2\Omega'}\right)^2 +    (e^{-i\eta\omega_L}+1)r(-\omega_L)\left(\frac{\Omega}{2\Omega'}\right)^2\right. \nonumber\\
&&+\frac{r(\omega_L-\Omega')}{2}\left(\frac{\delta-\Omega'}{2\Omega'}\right)^2+\frac{
r(-(\omega_L-\Omega'))}{2}\left(\frac{\delta-\Omega'}{2\Omega'}\right)^2 \nonumber\\
&&\left. +\frac{r(\omega_L+\Omega')}{2}\left(\frac{\delta+\Omega'}{2\Omega'}\right)^2\frac{1}{2}
   + \frac{r(-(\omega_L+\Omega'))}{2}\left(\frac{\delta+\Omega'}{2\Omega'}  \right)^2+(-1)^n\frac{i}{\hbar}\left(\epsilon_--\epsilon_+\right)l\right]p_n,
\end{eqnarray}
Because we want to calculate the inclusive work CF, Eq.~(\ref{CFinclusiveworkopensystemgeneralbasedheat}), the initial conditions are $e^{i\eta H_A(0)}\rho_A(0)$ and $\rho_A(0)=e^{-\beta H_A(0)}/Z(0)$.
Although these equations are a bit long and appear complicated, they are simply linear equations with constant coefficients. Hence, we can obtain their exact solutions.

\end{document}